\documentclass[11pt,a4paper]{article}
\usepackage[ngerman,english]{babel}
\pdfoutput=1 

\usepackage{jinstpub} 
\usepackage{bm}
\usepackage[usenames,dvipsnames]{xcolor}
\usepackage{comment}
\usepackage[version=4]{mhchem}
\usepackage{siunitx}
\usepackage{amssymb}
\usepackage{lineno}
\usepackage{tabularx}
\usepackage{array}
\usepackage{booktabs}
\usepackage[font={small},labelfont=bf]{caption}
\usepackage[font={small}]{subcaption}
\usepackage{afterpage}
\usepackage{tikz}
\usetikzlibrary{patterns}

\defineshorthand{"~}{\babelhyphen{nobreak}}
\useshorthands{"}

\DeclareSIUnit\permille{\text{\textperthousand}}

\title{Flow and thermal modelling of the argon volume in the DarkSide-20k TPC}

\collaboration{\includegraphics[height=17mm]{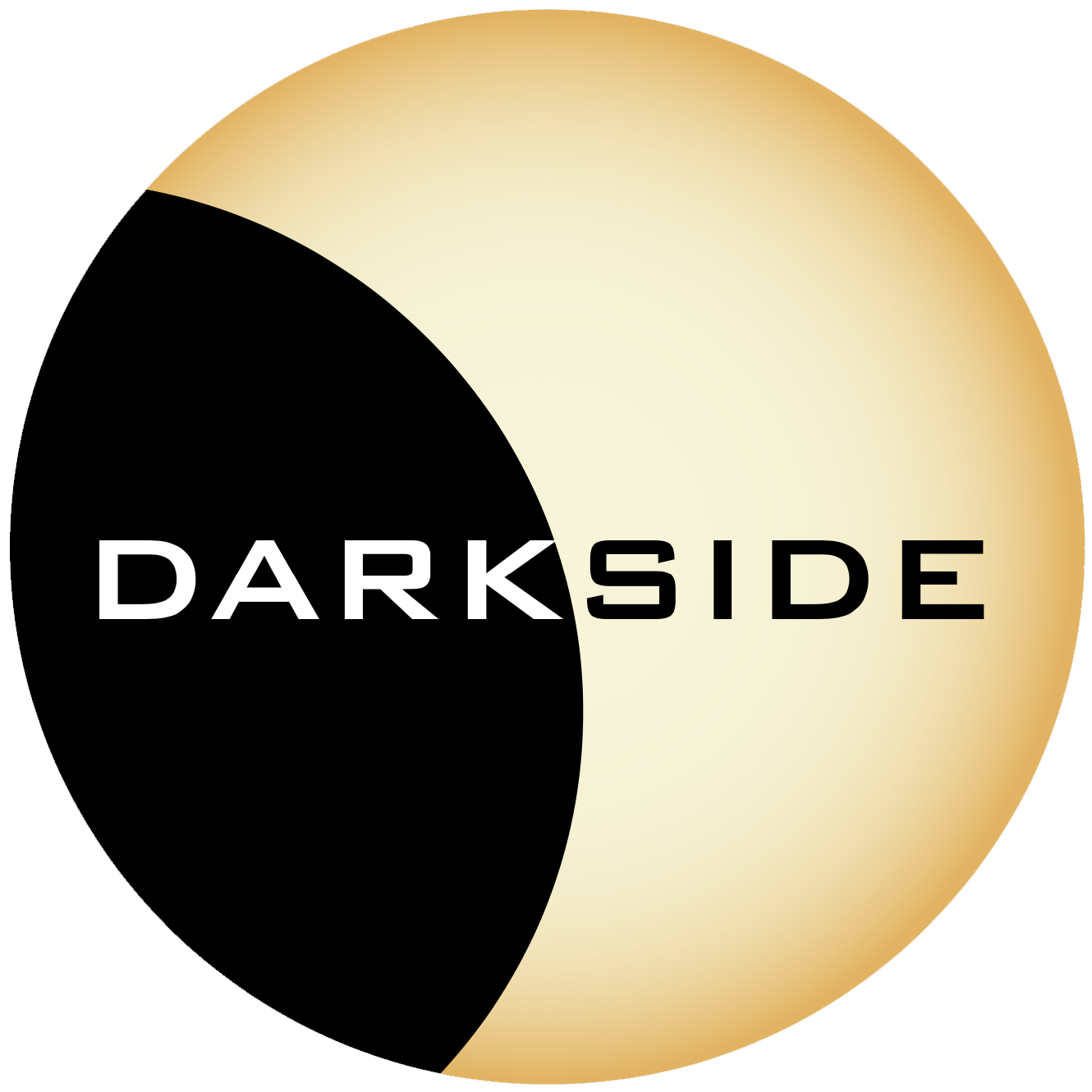}\\[6pt]
The DarkSide-20k Collaboration}
\newcommand{\Alberta}{Department of Physics, University of Alberta, Edmonton, AB T6G 2R3, Canada}
\newcommand{\APC}{APC, Universit\'e de Paris, CNRS, Astroparticule et Cosmologie, Paris F-75013, France}
\newcommand{\AQLNGS}{INFN Laboratori Nazionali del Gran Sasso, Assergi (AQ) 67100, Italy}
\newcommand{\AQGSSI}{Gran Sasso Science Institute, L'Aquila 67100, Italy}

\newcommand{\AstroCeNT}{AstroCeNT, Nicolaus Copernicus Astronomical Center of the Polish Academy of Sciences, 00-614 Warsaw, Poland}
\newcommand{\Augustana}{Physics Department, Augustana University, Sioux Falls, SD 57197, USA}
\newcommand{\Belgorod}{Radiation Physics Laboratory, Belgorod National Research University, Belgorod 308007, Russia}

\newcommand{\BINP}{Budker Institute of Nuclear Physics, Novosibirsk 630090, Russia}
\newcommand{\Birmingham}{School of Physics and Astronomy, University of Birmingham, Edgbaston, B15 2TT, Birmingham, UK}

\newcommand{\BOINFN}{INFN Bologna, Bologna 40126, Italy}
\newcommand{\BOUniPHY}{Department of Physics and Astronomy, Universit\`a degli Studi di Bologna, Bologna 40126, Italy}
\newcommand{\CAUniCHE}{Department of Mechanical, Chemical, and Materials Engineering, Universit\`a degli Studi, Cagliari 09042, Italy}
\newcommand{\CAUniEEE}{Department of Electrical and Electronic Engineering, Universit\`a degli Studi di Cagliari, Cagliari 09123, Italy}
\newcommand{\CAUniPHY}{Physics Department, Universit\`a degli Studi di Cagliari, Cagliari 09042, Italy}
\newcommand{\CAINFN}{INFN Cagliari, Cagliari 09042, Italy}
\newcommand{\Carleton}{Department of Physics, Carleton University, Ottawa, ON K1S 5B6, Canada}

\newcommand{\Columbia}{Physics Department, Columbia University, New York, NY 10027, USA}
\newcommand{\Chicago}{Department of Physics and Kavli Institute for Cosmological Physics, University of Chicago, Chicago, IL 60637, USA}
\newcommand{\BOCentroFermi}{Museo Storico della Fisica e Centro Studi e Ricerche Enrico Fermi, Roma 00184, Italy}

\newcommand{\CIEMAT}{CIEMAT, Centro de Investigaciones Energ\'eticas, Medioambientales y Tecnol\'ogicas, Madrid 28040, Spain}

\newcommand{\CPPM}{Centre de Physique des Particules de Marseille, Aix Marseille Univ, CNRS/IN2P3, CPPM, Marseille, France}
\newcommand{\CTINFN}{INFN Catania, Catania 95121, Italy}
\newcommand{\CTUNI}{Universit\`a of Catania, Catania 95124, Italy}
\newcommand{\CTLNS}{INFN Laboratori Nazionali del Sud, Catania 95123, Italy}

\newcommand{\ENUniCEE}{Engineering and Architecture Department, Universit\`a di Enna Kore, Enna 94100, Italy}
\newcommand{\ETHZ}{Institute for Particle Physics and Astrophysics, ETH Z\"urich, Z\"urich 8093, Switzerland}

\newcommand{\FortLewis}{Department of Physics and Engineering, Fort Lewis College, Durango, CO 81301, USA}
\newcommand{\GEUni}{Physics Department, Universit\`a degli Studi di Genova, Genova 16146, Italy}
\newcommand{\GEINFN}{INFN Genova, Genova 16146, Italy}
\newcommand{\Hawaii}{Department of Physics and Astronomy, University of Hawai'i, Honolulu, HI 96822, USA}
\newcommand{\Houston}{Department of Physics, University of Houston, Houston, TX 77204, USA}
\newcommand{\IHEP}{Institute of High Energy Physics, Chinese Academy of Sciences, Beijing 100049, China}
\newcommand{\INFN}{Istituto Nazionale di Fisica Nucleare, Roma 00186, Italia}

\newcommand{\JINR}{Joint Institute for Nuclear Research, Dubna 141980, Russia}
\newcommand{\Krakow}{M.~Smoluchowski Institute of Physics, Jagiellonian University, 30-348 Krakow, Poland}
\newcommand{\Kurchatov}{National Research Centre Kurchatov Institute, Moscow 123182, Russia}
\newcommand{\Laurentian}{Department of Physics and Astronomy, Laurentian University, Sudbury, ON P3E 2C6, Canada}
\newcommand{\Lancaster}{Physics Department, Lancaster University, Lancaster LA1 4YB, UK}
\newcommand{\Liverpool}{Department of Physics, University of Liverpool, The Oliver Lodge Laboratory, Liverpool L69 7ZE, UK}

\newcommand{\LNLINFN}{INFN Laboratori Nazionali di Legnaro, Legnaro (Padova) 35020, Italy}
\newcommand{\Lodz}{Institute of Applied Radiation Chemistry, Lodz University of Technology, 93-590 Lodz, Poland}

\newcommand{\Manchester}{Department of Physics and Astronomy, The University of Manchester, Manchester M13 9PL, UK}
\newcommand{\MEPhI}{National Research Nuclear University MEPhI, Moscow 115409, Russia}
\newcommand{\MendeleevUniverisity}{Mendeleev University of Chemical Technology, Moscow 125047, Russia}

\newcommand{\MIINFN}{INFN Milano, Milano 20133, Italy}
\newcommand{\MIPoliICA}{Civil and Environmental Engineering Department, Politecnico di Milano, Milano 20133, Italy}

\newcommand{\MIUni}{Physics Department, Universit\`a degli Studi di Milano, Milano 20133, Italy}
\newcommand{\MSU}{Skobeltsyn Institute of Nuclear Physics, Lomonosov Moscow State University, Moscow 119234, Russia}
\newcommand{\NAINFN}{INFN Napoli, Napoli 80126, Italy}
\newcommand{\NAUniPHY}{Physics Department, Universit\`a degli Studi ``Federico II'' di Napoli, Napoli 80126, Italy}
\newcommand{\NAUniCHE}{Chemical, Materials, and Industrial Production Engineering Department, Universit\`a degli Studi ``Federico II'' di Napoli, Napoli 80126, Italy}
\newcommand{\NAUniDIST}{Department of Structures for Engineering and Architecture, Universit\`a degli Studi ``Federico II'' di Napoli, Napoli 80126, Italy}
\newcommand{\NAUniPHARM}{Pharmacy Department, Universit\`a degli Studi ``Federico II'' di Napoli, Napoli 80131, Italy}

\newcommand{\McDonald}{Arthur B.~McDonald Canadian Astroparticle Physics Research Institute, Queen's University, Kingston, ON K7L 3N6, Canada}

\newcommand{\Oxford}{University of Oxford, Oxford OX1 2JD, United Kingdom}
\newcommand{\Petersburg}{Saint Petersburg Nuclear Physics Institute, Gatchina 188350, Russia}

\newcommand{\PIINFN}{INFN Pisa, Pisa 56127, Italy}
\newcommand{\PIUniPHY}{Physics Department, Universit\`a degli Studi di Pisa, Pisa 56127, Italy}
\newcommand{\PNNL}{Pacific Northwest National Laboratory, Richland, WA 99352, USA}
\newcommand{\Princeton}{Physics Department, Princeton University, Princeton, NJ 08544, USA}
\newcommand{\Queens}{Department of Physics, Engineering Physics and Astronomy, Queen's University, Kingston, ON K7L 3N6, Canada}
\newcommand{\RHUL}{Department of Physics, Royal Holloway University of London, Egham TW20 0EX, UK}
\newcommand{\RMTreINFN}{INFN Roma Tre, Roma 00146, Italy}
\newcommand{\RMTreUni}{Department of Mathematics and Physics – Roma Tre University, Roma 00146, Italy}
\newcommand{\RMUnoINFN}{INFN Sezione di Roma, Roma 00185, Italy}
\newcommand{\RMUnoUni}{Physics Department, Sapienza Universit\`a di Roma, Roma 00185, Italy}

\newcommand{\SNL}{Savannah River National Laboratory, Jackson, SC 29831, United States}

\newcommand{\SNOLAB}{SNOLAB, Lively, ON P3Y 1N2, Canada}

\newcommand{\STFCInterconnect}{Science \& Technology Facilities Council (STFC), Rutherford Appleton Laboratory, Technology, Harwell Oxford, Didcot OX11 0QX, UK}
\newcommand{\STFCppd}{Science \& Technology Facilities Council (STFC), Rutherford Appleton Laboratory, Particle Physics Department, Harwell Oxford, Didcot OX11 0QX, UK}

\newcommand{\Temple}{Physics Department, Temple University, Philadelphia, PA 19122, USA}
\newcommand{\TNFBK}{Fondazione Bruno Kessler, Povo 38123, Italy}

\newcommand{\TOINFN}{INFN Torino, Torino 10125, Italy}
\newcommand{\TOPoli}{Department of Electronics and Telecommunications, Politecnico di Torino, Torino 10129, Italy}

\newcommand{\TRIUMF}{TRIUMF, 4004 Wesbrook Mall, Vancouver, BC V6T 2A3, Canada}

\newcommand{\UCDavis}{Department of Physics, University of California, Davis, CA 95616, USA}
\newcommand{\UCRiverside}{Department of Physics and Astronomy, University of California, Riverside, CA 92507, USA}

\newcommand{\UCLA}{Physics and Astronomy Department, University of California, Los Angeles, CA 90095, USA}
\newcommand{\UCAS}{University of Chinese Academy of Sciences, Beijing 100049, China}
\newcommand{\UMass}{Amherst Center for Fundamental Interactions and Physics Department, University of Massachusetts, Amherst, MA 01003, USA}

\newcommand{\UniHAM}{Institute of Experimental Physics, University of Hamburg, 22761 Hamburg, Germany}
\newcommand{\UnivAQ}{Universit\`a degli Studi dell’Aquila, L’Aquila 67100, Italy}
\newcommand{\UniversityofEdinburgh}{School of Physics and Astronomy, University of Edinburgh, Edinburgh EH9 3FD, UK}

\newcommand{\USP}{Instituto de F\'isica, Universidade de S\~ao Paulo, S\~ao Paulo 05508-090, Brazil}
\newcommand{\VTech}{Virginia Tech, Blacksburg, VA 24061, USA}
\newcommand{\Warwick}{University of Warwick, Department of Physics, Coventry CV47AL, UK}
\newcommand{\Washington}{Center for Experimental Nuclear Physics and Astrophysics, and Department of Physics, University of Washington, Seattle, WA 98195, USA}
\newcommand{\WUT}{Institute of Radioelectronics and Multimedia Technology, Faculty of Electronics and Information Technology, Warsaw University of Technology, 00-661 Warsaw, Poland}
\newcommand{\WilliamsCollege}{Williams College, Department of Physics and Astronomy, Williamstown, MA 01267 USA}
\newcommand{\Zaragoza}{Centro de Astropart\'iculas y F\'isica de Altas Energ\'ias, Universidad de Zaragoza, Zaragoza 50009, Spain}

\author[1]{F.~Acerbi,}\affiliation[1]{\TNFBK}
\author[2]{P.~Adhikari,}\affiliation[2]{\Carleton}
\author[3,4]{P.~Agnes,}\affiliation[3]{\AQGSSI}\affiliation[4]{\AQLNGS}
\author[5]{I.~Ahmad,}\affiliation[5]{\AstroCeNT}
\author[6,7]{S.~Albergo,}\affiliation[6]{\CTUNI}\affiliation[7]{\CTINFN}
\author[8]{I.~F.~Albuquerque,}\affiliation[8]{\USP}
\author[9]{T.~Alexander,}\affiliation[9]{\PNNL}
\author[10]{A.~K.~Alton,}\affiliation[10]{\Augustana}
\author[11]{P.~Amaudruz,}\affiliation[11]{\TRIUMF}
\author[3,4]{M.~Angiolilli,}
\author[12]{E.~Aprile,}\affiliation[12]{\Columbia}
% ...
\emailAdd{ds-ed@lists.infn.it}

\abstract{The DarkSide-20k dark matter experiment, currently under construction at LNGS, features a dual-phase time projection chamber~(TPC) with a $\SI{\sim 50}{t}$ argon target from an underground well. At this scale, it is crucial to optimise the argon flow pattern for efficient target purification and for fast distribution of internal gaseous calibration sources with lifetimes of the order of hours. To this end, we have performed computational fluid dynamics simulations and heat transfer calculations. The residence time distribution shows that the detector is well-mixed on time-scales of the turnover time ($\SI{\sim 40}{\day}$). Notably, simulations show that despite a two-order-of-magnitude difference between the turnover time and the half-life of \ce{^{83\text{m}}Kr} of $\SI{1.83}{h}$, source atoms have the highest probability to reach the centre of the TPC $\SI{13}{min}$ after their injection, allowing for a homogeneous distribution before undergoing radioactive decay. We further analyse the thermal aspects of dual-phase operation and define the requirements for the formation of a stable gas pocket on top of the liquid. We find a best-estimate value for the heat transfer rate at the liquid-gas interface of $\SI{62}{W}$ with an upper limit of $\SI{144}{W}$ and a minimum gas pocket inlet temperature of $\SI{89}{K}$ to avoid condensation on the acrylic anode. This study also informs the placement of liquid inlets and outlets in the TPC. The presented techniques are widely applicable to other large-scale, noble-liquid detectors.}


\arxivnumber{2503.08468} 

\begin{document}

\maketitle
\flushbottom

\section{Introduction}
\label{Sec:Intro}
Liquid argon and xenon are used as target media in time projection chambers~(TPCs) to search for rare event interactions in astroparticle and neutrino physics~\cite{Majumdar:2021llu,EXO-200:2019rkq}. TPCs with a gas layer on top of the liquid are the leading detector technology in the quest for weakly interacting massive particles~(WIMPs), a prominent dark matter candidate~\cite{McDonald:2024osu,Baudis:2023pzu}. Such dual-phase TPCs allow for the reconstruction of the interaction energy and position, and for background discrimination based on the properties of a prompt and a delayed scintillation signal~(S1 and S2). The S1 signal is composed of the direct scintillation of the noble gas atoms in the liquid phase and of the light from recombining electrons and ions. The S2 signal is electroluminescence light from gas collisions of non-recombined ionisation electrons that are drifted and extracted into the gas phase by vertical electric fields. Both S1 and S2 signals are observed by photosensors, typically arranged in arrays at the bottom and top of the TPC.

The handling of the target fluid is an integral part of the cryogenic design considerations for such detectors and vital to their operation. The cryogenic fluid does not remain quiescent in the bulk reservoir of the detector. Heat loads in the system, such as those from the photosensor electronics, induce convective currents and evaporated gas is recondensed elsewhere to maintain a constant pressure. In addition, the continuous removal of electronegative contaminants is required, as their presence degrades the S2 signal by capturing the drifting electrons and affects the shape and size of the S1 signal through light absorption. The purification can either be performed in gas or in liquid phase but requires a closed-loop flow~\cite{Cennini:1993abz,Plante:2022khm,Vogl:2023hbg}. For low-energy searches on the order of $\SIrange{1}{100}{keV}$, radiogenic backgrounds from \ce{^{222}Rn} and \ce{^{85}Kr} are mitigated through the use of distillation columns during detector operation~\cite{XENON100:2017gsw,XENON:2021fkt,Murra:2022mlr}. Adsorbent purifiers with activated charcoal can be used to trap radon~\cite{ABE201250,Pushkin:2018wdl} and filters are deployed to reduce the concentration of particulate contamination. These purification, filtering or distillation processes necessitate a turnover of the target fluid in a closed recirculation loop that involves extracting, processing and returning the cryogenic fluid which possibly undergoes intermediate phase changes. Thereby, heat transfer and fluid motion are induced within the target volume, with the latter being superimposed by convective flows.    
Modelling the liquid flow pattern and heat transfers in the design process is crucial to ensure the achievement of stable operating conditions. A flow analysis aids in particular the understanding of the impact of heat loads onto the system through convection, and guides the placement and sizing of supply inlets. It furthermore guides the optimisation of flow-related design aspects of the TPC for good background mitigation and high signal yield by providing an assessment of the turnover efficiency. Turnover efficiency is the ability to remove impurities from all detector regions in a recirculation process on timescales of a few turnover times. The turnover time is the quotient of fluid mass and mass flow rate.
Background sources include dust particulates that are unavoidably introduced during assembly, even if carefully suppressed, and that migrate into the TPC, as well as \ce{^{222}Rn} atoms that are transported and attached to TPC surfaces. The latter can cause, due to electric field non-uniformities close to the vertical TPC boundaries, events with partially or completely lost S2~signal that can get misidentified or accidentally paired with uncorrelated S2~signals. The XENON1T and LUX-ZEPLIN~(LZ) experiments exploit coincident $\alpha$"~decays of \ce{^{222}Rn} and \ce{^{218}Po} in regions of laminar liquid flow to tag radon and discriminate $\beta$"~events from \ce{^{214}Pb}~\cite{XENON:2024lbh,LZCollaboration:2024lux}. In LZ, this is achieved through the injection of sub-cooled liquid xenon at the bottom of the TPC, yielding reduced convection. Conversely, liquid xenon slightly above the saturation temperature -- corresponding to the gas pressure -- is injected at the bottom to obtain spatially homogeneous events from distributed calibration sources through convective mixing\footnote{Due to heat transfer between supplied and returned argon, the temperature of the LAr injected into the DS"~20k TPC is fixed by the ullage pressure and cannot be modified independently~\cite{DarkSide-20k:UArCryo}.}. Simulations of the flow field in the DUNE far detector further aim at predicting the LAr purity in regions where no direct purity measurement is available, and at modelling ion movement~\cite{DUNE:2020txw,Tu:2023nyz}.
A model of the flow pattern lastly allows for the evaluation of the mixing of internally distributed calibration sources with the target fluid, ensuring efficient deployment within their lifetime. These aspects become increasingly important as detectors grow in size, soon utilising several tens of tonnes of target fluid~\cite{McDonald:2024osu,Aalbers:2022dzr,XLZD:2024nsu}.

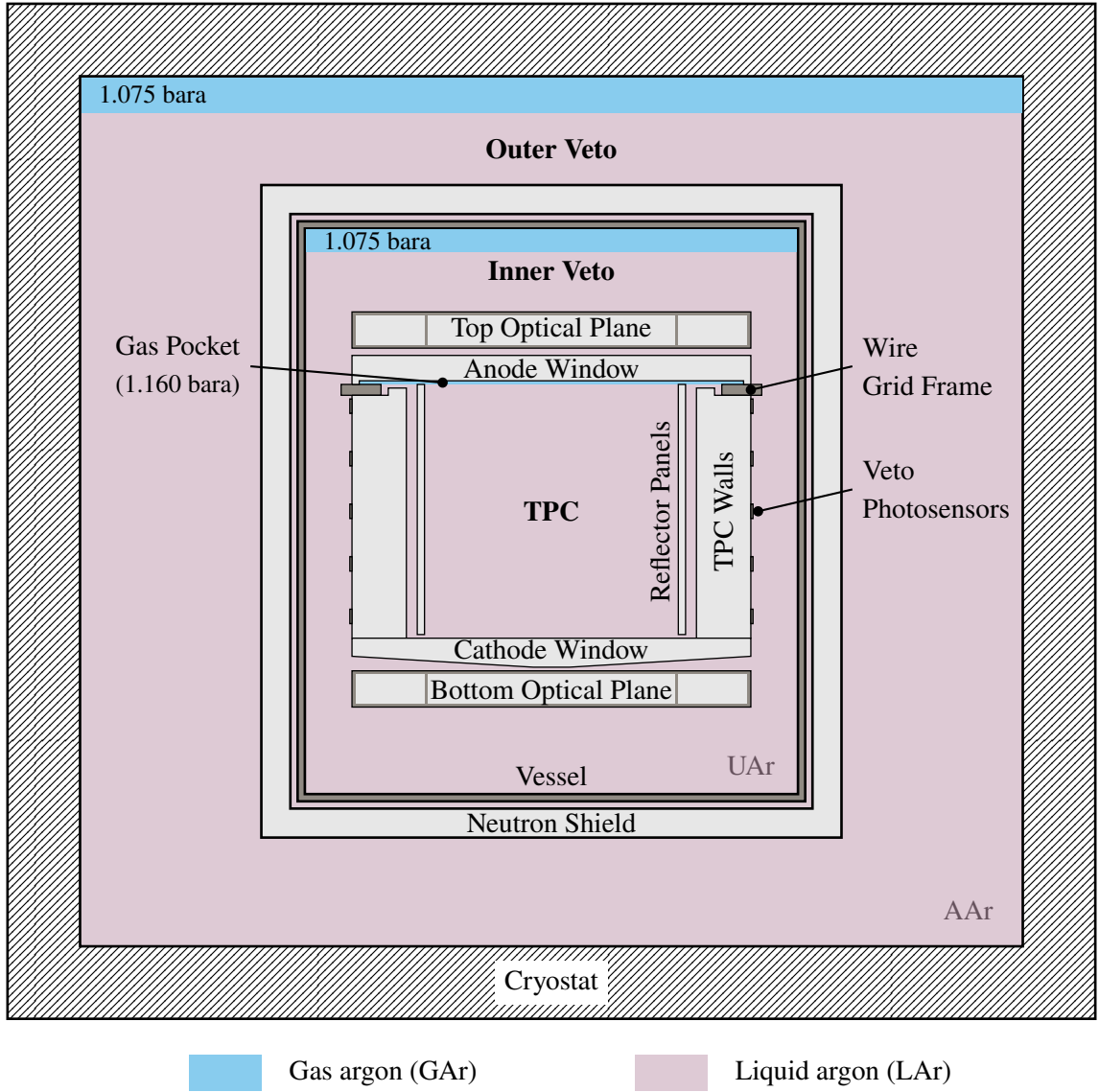
\begin{figure}[t]
\centering
\begin{tikzpicture}
\tikzstyle{every node}=[font=\normalsize]

\draw[pattern=north east lines, pattern color={rgb,255:red,0; green,0; blue,0}, line width=1pt] (0,0) rectangle (15,-14);
\draw [fill={rgb,255:red,222; green,202; blue,213}, line width=1pt] (1,-1) rectangle (14,-13);

\fill [fill={rgb,255:red,136; green,204; blue,238}] (1.02,-1.02) rectangle (13.98,-1.5);

\draw [fill={rgb,255:red,231; green,231; blue,231}, line width=1pt] (3.5,-2.5) rectangle (11.5,-11.5);
\draw [fill={rgb,255:red,222; green,202; blue,213}, line width=1pt] (3.9,-2.9) rectangle (11.1,-11.1);

\draw [fill={rgb,255:red,144; green,138; blue,131}, line width=1pt] (4,-3) rectangle (11,-11);
\draw [fill={rgb,255:red,222; green,202; blue,213}, line width=1pt] (4.1,-3.1) rectangle (10.9,-10.9); 

\fill [fill={rgb,255:red,136; green,204; blue,238}] (4.12,-3.12) rectangle (10.88,-3.425);

\fill [fill={rgb,255:red,136; green,204; blue,238}] (4.85,-5.2) rectangle (10.15,-5.25);

\draw [fill={rgb,255:red,144; green,138; blue,131}, line width=0.5pt ] (4.6,-5.25) rectangle (5.15,-5.4);
\draw [fill={rgb,255:red,144; green,138; blue,131}, line width=0.5pt ] (10.4,-5.25) rectangle (9.85,-5.4);

\draw [fill={rgb,255:red,231; green,231; blue,231}, line width=0.5pt] (4.75,-5.4) -- (4.75,-8.75) -- (5.5,-8.75) -- (5.5,-5.3) -- (5.25,-5.3)  -- (5.25,-5.4) -- (4.75,-5.4);
\draw [fill={rgb,255:red,231; green,231; blue,231},line width=0.5pt] (10.25,-5.4) -- (10.25,-8.75) -- (9.5,-8.75) -- (9.5,-5.3) -- (9.75,-5.3)  -- (9.75,-5.4) -- (10.25,-5.4);

\draw [fill={rgb,255:red,144; green,138; blue,131}, line width=0.5pt ] (4.75,-4.75) rectangle (10.25,-4.25);
\draw [fill={rgb,255:red,144; green,138; blue,131}, line width=0.5pt ] (4.75,-9.2) rectangle (10.25,-9.7);

\fill [fill={rgb,255:red,231; green,231; blue,231}] (4.8,-4.7) rectangle (5.75,-4.3);
\fill [fill={rgb,255:red,231; green,231; blue,231}] (5.8,-4.7) rectangle (9.2,-4.3);
\fill [fill={rgb,255:red,231; green,231; blue,231}] (9.25,-4.7) rectangle (10.2,-4.3);

\fill [fill={rgb,255:red,231; green,231; blue,231}] (4.8,-9.25) rectangle (5.75,-9.65);
\fill [fill={rgb,255:red,231; green,231; blue,231}] (5.8,-9.25) rectangle (9.2,-9.65);
\fill [fill={rgb,255:red,231; green,231; blue,231}] (9.25,-9.25) rectangle (10.2,-9.65);

\draw [fill={rgb,255:red,231; green,231; blue,231}, line width=0.5pt ] (5.65,-5.25) rectangle (5.75,-8.7);
\draw [fill={rgb,255:red,231; green,231; blue,231}, line width=0.5pt ] (9.25,-5.25) rectangle (9.35,-8.7);

\draw [fill={rgb,255:red,231; green,231; blue,231},line width=0.5pt] (4.75,-8.75) -- (4.75,-9.0) -- (7.25,-9.15) -- (7.75,-9.15)  -- (10.25,-9.0) -- (10.25,-8.75) -- (4.75,-8.75);

\draw [fill={rgb,255:red,231; green,231; blue,231},line width=0.5pt] (4.75,-5.25) -- (4.75,-4.85) -- (10.25,-4.85) -- (10.25,-5.25)  -- (10.15,-5.25) -- (10.15,-5.2) -- (4.85,-5.2) -- (4.85,-5.25) -- (4.75,-5.25);

\draw [fill={rgb,255:red,144; green,138; blue,131}, line width=0.5pt ] (4.72,-5.65) rectangle (4.75,-5.45);
\draw [fill={rgb,255:red,144; green,138; blue,131}, line width=0.5pt ] (4.72,-6.175) rectangle (4.75,-6.375);
\draw [fill={rgb,255:red,144; green,138; blue,131}, line width=0.5pt ] (4.72,-6.9) rectangle (4.75,-7.1);
\draw [fill={rgb,255:red,144; green,138; blue,131}, line width=0.5pt ] (4.72,-7.625) rectangle (4.75,-7.825);
\draw [fill={rgb,255:red,144; green,138; blue,131}, line width=0.5pt ] (4.72,-8.35) rectangle (4.75,-8.55);
\draw [fill={rgb,255:red,144; green,138; blue,131}, line width=0.5pt ] (10.25,-5.65) rectangle (10.28,-5.45);
\draw [fill={rgb,255:red,144; green,138; blue,131}, line width=0.5pt ] (10.25,-6.175) rectangle (10.28,-6.375);
\draw [fill={rgb,255:red,144; green,138; blue,131}, line width=0.5pt ] (10.25,-6.9) rectangle (10.28,-7.1);
\draw [fill={rgb,255:red,144; green,138; blue,131}, line width=0.5pt ] (10.25,-7.625) rectangle (10.28,-7.825);
\draw [fill={rgb,255:red,144; green,138; blue,131}, line width=0.5pt ] (10.25,-8.35) rectangle (10.28,-8.55);

\node [font=\normalsize, fill={rgb,255:red,255; green,255; blue,255}] at (7.5,-13.5) {Cryostat};
\node [font=\normalsize] at (7.5,-10.65) {Vessel};
\node [font=\normalsize] at (7.5,-11.3) {Neutron Shield};
\node [font=\normalsize] at (7.5,-9.5) {Bottom Optical Plane};
\node [font=\normalsize] at (7.5,-4.5) {Top Optical Plane};
\draw[-, thick,align=left] (10.25,-5.325) -- (11.65,-5) node[right] {Wire \\ Grid Frame};
\fill (10.25,-5.325) circle (2pt);
\draw[-, thick,align=left] (10.35,-7.0) -- (11.65,-6.698) node[right] {Veto \\ Photosensors};
\fill (10.35,-7.0) circle (2pt);
\draw[-, thick,align=right] (6,-5.225) -- (3.35,-5) node[left] {Gas Pocket \\ \small (1.160 bara)};
\fill (6,-5.225) circle (2pt);
\node [font=\normalsize, rotate around={90:(0,0)}] at (9,-7) {Reflector Panels};
\node [font=\normalsize, rotate around={90:(0,0)}] at (9.875,-7) {TPC Walls};
\node [font=\normalsize] at (7.5,-8.91) {Cathode Window};
\node [font=\normalsize] at (7.5,-5.025) {Anode Window};
\node [font=\normalsize,color={rgb,255:red,102; green,82; blue,93}] at (13.25,-12.5) {AAr};
\node [font=\normalsize,color={rgb,255:red,102; green,82; blue,93}] at (10.25,-10.5) {UAr};
\node [font=\normalsize] at (7.5,-7) {\textbf{TPC}};
\node [font=\normalsize] at (7.5,-2) {\textbf{Outer Veto}};
\node [font=\normalsize] at (7.5,-3.675) {\textbf{Inner Veto}};
\node [font=\small] at (2.0,-1.25) {1.075 bara};
\node [font=\small] at (5.1,-3.263) {1.075 bara};

\fill [fill={rgb,255:red,136; green,204; blue,238}] (2.5,-14.5) rectangle (3.5,-15);
\node [font=\normalsize,anchor=west] at (3.75,-14.75) {Gas argon~(GAr)};
\fill [fill={rgb,255:red,222; green,202; blue,213}] (8.65,-14.5) rectangle (9.65,-15);
\node [font=\normalsize,anchor=west] at (9.9,-14.75) {Liquid argon~(LAr)};

\end{tikzpicture}

\caption{Schematic view of the DS"~20k experiment (not to scale).}
\label{fig:DS-20k_overview}
\end{figure}

DarkSide"~20k~(DS"~20k) is a next-generation WIMP-search experiment~\cite{DarkSide-20k:2017zyg}, which is currently under construction underground
at the Laboratori Nazionali del Gran Sasso~(LNGS) in Italy~\cite{Zani:2024ybb}. A schematic overview of its nested design is shown in figure~\ref{fig:DS-20k_overview}. The inner detector of DS"~20k is contained inside a stainless steel vessel and composed of a dual-phase TPC, with the shape of a right octagonal prism, surrounded by an inner neutron veto. The active TPC volume (diameter of inscribed circle $\times$ height $= \SI{3.5}{m} \times \SI{3.5}{m}$) is defined by the reflector panels on the sides, the cathode window at the bottom and the anode window at the top. Like the TPC walls, these components are made from acrylic (poly(methyl methacrylate), PMMA). We use the terms \textit{anode} and \textit{cathode windows} to refer to the solid PMMA objects. The coated electrically conductive surfaces facing the inside of the TPC, i.e.~the actual electrodes, are called \textit{anode} and \textit{cathode}. The gate electrode, which divides the drift and the extraction field region, is a wire grid that is supported by a stainless steel frame. The inner veto is defined by the vessel on the outside and the TPC boundaries on the inside. The optical planes are stainless steel structures filled with PMMA bricks and with surface-mounted photosensors facing the TPC and the inner veto volume. Additional photosensors are mounted around the TPC on the outer TPC wall surfaces.
The inner detector contains a total of $\SI{\sim 100}{t}$ of argon from an underground source~(UAr) of which half is inside the active TPC volume. UAr has a reduced content of the isotope \ce{^{39}Ar} compared to regular argon of atmospheric origin~(AAr)~\cite{DarkSide:2015}. At the nominal recirculation flow rate of $\SI{1000}{slpm}$ ($\SI{29.3}{g/s}$), the entire UAr mass is circulated in a turnover time of $\SI{\sim 40}{\day}$. Further details on the UAr cryogenics system are found in reference~\cite{DarkSide-20k:UArCryo}. The vessel is surrounded by a high-density polyethylene neutron shield and located in a ProtoDUNE-like~\cite{DUNE:2021hwx} membrane cryostat, which is filled with $\SI{\sim 650}{t}$ of AAr serving as outer muon veto and radiogenic shield. 
In equilibrium, the temperatures in both the AAr and UAr volumes are defined by the gas ullage pressures at saturation. Due to the large heat transfer surface of the vessel (liquid can flow across the neutron shield), the two pressures cannot be offset at reasonable cooling power expense and must thus be matched.

It is a complex challenge to characterise the internal TPC flow with respect to the aforementioned aspects at the scale of DS"~20k. To this end, we have performed 3D~fluid flow and thermal simulations using the computational fluid dynamics~(CFD) software ANSYS\textsuperscript{\textregistered}~Fluent~2020~R2~\cite{ANSYS}. Throughout this work, CFD results are presented as 2D~slices of the 3D~simulations.
This work is structured as follows: in section~\ref{Sec:Motivation}, we discuss the goals of this study and its relevance to DS"~20k. We detail the methodology and results of the single-phase modelling in section~\ref{sec:single_phase}. The dual-phase simulation and thermal aspects of the gas pocket formation are found in section~\ref{sec:dual_phase_gas_pocket}. The simulation framework, software settings and the technical details of the individual cases are presented at the beginning of each section. We end with a summary of our findings in section~\ref{Sec:Conclusions}.
\section{Motivation}
\label{Sec:Motivation}

This work focuses on flow and thermal aspects of the DS"~20k TPC related to: argon purification, detector calibration and the dual-phase operation. Purification relies on continuous gas circulation through a zirconium-based getter and an inline radon trap~\cite{DarkSide-20k:UArCryo}. To fully exploit their performance, it must be assured that the entire UAr volume is efficiently recirculated. Dead zones, i.e.~regions with stagnant flow or steady circulation that do not participate in the exchange, must be mitigated. 
Efficient impurity removal relies on a flow pattern that continuously replaces impure argon with pure argon. Ideally, such a process exchanges the highest-age by lowest-age argon rather than mixing the two. This is the case if the purification process is described by a plug flow reactor model in which the age of the fluid inside the volume is strictly monotonically increasing from the inlet to the outlet~\cite{Levenspiel1999} (chapter~5). At the outlet it is equal to the turnover time.

The internal source \ce{^{83\text{m}}Kr} will be deployed in the DS"~20k TPC~\cite{vanUffelen:2023tkl}, allowing for the calibration of the energy response at $\SI{41.6}{keV}$~\cite{Lippincott:2009ea,McCutchan:2015vcl} and the correction of the non-uniformity of the S2 signal yield as well as electric field inhomogeneity. This gaseous source is added to the recirculation flow at room temperature and thereby mixed with the argon. The source is then introduced into the TPC along with the LAr~\cite{DarkSide-20k:UArCryo}. Due to its half-life of $T_{1/2}=\SI{1.83}{h}$~\cite{McCutchan:2015vcl}, \ce{^{83\text{m}}Kr} must be distributed in the liquid target volume within a few hours such that good mixing is attained. Blind spots, i.e.~those not reached by the source, should be mitigated. At a minimum, assurance is needed that delivery of \ce{^{83\text{m}}Kr} to the horizontal centre of the TPC is achievable, as required for S2 signal calibration in the horizontal plane. If the calibration source is also well distributed in the vertical direction, it can be used to measure the free electron lifetime, i.e.~to calibrate the dependence of the S2 signal on the interaction depth, which serves as a metric for argon purity. Homogeneously distributed \ce{^{83\text{m}}Kr} events at large radii are useful to map the electric field in regions of field inhomogeneity. Unlike good purification performance, effective source distribution requires fast mixing, as is the case in a continuous stirred-tank or mixed flow reactor~\cite{Levenspiel1999} (chapter~5).

Mixing on the timescale of the turnover time can be quantified with the \emph{residence time distribution}~(RTD) at the outlet~\cite{DANCKWERTS:1953}. The residence time is the duration that a tracer particle has spent inside a volume since its injection. At the outlet of the volume, the exit age of the particle can be measured. This yields, for a high number of tracer particles, a good approximation of the characteristic RTD function of the volume. A fundamental property of the RTD at the outlet is that its mean, the mean residence time, is equal to the turnover time\footnote{Numerically, however, the mean residence time can be different from the turnover time due to incomplete tracking of trapped particles, insufficient statistics required to map the entire volume appropriately or limitations of the particle tracking accuracy at very low fluid speeds.}~\cite{Fogler} (section~13.3.2). We have introduced two ideal models above: the plug flow reactor, which features no mixing along the flow direction but perfect radial mixing, and the mixed flow reactor with perfect omnidirectional mixing. The corresponding RTDs of these are a Dirac delta distribution and a negative exponential distribution, respectively. When applying the concept of the RTD at the outlet to our case it is important to observe that the timescale of the \ce{^{83\text{m}}Kr} calibration, defined by its half-life, and that of the purification, given by the turnover time, differ by a factor of more than $500$. Thus, while being an appropriate measure for the purification performance, the RTD at the outlet will give limited insight into the source distribution. As we see in section~\ref{sec:single_phase}, the first tracer particles that reach the outlet have ages of tens of half-lives. However, an RTD evaluated in the TPC interior can be meaningful for the deployment of \ce{^{83\text{m}}Kr}. 

\section{Single-phase case} 
\label{sec:single_phase}

This section focuses on the single-phase case, assuming that the liquid-gas interface beneath the gas pocket is replaced by a solid, unheated TPC boundary. While this is a simplification of the dual-phase case presented in the next section, this first step allows us to address the mixing behaviour and dominant convection pattern in greater geometrical detail at a moderate computational cost. The placement of the LAr TPC inlets and outlets has been optimised by studying the RTD properties of various configurations in the isothermal flow-only case (see appendix~\ref{app:configurations}). The results presented below are for the preferred inlet/outlet configuration. It features 2~rings with 8~inlets each, located at the top and the middle of the TPC (see figure~\ref{fig:CFD_geometry}). The 8~outlets are located at the bottom behind the reflector panels and run through the cathode window. The inlets (outlets) have diameters of $\SI{9}{mm}$ ($\SI{12.7}{mm}$), each with a combined cross-sectional area equivalent to twice that of a circle with a $\SI{1}{inch}$ diameter. Starting from the isothermal flow-only case in subsection~\ref{subsec:turnover1}, we assess the mixing behaviour and the turnover efficiency in terms of the RTD and the spatial distribution of the mean residence time. From subsection~\ref{subsec:BOP_heat} onward, we add thermal modelling to the simulation, which allows the formation of convective currents. Convection inside the TPC is driven by heat from the bottom photoelectronics and the gas pocket supply. The former is discussed in this section, while the latter is addressed in the next. 

\subsection{Framework}
\label{subsec:framework1}
\paragraph{Geometry and mesh}

\begin{figure}[t]
\centering
\includegraphics[width=1\textwidth]{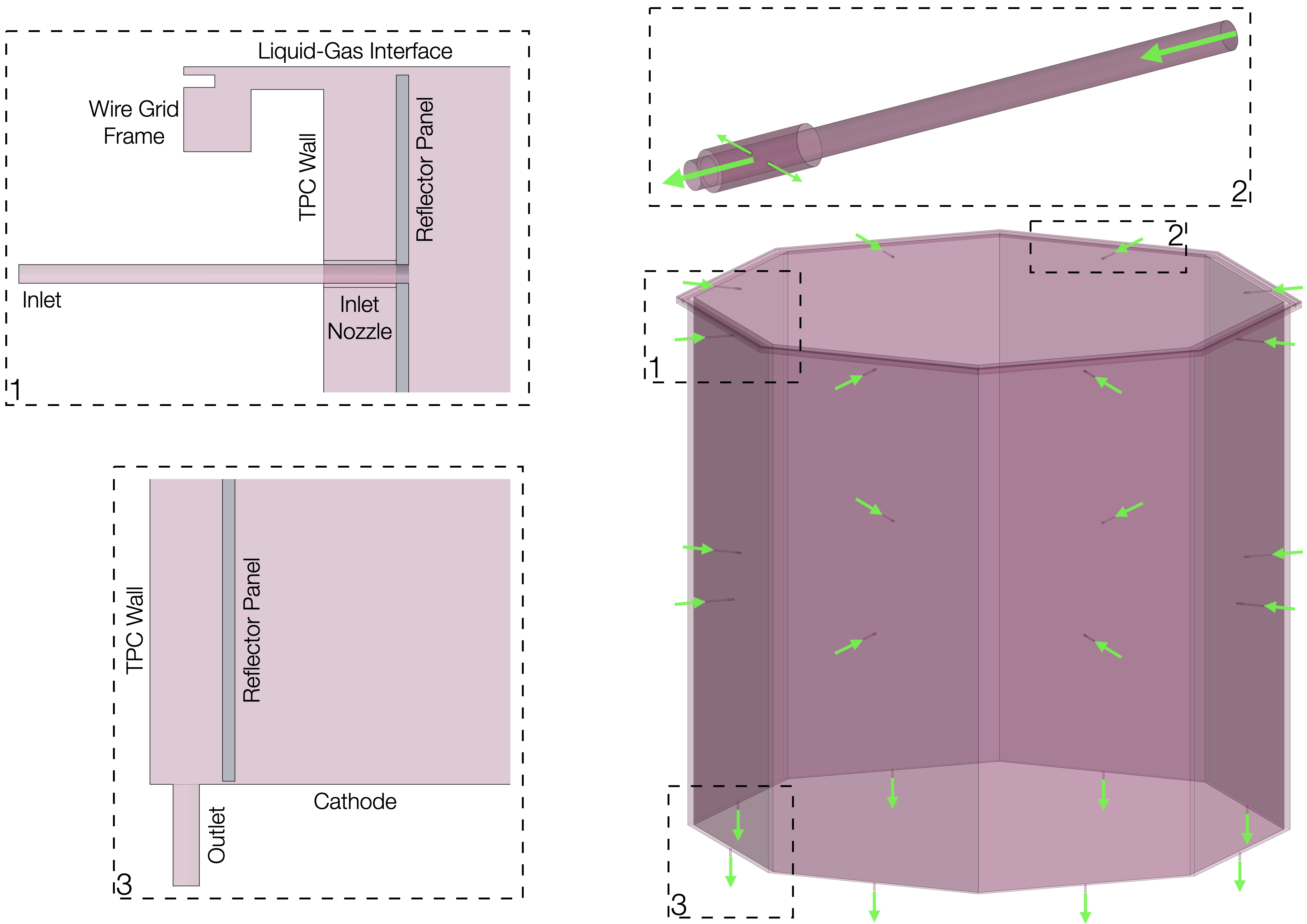}
\caption{Liquid volume in the TPC. The arrows indicate the flow direction. (Detail~1): Section view of a top edge with an inlet. (Detail~2): View of an inlet. The majority of the inlet flow is injected directly into the active volume, while a small fraction is supplied to the intermediate space between the walls and the reflector panels. The cross section ratio of the two small holes and the main tube corresponds to the volume ratio of the intermediate space and the active region. (Detail~3): Section view of a bottom edge with an outlet.}
\label{fig:CFD_geometry}
\end{figure}

To keep the computational complexity moderate, the simulations in this work are based on a simplified but full-size 3D~geometry of the TPC. The model includes an abstract TPC volume, defined by the boundaries of the TPC walls, cathode, anode, and reflector panels (see figure~\ref{fig:CFD_geometry}). Details such as field-shaping rings, the gate wire grid, and fasteners are omitted as they are expected to have a negligible impact on the macroscopic flow pattern and would unnecessarily increase mesh complexity. While the geometry used for the isothermal simulations only includes the fluid body of the argon, the thermal simulations are extended by the solid body of the reflector panels. This internal volume accounts for thermal conduction. The liquid domain has a height of $\SI{3484}{mm}$ and a total fluid volume of $\SI{36.81}{m^3}$. The gas pocket of $\SI{7}{mm}$ thickness is not simulated here and thus not shown in figure~\ref{fig:CFD_geometry}. The inscribed circle of the octagon has a diameter of $\SI{3582}{mm}$, the reflector panels are $\SI{6}{mm}$ thick and separated by a $\SI{35}{mm}$ gap from the TPC wall. The gap between the reflector panels and the cathode (liquid surface) is $\SI{1.5}{mm}$ ($\SI{4}{mm}$). 

We use an unstructured tetrahedral fluid mesh without defeaturing with a minimum (maximum) side length of $\SI{0.1}{mm}$ ($\SI{200}{mm}$) and a growth rate of $1.2$ between cells (typical cell side length: $\SI{100}{mm}$). Depending on the inlet/outlet configuration, the mesh includes $\SIrange{22}{28}{}\times 10^{6}$ fluid cells. This yields a good mesh quality\footnote{Changes to the mesh properties do not perceptibly affect the flow results.} with an orthogonality of $0.73\,\mathrm{(av.)}\pm 0.13\,\mathrm{(std.)}$ (min.~$0.11$), a skewness of $0.27\,\mathrm{(av.)}\pm 0.13\,\mathrm{(std.)}$ (max.~$0.89$), and an aspect ratio of $1.9\,\mathrm{(av.)}\pm 0.5\,\mathrm{(std.)}$ (max.~$11.3$). The ANSYS\textsuperscript\textregistered Fluent User's Guide recommends the following mesh metric values: a minimum orthogonality of $0.01$, a maximum skewness of $0.95$, and a maximum aspect ratio of $5$ in the bulk and $10$ in the boundary layers. While the stability of the flow solution is not affected by cells with high aspect ratios, the stability of the energy solution benefits from keeping the maximum aspect ratio below $35$. For further details on mesh quality, refer to reference~\cite{UserGuide}.

\paragraph{Physics and solver setup}

We perform steady state fluid flow simulations with the pressure-based solver of ANSYS\textsuperscript{\textregistered}~Fluent using the coupled pressure-velocity scheme and second-order spatial discretisation~\cite{UserGuide,TheoryGuide}. These choices are appropriate for low-speed incompressible flows, ensure high convergence speed through the simultaneous solving of momentum and pressure-based continuity equations, and yield high accuracy results for complex flows that cross the mesh lines obliquely as it is the case in a tetrahedral mesh. Turbulence, which can occur particularly at the inlets and outlets, is described with the widely used Reynolds-averaged Navier-Stokes $k-\omega$ shear stress transport~(SST) turbulence model, which offers an economic and robust approach that yields accurate results close and far from walls. RTDs can be obtained with ANSYS\textsuperscript{\textregistered}~Fluent via particle tracking in the Lagrangian reference frame\footnote{In the Lagrangian specification the simulation nodes follow the fluid parcel as it moves in the velocity flow field.} in a post-processing step~\cite{Li:2010}. For such tracer simulations, we use the discrete phase model with high-resolution tracking and the injection of massless particles from the inlet surface. Turbulent velocity fluctuations are determined by stochastic tracking with the discrete random walk model and with the option of random Eddy time enabled. The massless particles follow the flow of the continuous phase without affecting it. The use of such tracers is justified when their number density in the fluid volume is small. For a \ce{^{83\text{m}}Kr} calibration of the active volume at an event rate of the order of $\SI{10}{Hz}$, a krypton concentration of $\SI{e-25}{mol/mol}$ is required, which means that individual krypton atoms are surrounded solely by argon. Using massless tracers in a continuous phase neglects atomic properties and resulting interactions of the tracers with the fluid atoms. In the discrete phase model, the maximum possible step number of $10^{9}$ is used with a step length factor of $5$. The latter controls the time step size used to integrate the equations of motion and defines the number of steps required to traverse a cell. This choice allows tracers to traverse $\num{\sim 2e8}$ cells, which is seven to nine times the total cell count of the considered configurations. Spatial distribution maps of the mean residence time are calculated at run time of the CFD solver in the Eulerian frame\footnote{In the Eulerian specification the simulation is based on the fixed fluid mesh.} with a user-defined scalar following reference~\cite{UDFManual} (section 2.3.3). For this, we calculate the effective diffusivity with a user-defined function, using a mass self-diffusivity of $\SI{1.53e-9}{m^2/s}$~\cite{Cini-Castagnoli:1960,Fisher:1972} and setting the turbulent Schmidt number, i.e.~the ratio of turbulent momentum and mass transport rates, to unity (Reynolds analogy)~\cite{Baleo:2000,Liu:2010}. Thermal simulations are performed by solving the built-in energy equation with second-order spatial discretisation~\cite{UserGuide,TheoryGuide}. The argon fluid properties such as mass density, viscosity, specific heat capacity and thermal conductivity are implemented according to temperature-dependent isobaric data from the NIST Standard Reference Database~\cite{NIST}, as these properties are insensitive to the pressure variations present in the model. We use an operating ullage pressure of $\SI{1.075}{bara}$~\cite{DarkSide-20k:UArCryo} with the LAr being at saturation at the top. The fluid is subject to the downward gravitational force to account for the hydrostatic pressure profile and to allow for density-driven free convection processes. The following material properties at LAr temperature are used for PMMA (stainless steel): a specific heat capacity of $\num{510}\,(\num{239.6})\,\si{J/(kg\cdot K)}$ and a thermal conductivity of $\num{0.15}\,(\num{8.6})\,\si{W/(m\cdot K)}$~\cite{Gaur:1982,Pyda2014:sm_ptd_athas_0010,STEPHENS:1972,hartwig:1982,Bradley:2013}.

\paragraph{Operating and boundary conditions}

\begin{table}
\caption[Operating and Boundary Conditions]{Operating and boundary conditions for the single-phase TPC simulations with bottom heat.}
\centering
\begin{tabularx}{\textwidth}{>{\setlength{\baselineskip}{0.85\baselineskip}}Xc >{\setlength{\baselineskip}{0.85\baselineskip}}X >{\setlength{\baselineskip}{0.85\baselineskip}}X}
\toprule
\textbf{Condition}   & \textbf{Value}  & \textbf{Setup}   & \textbf{Comment}\\
\midrule
\textbf{Operating}&&&\\
Gravity     &   $\SI{-9.81}{m/s^2}$ & Downward &\\
&&&\\[-10pt]
Top boundary pressure & $\SI{1.160}{bara}$ & & At $\SI{620}{mm}$ depth\\
\textbf{Boundary}&&&\\
Nominal TPC mass flow rate & $\SI{14.9}{g/s}$ & Mass flow inlet, distributed equally over inlet surfaces, $\SI{5}{\%}$ turbulent intensity at $\SI{1}{mm}$ length scale & $\hat{=} \, \SI{\sim 500}{slpm}$\\
&&&\\[-10pt]
Outlet pressure & $\SI{0}{barg} $ & Pressure outlet &  $\hat{=} \, \SI{1.642}{bara}$, driven by operating density\\
Inlet temperature & $\SI{87.87}{K}$ & Inlet with fixed temperature & Saturation temperature at ullage pressure\\
&&&\\[-10pt]
Cathode window underside temperature & $\SIrange{88}{92}{K}$ & Wall with fixed temperature and thickness & $\SI{92.14}{K}$ is saturation temperature\\
&&&\\[-10pt]
TPC outer surfaces temperature & $\SI{87.87}{K}$ & Walls with fixed temperature and thicknesses & Saturation temperature at ullage pressure\\
&&&\\[-10pt]
Reflector body thermal & -- & Thermally coupled to fluid &\\
\bottomrule
\end{tabularx}
\label{tab:conditions}
\end{table}

An overview of all operating and boundary conditions used in this section can be found in table~\ref{tab:conditions}. For an operating LAr density of $\SI{1392}{kg/m^3}$~\cite{NIST}, considered constant throughout the isothermal case model, the total pressure at the depth of the anode is $\SI{1.160}{bara}$. Accordingly, we obtain a total pressure of $\SI{1.642}{bara}$ at the TPC outlets and use $\SI{0}{barg}$ as outlet pressure boundary condition. Mass flow inlets with a nominal value of $\SI{14.9}{g/s}$ are used, distributed equally over all inlet surfaces. This corresponds to a flow rate of approximately $\SI{500}{slpm}$\footnote{More accurately $\SI{500}{slpm}$ equals $\SI{14.7}{g/s}$ at $\SI{1}{bara}$ and $\SI{273.15}{K}$. The value $\SI{14.9}{g/s}$ comes from the conversion according to the old definition of slpm at $\SI{1.01325}{bara}$ and $\SI{273.15}{K}$.}, which is half of the nominal gas recirculation rate of the UAr cryogenics system~\cite{DarkSide-20k:UArCryo}. In this scheme, the other half would be injected into the inner veto volume. However, in appendix~\ref{app:configurations} we also show results of simulations at altered TPC flow rates in the range $\SIrange{125}{1000}{slpm}$. The inlet flows are fully developed and assumed to be almost laminar, with a turbulent intensity of $\SI{5}{\%}$ on a length scale of $\SI{1}{mm}$. Additionally, we enforce thermal boundary conditions, which we motivate as follows. As evidenced from the cryogenic performance of the ProtoDUNE LAr TPC~\cite{DUNE:2021hwx} and from CFD simulations~\cite{DUNE-doc-17481}, the temperature variations inside the large membrane cryostat volume are expected to be small and of the order of $\SI{10}{mK}$. The operating ullage pressure of $\SI{1.075}{bara}$ corresponds to a saturation temperature of $\SI{87.87}{K}$~\cite{NIST}. It is a reasonable assumption that the UAr in the inner veto volume, at an appropriate distance from the photoelectronics, is at the same temperature. The following boundary conditions are therefore applied: all TPC surfaces that separate the TPC and the inner veto fluid volumes are set to a fixed temperature and are assigned a wall-thickness. That is, the heat transfer through those external walls is not simulated. Instead, the heat conduction rate expected from the temperature difference, the wall thickness and the material's thermal conductivity is used. Boundary TPC components with irregular shapes, like the cathode window, are treated as walls with averaged constant thicknesses (cathode window: $\SI{57}{mm}$). We discuss the impact of this geometrical abstraction of the cathode window below in the context of the resulting convection cell. We apply $\SI{87.87}{K}$ to all walls except the cathode. The cathode window is located just above the bottom optical plane and is subject to a heat load from the photoelectronics. It is designed as an upside-down truncated octagonal pyramid to allow the dislodgement of GAr bubbles that may form below (see figure~\ref{fig:DS-20k_overview}). As a result, a static gas layer cannot form underneath the cathode window, and no gas heating is involved. The maximum temperature that can be reached at the underside of the cathode window is therefore given by the saturation temperature ($\SI{92.14}{K}$) at the respective total pressure of $\SI{1.642}{bara}$. We perform TPC flow simulations at various cathode window underside temperatures in the range $\SIrange{89}{92}{K}$. The temperature of the inlet flow is $\SI{87.87}{K}$ at the operating ullage pressure due to the design of the UAr cryogenics system~\cite{DarkSide-20k:UArCryo}.

\subsection{Turnover efficiency}
\label{subsec:turnover1}

\begin{figure}[t]
\centering
\begin{subfigure}[b]{0.48\textwidth}
\centering
\includegraphics*[width=\textwidth]{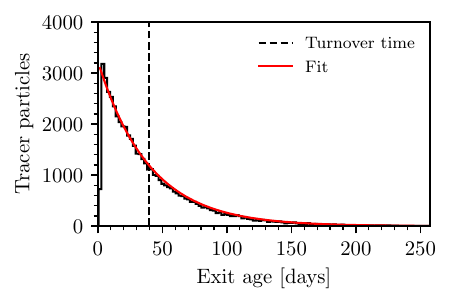}
\end{subfigure}
\quad
\begin{subfigure}[b]{0.48\textwidth}
\centering
\includegraphics*[width=\textwidth]{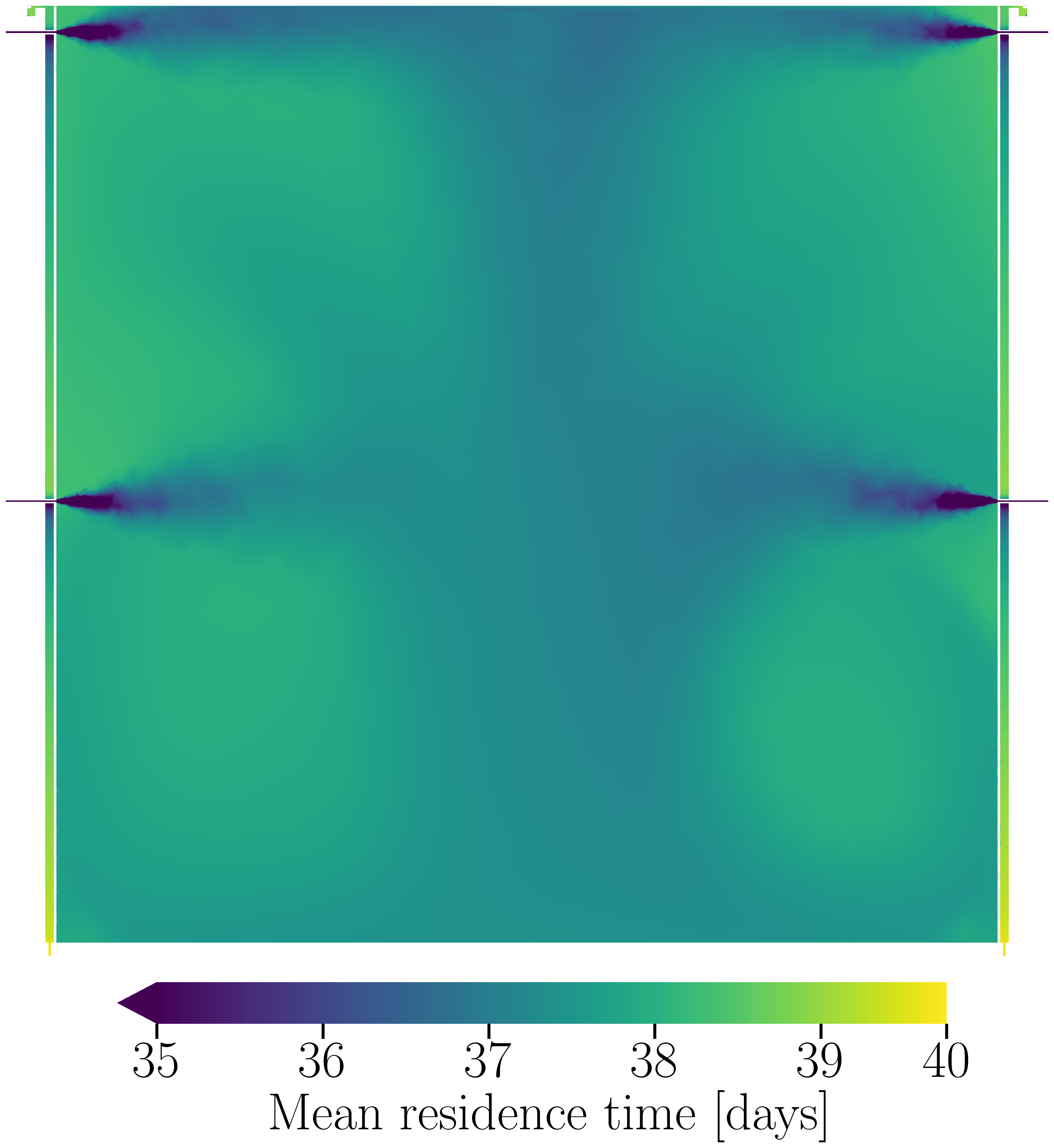}
\end{subfigure}
\caption[]{(Left): RTD at the outlet. The red curve is a fit with the expected RTD function for a mixed flow reactor with delayed arrival of the tracers, i.e.~the delay is a free parameter of the exponential fit while the time constant is fixed to the turnover time. The first bin of the histogram is excluded in the fit. (Right): Spatial mean RTD in a vertical cut view through the inlets and outlets.}
\label{fig:RTD_2x8_double_no_heat_MeanRT_spatial_distribution}
\end{figure}

Figure~\ref{fig:RTD_2x8_double_no_heat_MeanRT_spatial_distribution}~(left) shows the simulated RTD in the isothermal flow-only case (see appendix~\ref{app:flowpattern1} for the corresponding flow pattern). On timescales of the turnover time, the RTD closely resembles the one of a well-mixed system for which an exponential decrease of the number of tracers $n(t)$ with exit age is expected~\cite{Fogler} (section 13.3.4):
\begin{equation}
n(t)=n_{0} \cdot e^{-\frac{t-t_{0}}{\tau}} \quad .
\end{equation}
Here, $n_{0}$ is the number of tracer particles in the bin with the most entries -- representing the most probable residence time -- and $\tau$ is the turnover time of $\SI{\sim 40}{\day}$. Due to the system's size and the small flow speeds, the RTD exhibits a delayed arrival of $\SI{\sim 3.5}{\day}$ for the tracers in the peak of the histogram. With the root mean square~(RMS) speed (see appendix~\ref{app:flowpattern1}), one can estimate that the tracers travel a distance of $\SI{\sim 34}{m}$ within this period, corresponding to around $\num{10}$ traverses of the TPC dimensions. This delay is accounted for by the free parameter $t_{0}$ in the model. The distribution is narrow, with a full width at tenth maximum of approximately twice the turnover time. Within five turnover times, $\SI{99}{\%}$ of the tracers are collected at the outlet. This implies an efficient recirculation and suggests no significant fluid particle trapping (see table~\ref{tab:RTDs_no_heat} in appendix~\ref{app:configurations} for the RTD properties). A fraction of $(63.0 \pm 0.3)\,\si{\%}$ of the tracers arrive ahead of the turnover time ($1-1/e=\SI{63.2}{\%}$ is expected for an exponential RTD). This asymmetry (compared to e.g.~the centred symmetric distribution of a realistic plug flow reactor that features limited longitudinal mixing) is the result of an almost perfect mixing. Tracers in the tail of the exponential have long residence times, which are compensated by a large number of tracers with shorter-than-average residence times. 

As we will see in subsection~\ref{subsec:temperature_map_flowpattern2}, the fluid speed increases significantly when liquid-phase convection is induced by the bottom photoelectronics heat. The higher speed increases the computational effort to reproduce an RTD at the outlet enormously. Since the turnover time fixes the residence timescale, the number of steps through the fluid cells, and thus the computational time, required for a tracer particle to reach from an inlet to an outlet scales roughly with the mean speed. Depending on the inlet/outlet configuration we typically track $\num{\sim 26500}$ tracers in the flow-only case, of which on average $\SI{99.3}{\%}$ reach the outlet. The tracing of the remaining samples is incomplete, i.e.~an average fraction of $\SI{0.7}{\%}$ does not reach the outlet within the maximum possible steps of $10^{9}$. As this fraction increases, the resulting RTD is progressively distorted. It is found that the upper limit of possible steps in ANSYS\textsuperscript{\textregistered}~Fluent is insufficient to sample properly the RTD for the case with bottom photoelectronics heat. The number of cells cannot be reduced enough with coarser meshing to compensate for this effect while maintaining an acceptable mesh quality.

A practical technique for identifying dead zones is to consider a mean residence time map of the detector~\cite{Li:2010,Baleo:2000,Liu:2010}, which is shown in figure~\ref{fig:RTD_2x8_double_no_heat_MeanRT_spatial_distribution}~(right). The mean residence time is homogeneous throughout the model with minor fluctuations. This includes the remote region in proximity of the wire grid frame, where an increased mean residence time could be anticipated due to its isolation from the bulk volume, cf.~figure~\ref{fig:CFD_geometry}~(detail~1). Since the highest mean residence time is found at the outlets, we conclude that on timescales of the turnover time no macroscopic dead zones, as relevant for purification, can be identified. Fluid that is locally trapped in steady recirculation zones or hidden detector regions for periods comparable or longer than the turnover time would feature a mean residence time above the one at the outlet. An important continuity check of the simulation is the equivalence of the mean residence time at the outlet and the turnover time. In fact, they differ by only $\SI{301}{s}$, i.e.~the relative error is $<\num{e-4}$.

\subsection{Bottom photoelectronics heat}
\label{subsec:BOP_heat}

The photosensors underneath the cathode window warm the LAr in proximity of the cathode window underside\footnote{Due to its position below the top photosensors, the temperature of the anode window's top surface is minimally affected by the photosensor heat. Regardless, the impact of this top heating on the TPC flow is not addressed here within the scope of the single-phase discussion. The presence of a gas pocket below the anode establishes the most stringed temperature boundary condition on top of the liquid (saturation temperature), as discussed in section~\ref{sec:dual_phase_gas_pocket}. We likewise disregard the effect of the sparsely placed veto photosensors on the outside of the TPC walls, as their heat is efficiently removed through unobstructed convective currents in the veto volume.}. To minimise the computational complexity of the problem of obtaining the impact of the bottom photoelectronics heat load on the TPC flow, we follow a two-step approach: instead of solving the combined volume of the TPC and the inner veto, we first derive the temperature of the cathode window underside in a separate model, neglecting the interaction with the TPC fluid volume. In a second step, we apply the obtained temperature as boundary condition on the cathode in the TPC model and study the resulting convection. The simulation framework of the dedicated model for the first step is described below.

\paragraph{Geometry and mesh} 

\begin{figure}[t]
\centering
\includegraphics[width=0.618\textwidth]{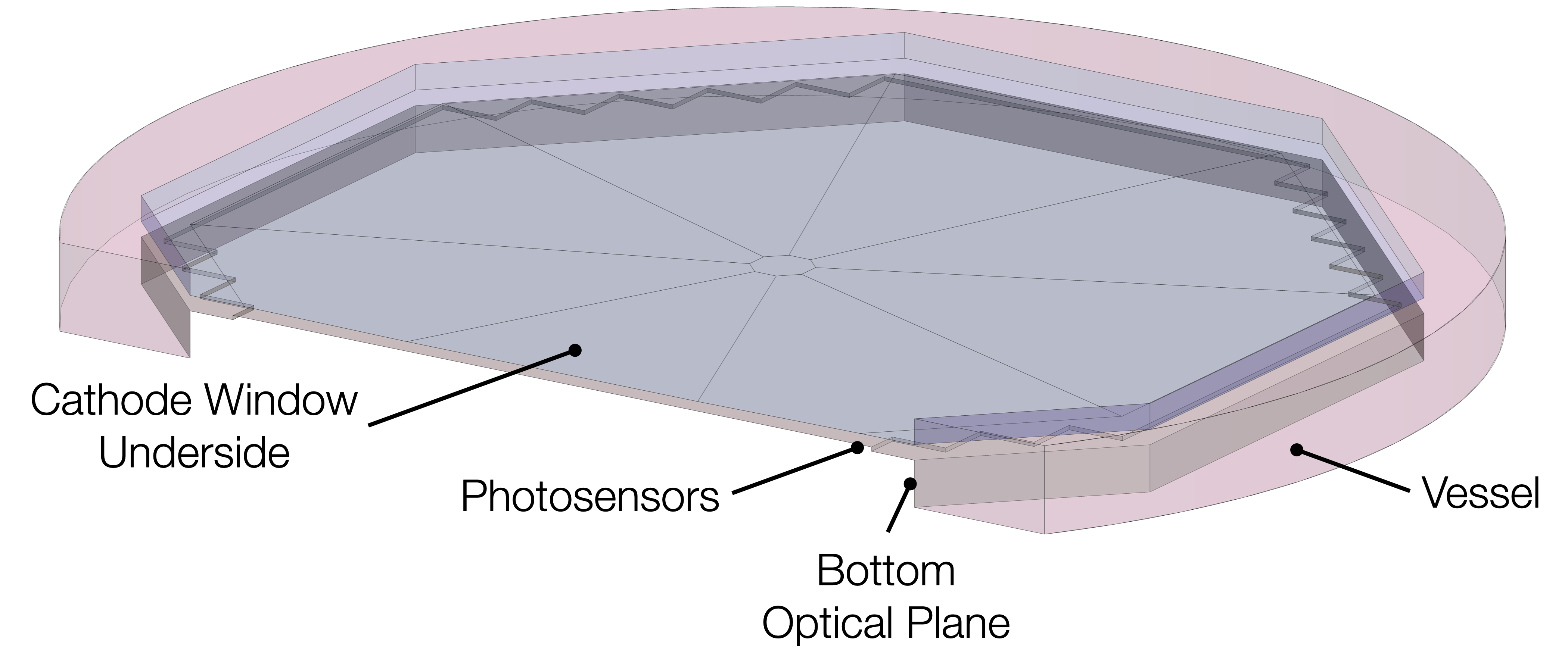}
\caption{Break view of the fluid geometry used to simulate the cathode window underside temperature. For improved visibility, the surfaces of the fluid volume in contact with the cathode window are highlighted in blue, while those in contact with the bottom optical plane or the photosensors are highlighted in grey.}
\label{fig:cathode_geometry}
\end{figure}

\begin{table}
\caption[Operating and Boundary Conditions]{Operating and boundary conditions for the bottom photoelectronics simulation.}
\centering
\begin{tabularx}{\textwidth}{>{\setlength{\baselineskip}{0.85\baselineskip}}Xc >{\setlength{\baselineskip}{0.85\baselineskip}}X >{\setlength{\baselineskip}{0.85\baselineskip}}X}
\toprule
\textbf{Condition}   & \textbf{Value}  & \textbf{Setup}   & \textbf{Comment}\\
\midrule
\textbf{Operating}&&&\\
Gravity     &   $\SI{-9.81}{m/s^2}$ & Downward &\\
&&&\\[-10pt]
Bottom boundary pressure & $\SI{1.671}{bara}$ & &\\
\textbf{Boundary}&&&\\
Outlet pressure & $\SI{0}{barg} $ & Pressure outlet & No directed mass flow\\
&&&\\[-10pt]
Nominal photoelectronics heat flux & $\SI{45.2}{W/m^2}$ & Wall with heat flux & $\hat{=} \, \SI{500}{W}$ total heat load\\
&&&\\[-10pt]
Cathode window underside heat flux & $\SI{0}{W/m^2}$ & Adiabatic wall & \\
&&&\\[-10pt]
Temperature of other surfaces & $\SI{87.87}{K}$ & Walls with fixed temperature & Saturation temperature at ullage pressure, heat sink\\
&&&\\[-10pt]
Outlet backflow temperature & $\SI{87.87}{K}$ &  & Saturation temperature at ullage pressure\\
\bottomrule
\end{tabularx}
\label{tab:conditions2-photoelectronics}
\end{table}

We use a separate geometry for the region around the bottom optical plane (see figure~\ref{fig:cathode_geometry}). Its fluid volume is bounded by the inner shell of the vessel and does not include the TPC fluid. The lower surface of the cathode window is visible on top of the fluid volume and the photosensor plane is located underneath, visible by its zigzag shape on four opposite sites. This is the surface to which the heat load is applied. The void underneath the photosensors corresponds to a simplified version of the optical plane. Due to the shape of the cathode window, the thickness of the fluid layer in between the photosensors and the cathode window underside is $\SIrange{18}{38}{mm}$, increasing from the centre towards the outside. The geometry consists of $\SI{2.8e6}{}$ fluid cells with a good mesh quality (orthogonality: $0.74\pm 0.13\,\mathrm{(std.)}$ (min.~$0.20$); skewness: $0.26\pm 0.13\,\mathrm{(std.)}$ (max.~$0.80$); aspect ratio: $1.9\pm 0.5\,\mathrm{(std.)}$ (max.~$10.4$)).

\paragraph{Operating and boundary conditions}

The thermal conditions in the separate model are as follows (see table~\ref{tab:conditions2-photoelectronics}): we assign a power of $\SI{500}{W}$ to the photosensor plane, treat the cathode window underside with an adiabatic thermal condition and assign a fixed temperature of $\SI{87.87}{K}$ to all other surfaces as motivated above for the TPC simulation. The top and bottom caps of the fluid volume, i.e.~the boundaries to the rest of the (non-simulated) inner veto volume, are used as pressure outlets with respect to the depth of the model. These allow for backflow with fluid at $\SI{87.87}{K}$.

\paragraph{Results}

\begin{figure}[t]
\centering
\begin{subfigure}[b]{0.48\textwidth}
\centering
\includegraphics*[width=\textwidth]{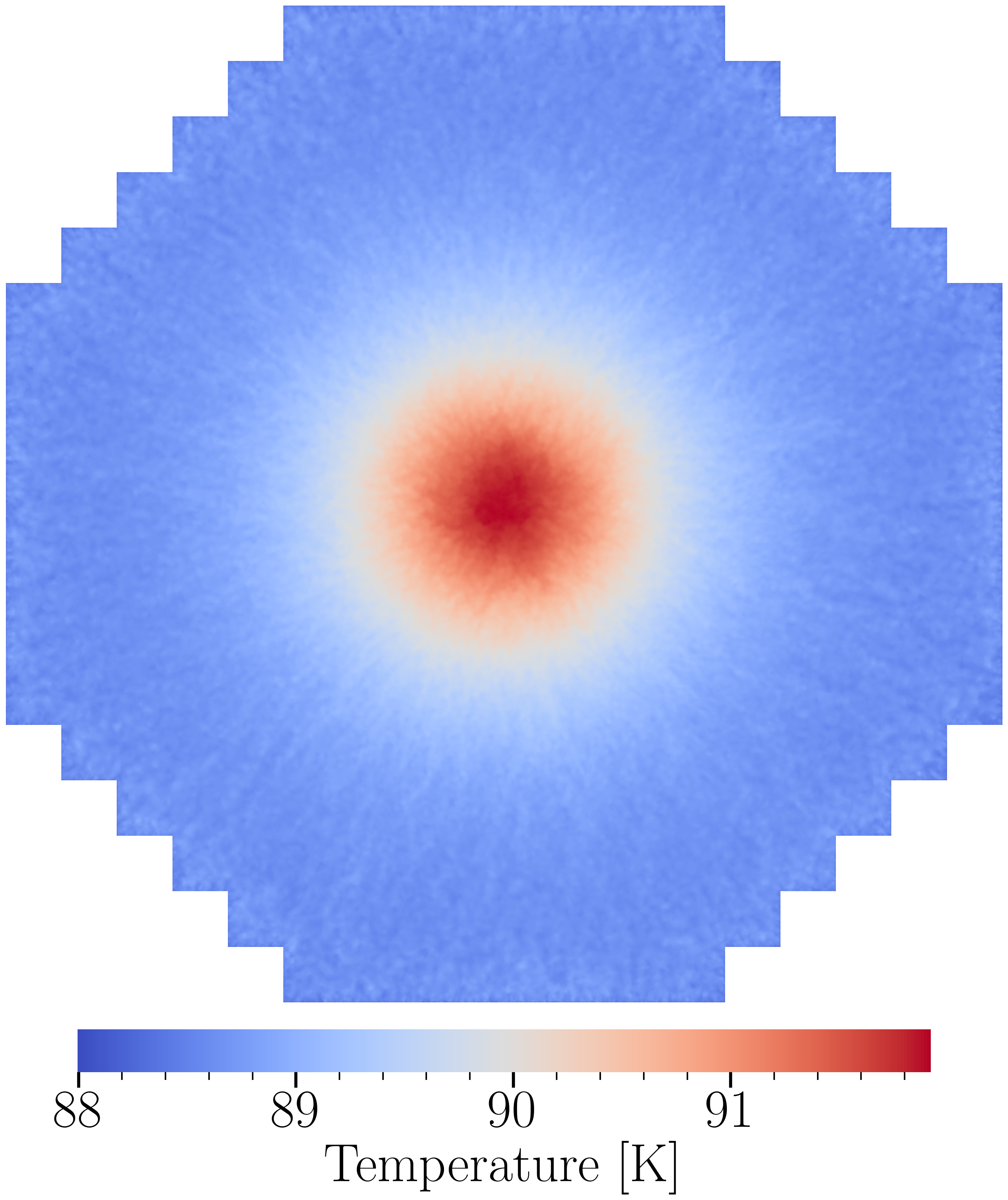}
\end{subfigure}
\quad
\begin{subfigure}[b]{0.48\textwidth}
\centering
\includegraphics*[width=\textwidth]{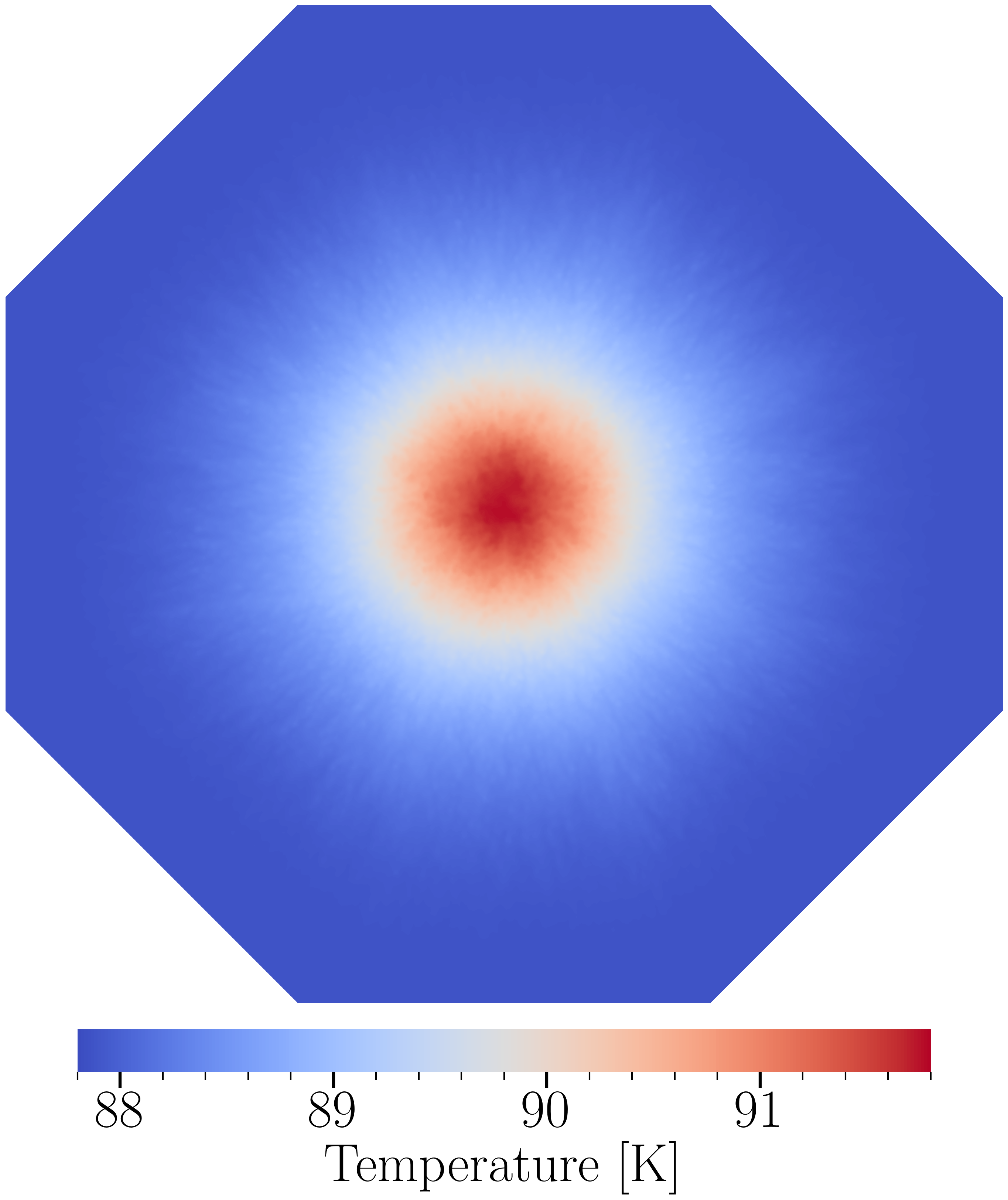}
\end{subfigure}
\caption[]{Temperature maps obtained with the separate model shown in figure~\ref{fig:cathode_geometry} at a photoelectronics heat load of $\SI{500}{W}$. (Left): Photosensor surface. (Right): Cathode window underside.}
\label{fig:cathode_temperatures}
\end{figure}

Figure~\ref{fig:cathode_temperatures} shows the resulting temperature maps of the photosensor surface (mean temperature: $\SI{89.2}{K}$) and the cathode window underside (mean temperature: $\SI{88.8}{K}$). This result is obtained neglecting the interaction with the fluid inside the TPC through heat conduction. This simplification can be made due to the low thermal conductivity of PMMA. It is thus reasonable to assume that the temperature of the cathode window underside is equal to that of the proximate liquid layer. Despite the equally distributed heat load that is applied on the photosensor plane, we observe a radial temperature gradient over this surface. The thin fluid disk in the centre, particularly in the region where the cathode window underside surface is flat, hinders radial convection, leading to the observed hotspot. At larger radii, the layer thickens and the flared shape of the cathode window allows the fluid to rise and move radially towards the edges where it contacts colder fluid. However close, the maximum temperature does not exceed the boiling temperatures at the pressure corresponding to the model's depth. On the cathode window underside for instance, the maximum temperature found is approximately $\SI{0.4}{K}$ below the boiling point. Note that we consider an abstract version of the photosensor plane as far as relevant for the purpose of obtaining the cathode window underside temperature. Conclusions on the thermodynamics of the photoelectronics and the proximate microscopic flow pattern depend on geometric details that are not implemented here.

\subsection{Temperature map and flow pattern}
\label{subsec:temperature_map_flowpattern2}

\begin{figure}[t]
\centering
\begin{subfigure}[b]{0.48\textwidth}
\centering
\includegraphics*[width=\textwidth]{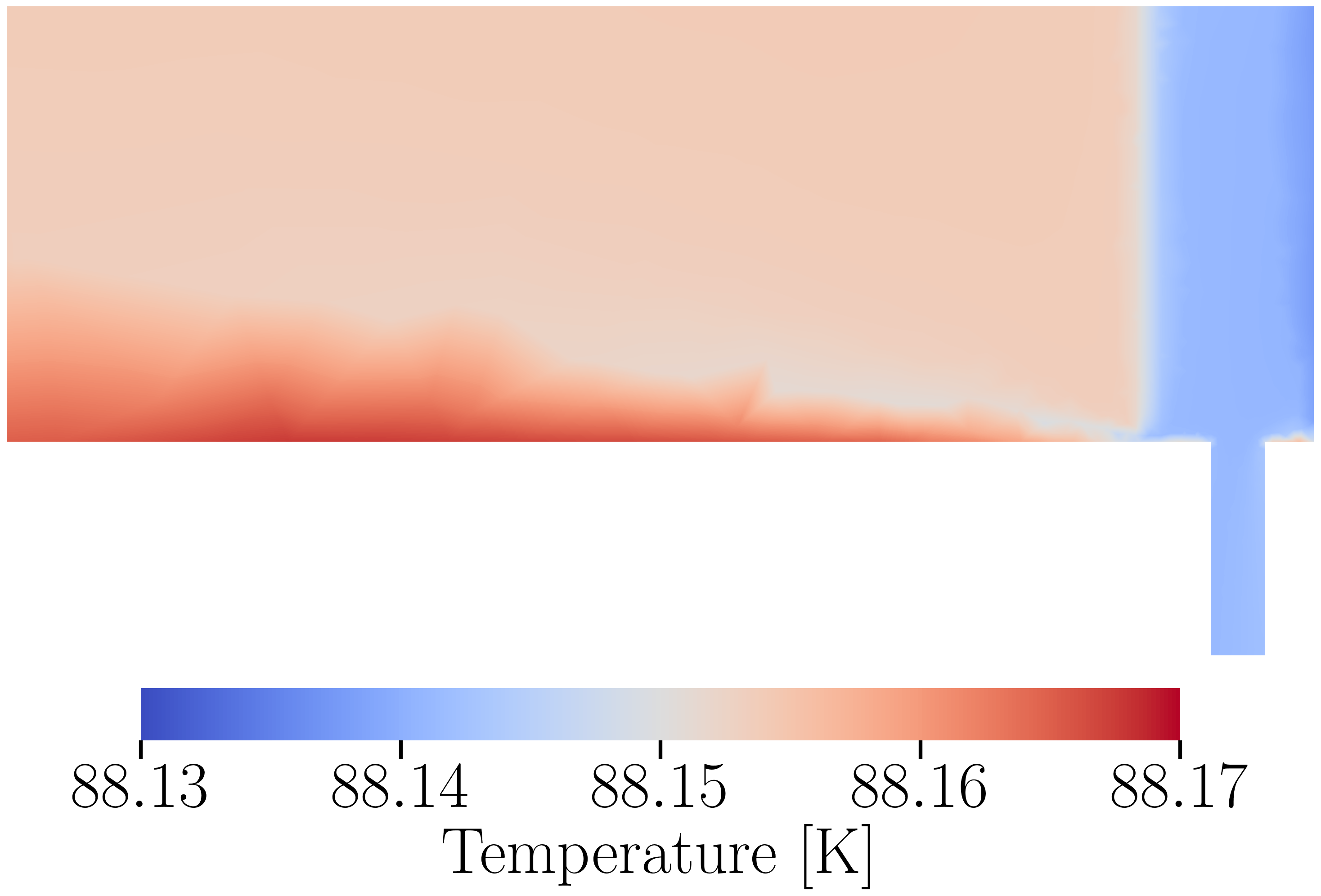}
\end{subfigure}
\quad
\begin{subfigure}[b]{0.48\textwidth}
\centering
\includegraphics*[width=\textwidth]{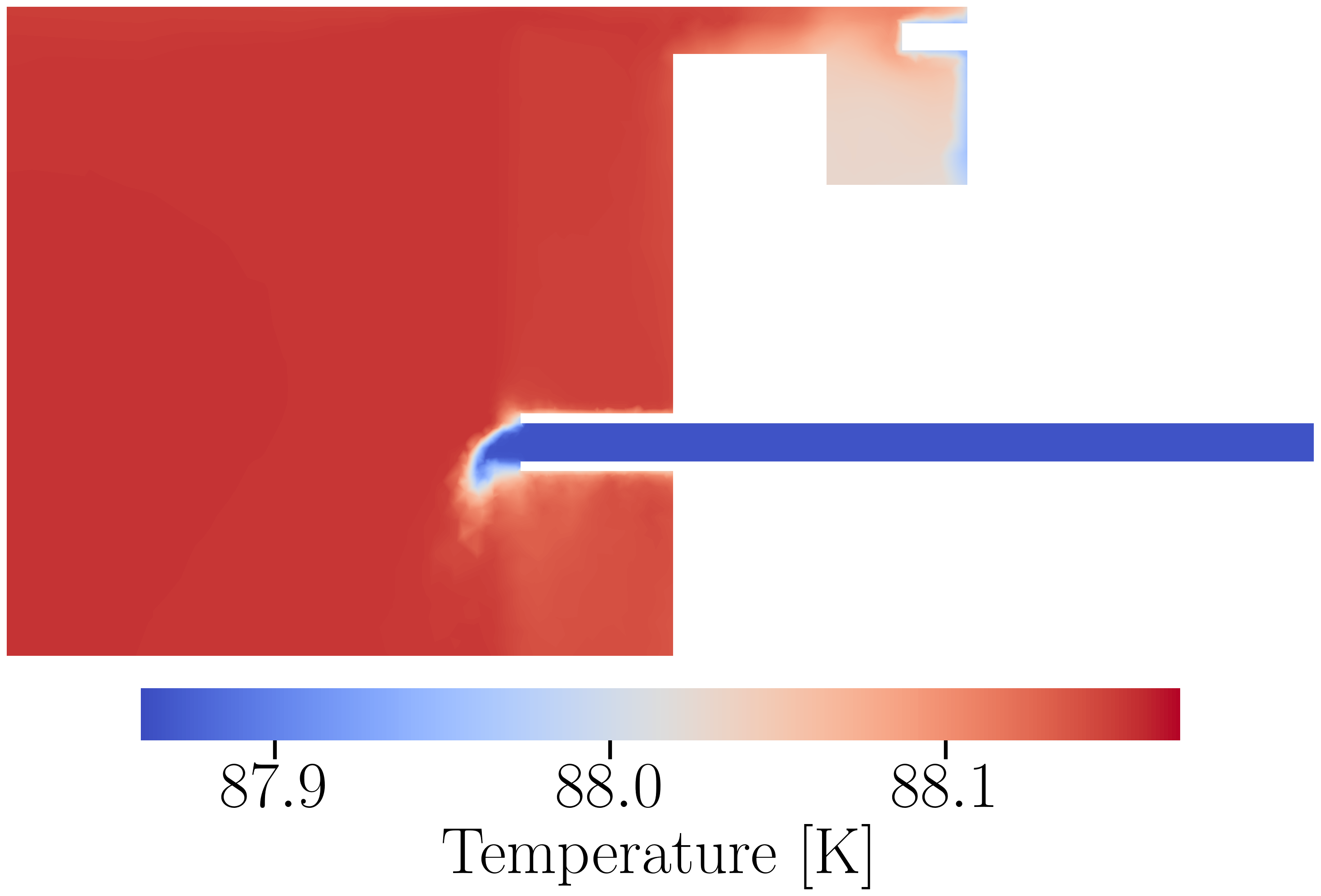}
\end{subfigure}
\quad
\begin{subfigure}[b]{0.48\textwidth}
\vspace{0.3cm}
\centering
\includegraphics*[width=\textwidth]{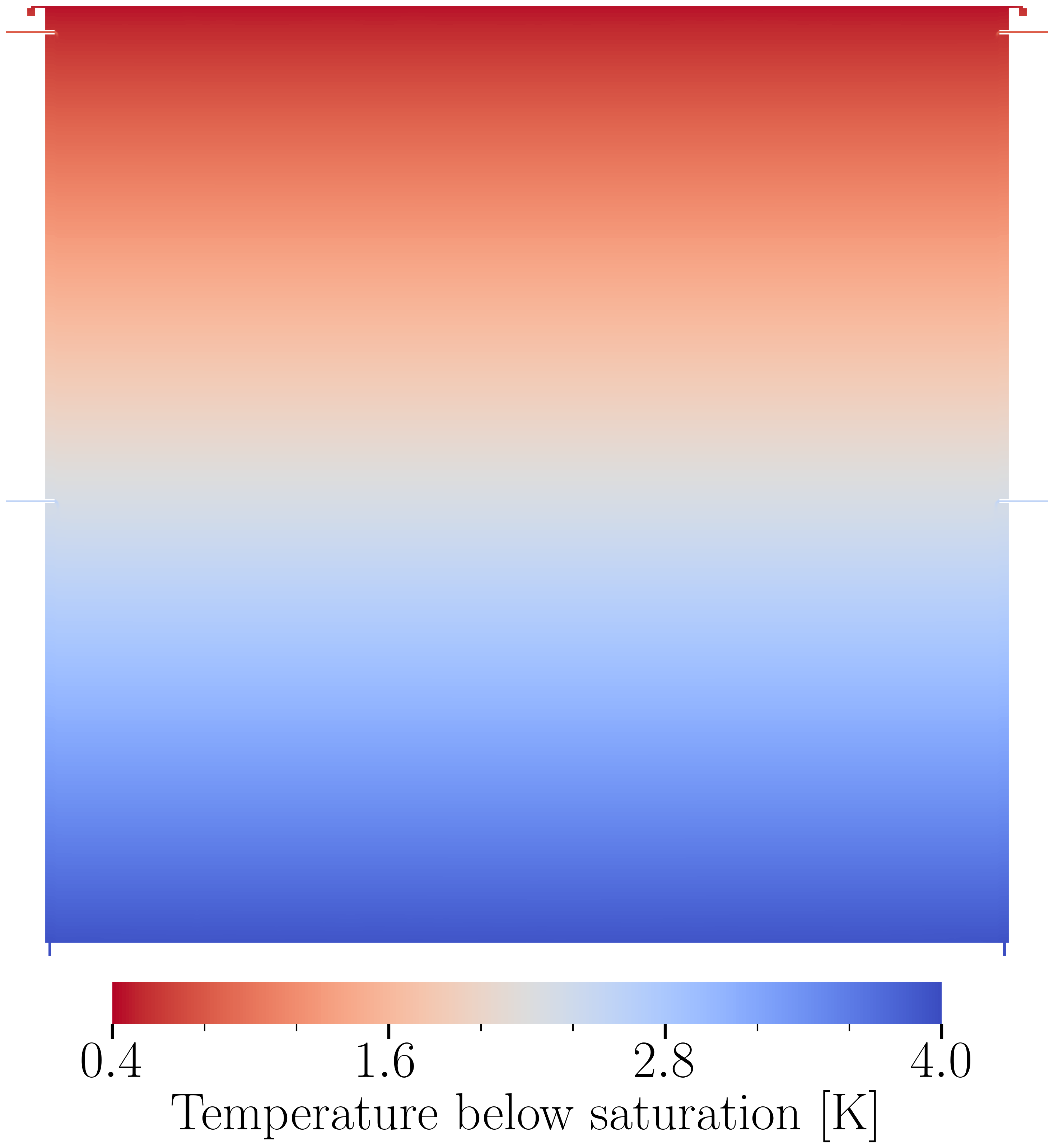}
\end{subfigure}
\caption[]{Temperature maps of the TPC at a cathode window underside temperature of $\SI{89}{K}$. (Top left): Vertical cut view through an outlet (cf.~figure~\ref{fig:CFD_geometry}~(detail~3)). (Top right): Vertical cut view through a top inlet (cf.~figure~\ref{fig:CFD_geometry}~(detail~1)). (Bottom): Temperature difference to the boiling temperature at the respective depths.}
\label{fig:Temperature_spatial_distribution_with_heat}
\end{figure}

To further simplify the complexity of the problem, we proceed with a uniform mean temperature for the cathode window underside instead of using the actual temperature distribution visible in figure~\ref{fig:cathode_temperatures}~(right): the rounded nominal cathode window underside temperature used in the following is $\SI{89}{K}$. This impacts the resulting convection cell in the TPC just as the constant-thickness treatment of the cathode window. We discuss this in the context of the convection pattern below. 

In figure~\ref{fig:Temperature_spatial_distribution_with_heat}~(top) the TPC temperature maps at nominal bottom photoelectronics heat are presented. As anticipated from experience with large LAr volumes~\cite{DUNE:2021hwx}, the result shows almost no temperature inhomogeneity in the bulk, featuring fluctuations of a few millikelvin. Only in proximity of boundary interfaces, in particular at the cathode, the inlets, around the wire grid frame and behind the reflector panels, gradients of a few tens of millikelvin are visible. In proximity of the cathode, the temperature rises towards the centre of the cathode. Isolated by the reflector panels, a slight temperature separation of the active volume and the space behind the reflectors is visible. The argon injected through the inlets sinks because it has a lower temperature with respect to the bulk and thus a higher mass density. The region around the wire grid frame is slightly colder than the bulk because the stainless steel of the frame carries heat efficiently towards the inner veto volume\footnote{The total heat conduction rate through the gate grid wires in this region of highest temperature gradient is estimated to be only a few milliwatts, thereby justifying their exclusion from the model from a thermal perspective.}. All other boundaries are well insulated with thick PMMA sheets. The volume-averaged temperature of the fluid is $\SI{88.15}{K}$. In figure~\ref{fig:Temperature_spatial_distribution_with_heat}~(bottom) we provide a map of the temperature difference of the simulation result to the saturation point. Although the temperature in the model is largely uniform, this map shows a gradient because the saturation temperature decreases upward due to hydrostatic pressure. While at the cathode depth the liquid is sub-cooled by $\SI{4}{K}$, the temperature at the top of the model reaches as close as $\SI{0.4}{K}$ to the boiling point. However, even at the highest considered temperature of $\SI{92}{K}$ on the underside of the cathode window, the temperature throughout the model remains at least $\SI{55}{mK}$ below the boiling point. Note that this simulation does not include the gas phase and thus also no interaction with the gas pocket through phase change.

\begin{figure}[t]
\centering
\begin{subfigure}[b]{0.48\textwidth}
\centering
\includegraphics*[width=\textwidth]{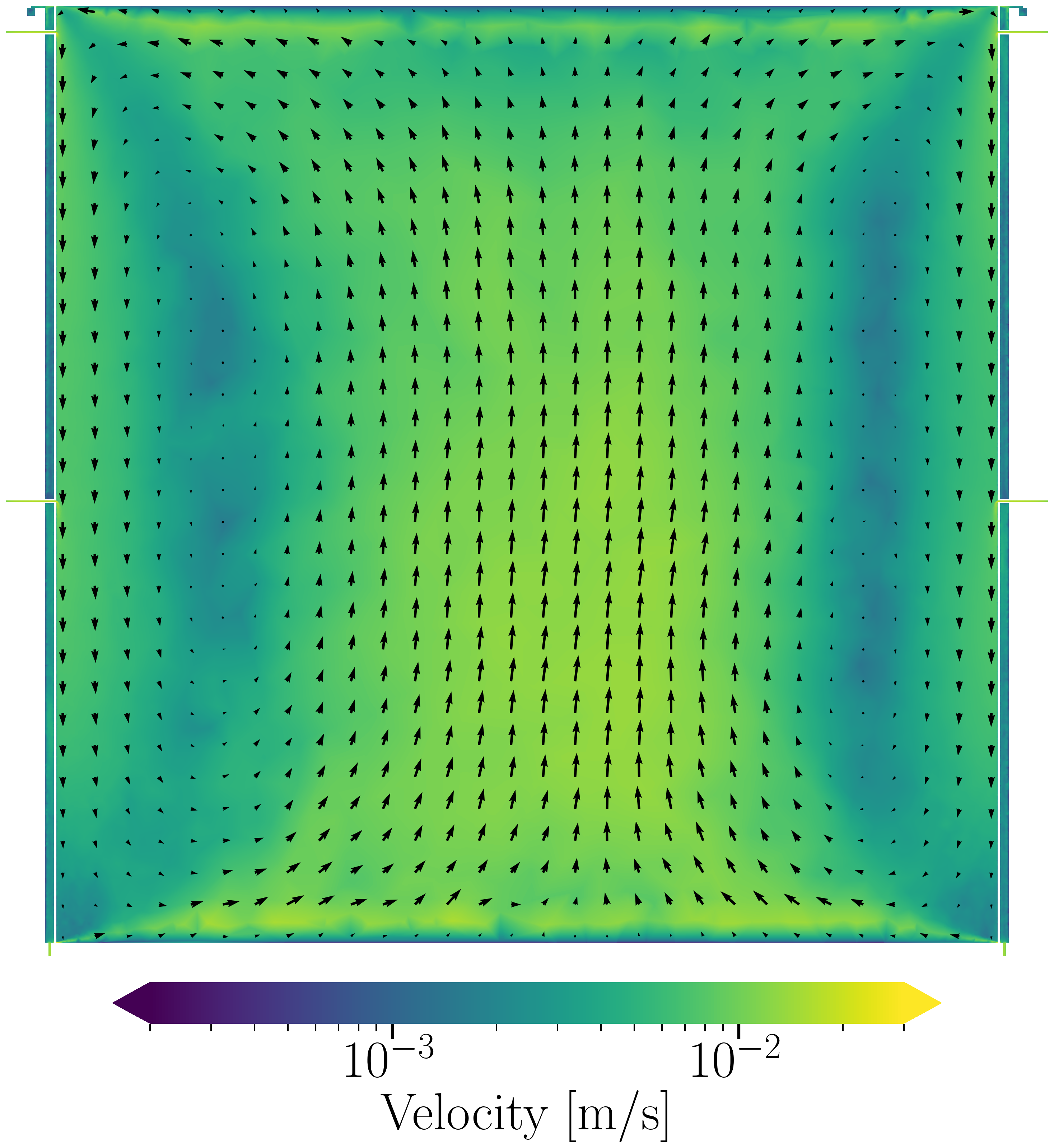}
\end{subfigure}
\quad
\begin{subfigure}[b]{0.48\textwidth}
\centering
\includegraphics*[width=\textwidth]{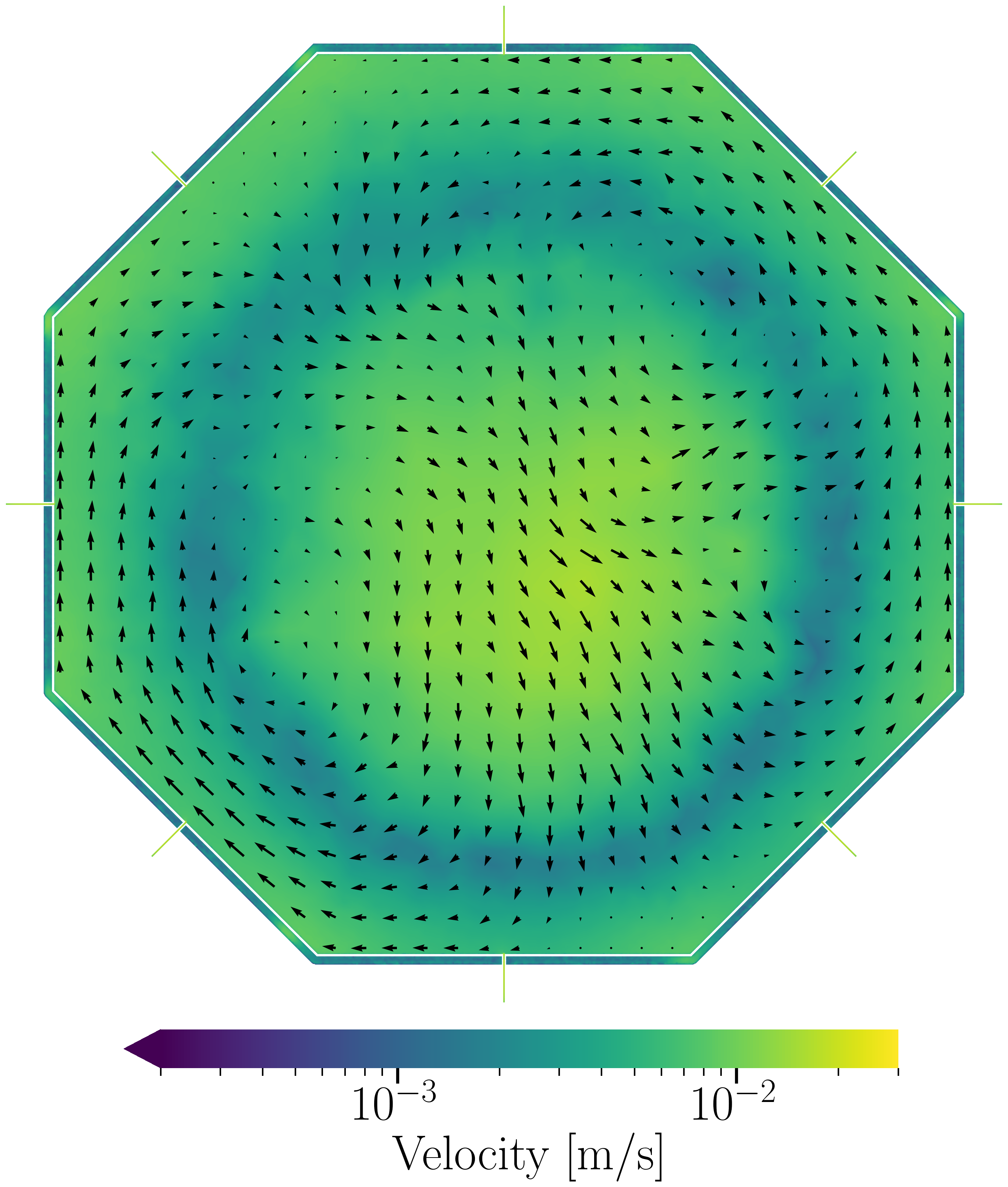}
\end{subfigure}
\quad
\begin{subfigure}[b]{0.48\textwidth}
\vspace{0.3cm}
\centering
\includegraphics*[width=\textwidth]{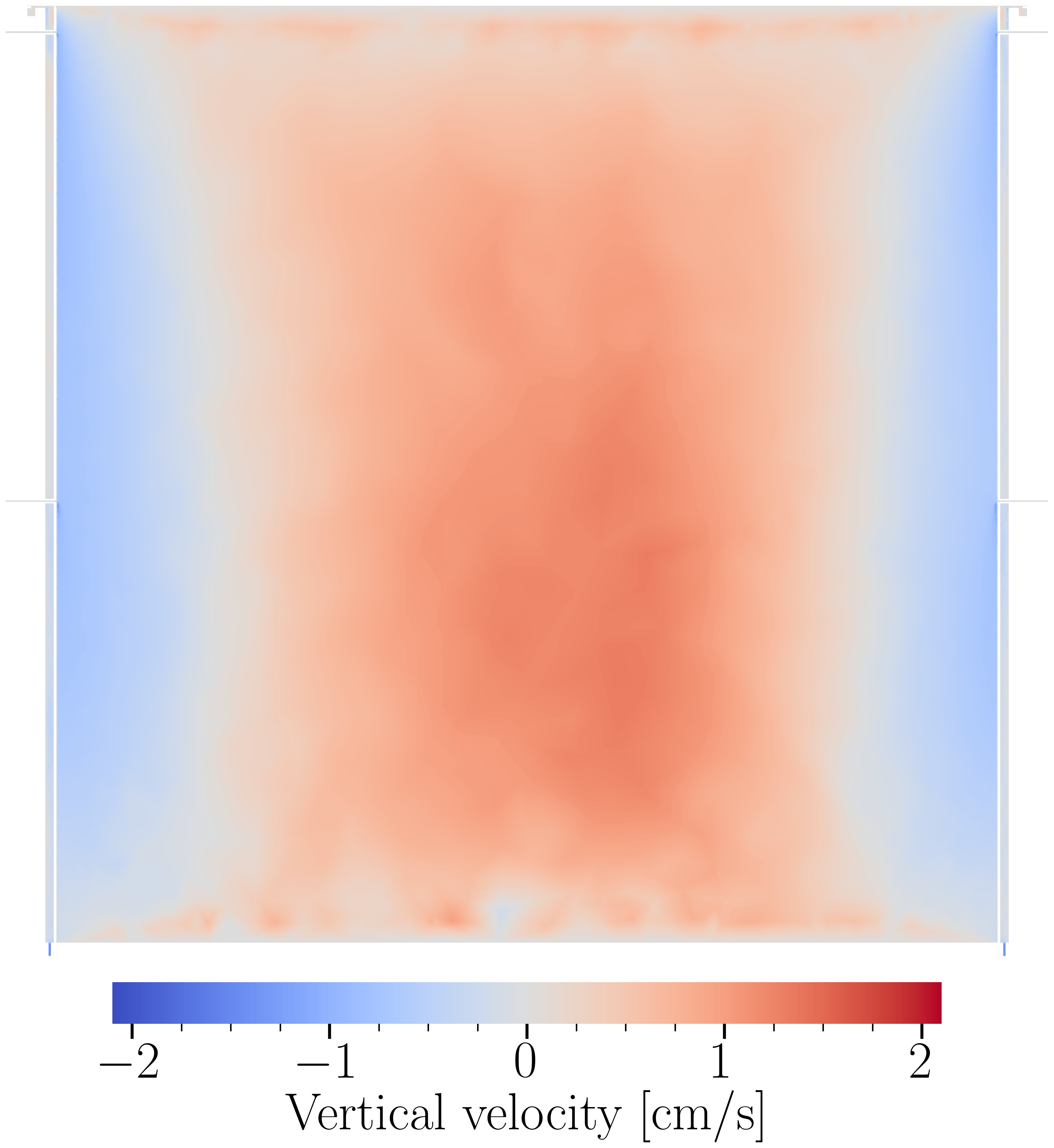}
\end{subfigure}
\caption[]{Flow velocity maps with a cathode window underside temperature of $\SI{89}{K}$. (Top left): Vertical cut view through the inlets and outlets. (Top right): Horizontal cut view through the lower row of inlets. The velocity vectors are scaled (relative to the left plot) for improved visibility. (Bottom): Vertical cut view through the inlets and outlets. Shown is the velocity in vertical direction with red indicating upwards and blue indicating downwards flow.}
\label{fig:Velocity_spatial_distribution_with_heat}
\end{figure}

\begin{figure}[t]
\centering
\includegraphics[width=1\textwidth]{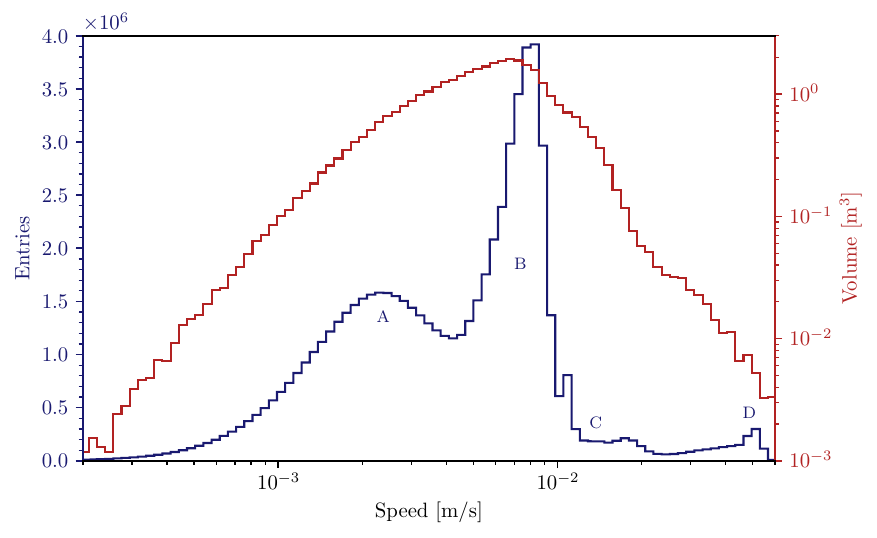}
\caption{Fluid speed distribution with the cathode window underside at $\SI{89}{K}$. The blue histogram shows the speeds of the individual fluid cells, providing a good visualisation of the speeds of the individual detector regions: A~--~Slow components of the bulk convection cell, behind the reflector panels, in proximity of wire grid frame and surfaces. B~--~Fast components of the bulk convection cell. C~--~Inlets and outlets. D~--~Gap between the cathode and the reflector panels. The red histogram is weighted by the volume of the fluid cells.}
\label{fig:SpeedDistribution_with_heat}
\end{figure}

\begin{figure}[t]
\centering
\includegraphics[width=0.618\textwidth]{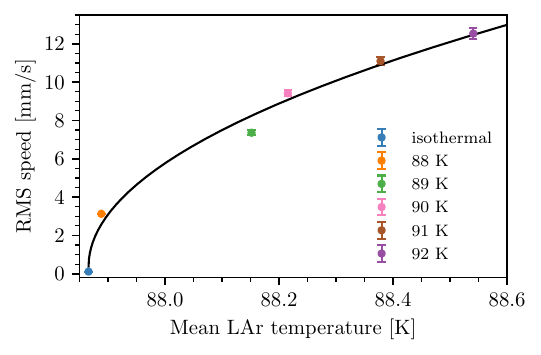}
\caption{Volume-averaged RMS speed at various cathode window underside temperatures as a function of the volume-averaged LAr bulk temperature. The error bars represent the standard errors on the RMS speeds.}
\label{fig:mean_speed_with_heat}
\end{figure}

The temperature gradients in the model, although small, drive a convection cell in the TPC, with liquid rising in the centre and falling on the sides (see figure~\ref{fig:Velocity_spatial_distribution_with_heat}). The flow direction in this pattern is opposite to the case without heat (cf.~figure~\ref{fig:Velocity_spatial_distribution_no_heat}~(bottom right) in appendix~\ref{app:flowpattern1}). The convection cell is located slightly off-centre (see figure~\ref{fig:Velocity_spatial_distribution_with_heat} (top right)). Stable CFD solutions to the problem that feature slight asymmetries in the flow pattern exist. These are spontaneously assumed and, although the solutions are sufficiently converged and stable, those asymmetries arise or vanish while the solver further iterates. In cylindrical geometries with adiabatic walls and diameter-to-height aspect ratios of 1 or less, a single convection roll is expected, characterised by rising fluid on one side of the cylinder and falling fluid on the other~\cite{PhysRevFluids.6.090502}. While in adiabatic enclosures observed only at aspect ratios close to 2, we find an almost axially symmetric cell at an aspect ratio slightly above 1. This phenomenon arises due to heat dissipation through the wall boundaries, which induces a downward fluid movement close to the side walls and establishes an almost axially symmetric cell. 

Based on the mean temperature of $\SI{88.15}{K}$ observed in the model, the constant temperature boundary on the cathode window underside, and the assumed average thickness of the cathode window, we calculate a typical vertical temperature gradient of $\SI{15}{mK/mm}$ in the cathode window. Assuming the same bulk temperature in case the actual cathode window underside temperature distribution (figure~\ref{fig:cathode_temperatures}~(right)) had been used, it would have resulted in a gradient of $\SI{51}{mK/mm}$ in the centre and a negligible one at the largest radii. This estimation demonstrates that the convection cell would in this case be driven by an even stronger central heat source with larger temperature gradients in the radial direction in proximity of the cathode, likely resulting in more pronounced convective currents. Since the constant temperature boundary and the constant-thickness cathode window cannot accurately represent the actual heat flux distribution over the cathode plane, we scan through other cathode window underside temperatures: $88, \, 90, \, 91, \, \SI{92}{K}$. This allows us to study the convective behaviour around the actual total heat flux. We highlight results from those simulations below in the context of the RMS fluid speed.

Figure~\ref{fig:SpeedDistribution_with_heat} provides the speed distribution of the entire fluid volume at a cathode window underside temperature of $\SI{89}{K}$, highlighting the typical speeds that are present in certain detector regions. Compared to the flow-only case (cf.~figure~\ref{fig:Velocity_spatial_distribution_no_heat} in appendix~\ref{app:flowpattern1}), the speeds in the bulk of the liquid are increased by two orders of magnitude due to the added convection. The fluid speeds in essentially the entire fluid volume are higher than $\SI{0.2}{mm/s}$, i.e.~higher than the majority of the speeds found in the bulk of the flow-only case. The region of the highest speeds, higher than in the inlets and outlets, is found in the $\SI{1.5}{mm}$ gap between the reflector panels and the cathode. This is likely due to the temperature difference between the bulk and the region behind the reflector panels that was mentioned above. The strong rise of the overall speed due to the bottom photoelectronics heat can also be seen in figure~\ref{fig:mean_speed_with_heat}, in which the RMS speed of the fluid volume $v_{\mathrm{RMS}}$, i.e.~the square root of the volume-weighted sum of the quadratic speeds of all fluid cells, as a function of the volume-weighted mean of the fluid temperature $\bar{T}$ is shown for the studied cathode window underside temperatures. The RMS speed is $(\SI{7.4}{}\pm\SI{0.1}{})\,\si{mm/s}$ at a cathode window underside temperature of $\SI{89}{K}$ and reaches $(\SI{12.5}{}\pm\SI{0.3}{})\,\si{mm/s}$ at $\SI{92}{K}$. This is of the same order of magnitude of the RMS flow speeds of $\SIrange{3}{7}{mm/s}$ that are observed in dual-phase xenon TPCs for which the flow pattern is convection-dominated~\cite{XENON:2024lbh,XENON:2016rze,Malling:2014oxk}. If we interpret the additional thermal energy of the fluid (relative to the isothermal flow-only case) as the source of the convective fluid motion, we can identify:
\begin{equation}
\frac{N\cdot m_{\mathrm{Ar}}}{2}v_{\mathrm{RMS}}^{2} = \frac{N}{2}k_{\mathrm{B}}\bar{T} \quad .
\end{equation}
Here $N$ denotes the number of argon atoms in the fluid volume and $m_{\mathrm{Ar}}$ the atomic mass of argon. This is a non-standard application of the equipartition theorem, in that it does not describe random thermal atomic motion but rather the vectored flow of atoms grouped into fluid cells with one degree of freedom. This interpretation of a thermal kinetic energy motivates a square-root function as fit model in figure~\ref{fig:mean_speed_with_heat}:
\begin{equation}
v_{\mathrm{RMS}} = \alpha \cdot \sqrt{\bar{T} - \bar{T_{0}}}+\beta \quad .
\end{equation}
The offset $\beta$ accounts for the non-zero fluid speeds in the isothermal flow-only case due to recirculation. We can then further identify $m_{\mathrm{Ar}}=k_{\mathrm{B}}/\alpha^2$. We find $\alpha=(\SI{14.7}{}\pm \SI{0.8}{})\,\si{ms^{-1}K^{-1/2}}$ and thus $m_{\mathrm{Ar}}=(\SI{6.4}{}\pm\SI{0.7}{})\times\SI{e-26}{kg}$, which is compatible with the literature value of $\SI{6.6e-26}{kg}$~\cite{NIST}, providing a good sanity check of the thermal simulation. Comparing the fluid kinetic energies using the RMS speeds, we find an increase of factor $\num{\sim 4300}$ from the isothermal flow-only case to the case with $\SI{89}{K}$ cathode window underside temperature.

\subsection{Krypton-83m calibration prospects}
\label{subsec:Kr_calibration2}

While the RTD at the outlet provides only limited information on the reach of the \ce{^{83\text{m}}Kr} source due to the long turnover time, we can determine the time required for the liquid to laterally reach the region near the central symmetry axis of the TPC\footnote{Since advection, rather than diffusion, is the dominant mass transport mechanism in the system, a suitable typical mixing time scale is the advection time, defined as the characteristic length divided by the typical flow speed. Using the radius of the inscribed circle of the active volume octagon along with the RMS speed, the advection time is $\SI{4}{min}$.}, which is relevant for calibrations discussed in section~\ref{Sec:Motivation}.
\begin{figure}[t]
\centering
\begin{subfigure}[b]{0.41\textwidth}
\centering
\includegraphics*[width=1\textwidth]{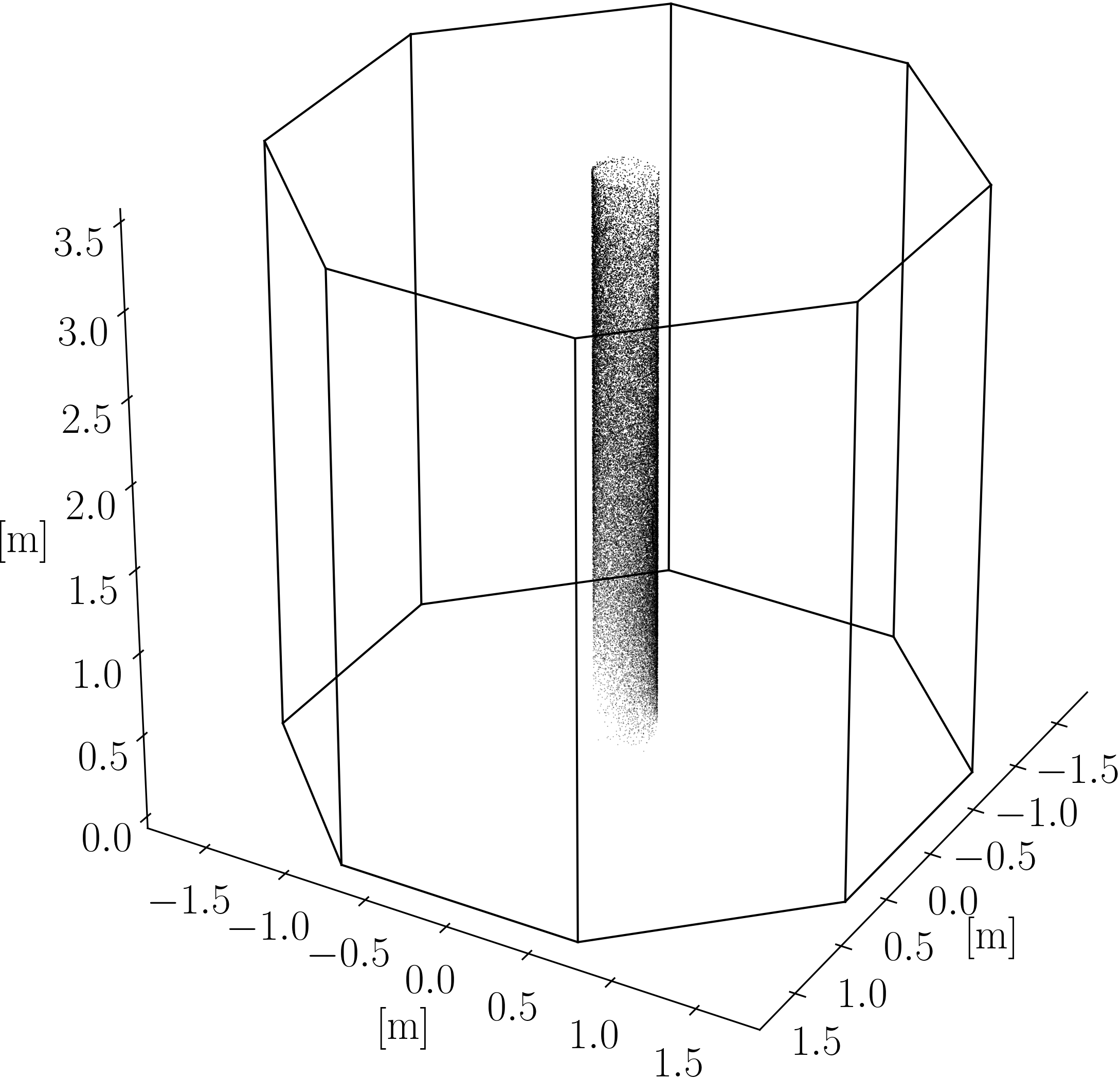}
\end{subfigure}
\quad
\begin{subfigure}[b]{0.55\textwidth}
\centering
\includegraphics*[width=1\textwidth]{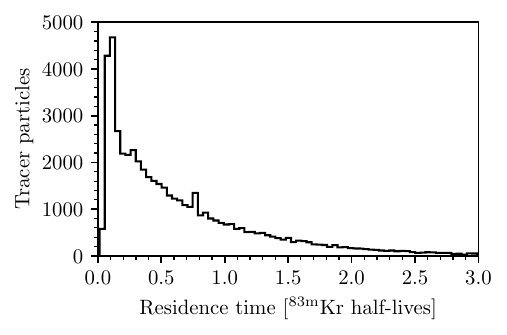}
\end{subfigure}
\caption[]{Tracer particles impacting a cylindrical shell in the TPC centre at $\SI{89}{K}$ cathode window underside temperature. (Left): Spatial distribution of tracers arriving within $10 \cdot T_{1/2}$ of \ce{^{83\text{m}}Kr}. The octagonal prism shown represents the active liquid volume of the TPC. (Right): RTD in units of \ce{^{83\text{m}}Kr} half-lives ($\SI{1.83}{h}$).}
\label{fig:Cylinder_with_heat}
\end{figure}
To assess the horizontal reach of the source along the entire vertical direction, we define a central cylinder shell that is collinear to the TPC's vertical symmetry axis and record the impacting tracer particles on this surface. Tracers are removed from the simulation upon their first impact on the cylinder shell. In figure~\ref{fig:Cylinder_with_heat} we present both spatial and residence time distributions of the tracers that arrived on a cylinder with a radius of $1/10$ of the radius of the inscribed circle of the active volume octagon ($\SI{17.5}{cm}$) for the nominal $\SI{89}{K}$ case. While the spatial distribution in figure~\ref{fig:Cylinder_with_heat}~(left) shows the tracers that arrived  within $10 \cdot T_{1/2}$ of \ce{^{83\text{m}}Kr}, the RTD in figure~\ref{fig:Cylinder_with_heat}~(right) is cut off at $3 \cdot T_{1/2}$. Except near the vertical boundaries of the fluid volume, the tracers are distributed homogeneously over the cylinder's shell due to the strong convection inside the TPC. The RTD shows that the tracers arrive with negligible delay, compared to the \ce{^{83\text{m}}Kr} half-life. The earliest tracers arrive $\SI{1.5}{min}$ after injection and the RTD peaks at $0.12 \cdot T_{1/2}$ ($\SI{13}{min}$). The residence time quickly falls off as expected in a well-mixed volume. A fraction of $\SI{99}{\%}$ of the tracked particles, released at the model inlets, traverse the cylinder's shell within the simulated period of $10 \cdot T_{1/2}$. A fraction of $\SI{94}{\%}$ of those collected tracers are impinging the cylinder within $2 \cdot T_{1/2}$, and $\SI{98}{\%}$ within $3 \cdot T_{1/2}$. This result demonstrates the promising calibration prospects with \ce{^{83\text{m}}Kr}. Note that this method can be applied to other surfaces in the TPC to identify potential blind spots. The cylinder surface used here, however, is sufficient to demonstrate the prospects of the source for the S2 signal calibration of the experiment.
\section{Dual-phase case} 
\label{sec:dual_phase_gas_pocket}

In this section we additionally consider a gas pocket on top of the liquid volume of the TPC. A gas pocket with a volume of $\SI{75}{L}$ can be created below the anode through the injection of GAr. A constant liquid level is maintained by ejecting excess gas through a small diving bell outside of the TPC at the height of the gas pocket.
The gas pocket and the liquid bulk below are not in thermal equilibrium, i.e.~the LAr is sub-cooled with respect to the saturation temperature at this depth and thus the gas pocket would vanish by condensation on the liquid surface if the supply were cut. Injected GAr is at or only slightly above this saturation temperature when created via boiling of proximity LAr or supplied by the cryogenics system through non-insulated tubing~\cite{DarkSide-20k:UArCryo}. In this section, we estimate the condensation rate of GAr in the gas pocket on the LAr surface and demonstrate the influence of the gas pocket on the internal TPC liquid flow. The condensation rate is relevant for the DS"~20k design because it informs the required GAr flow towards the gas pocket. This allows for an appropriate dimensioning of the required tubing to guarantee a minimal pressure drop over the gas pocket formation system, as required for a uniform gas pocket thickness and S2 signal uniformity. It further guides the dimensioning of the GAr source. We also assess the thermal conditions in the gas pocket that are necessary to prevent condensation on the anode.

The mass transfer rate across the interface of two-phase fluids can be modelled with the Hertz-Knudsen-Schrage equation that is derived solely from statistical mechanics~\cite{Vaartstra:2022,Zhang_2017}. However, this approach is impractical for our application for two reasons. First, it requires precise mass accommodation coefficients for condensation and evaporation. These describe the probability of an atom to transition into the other phase upon impact on the interface. Existing measurements of argon mass accommodation coefficients are insufficient for a precise estimation of the mass transfer~\cite{Yasuoka:1994}. Second, using the bulk temperature as value for the liquid temperature in proximity of the interface leads to an overestimation of the mass transfer. Instead, a boundary layer of an a priori unknown thickness is established which limits the heat transfer and in turn the mass transfer. For these reasons, we adapt a thermodynamically motivated approach instead. In a first step, we simulate the cooling power provided via convection and conduction from the liquid bulk to the interface, while neglecting any phase interaction. In a second step and deploying a mock-up geometry, we simulate the condensation rate of GAr, injected at a typical flow rate, on a surface that provides the obtained cooling power.
This demonstrates that heat transfer from the GAr to the liquid-gas interface via phase transition is indeed possible at this rate. 

\subsection{Framework}
\label{subsec:framework3}

\begin{figure}[t]
\centering
\begin{subfigure}[c]{0.342\textwidth}
\centering
\includegraphics*[width=\textwidth]{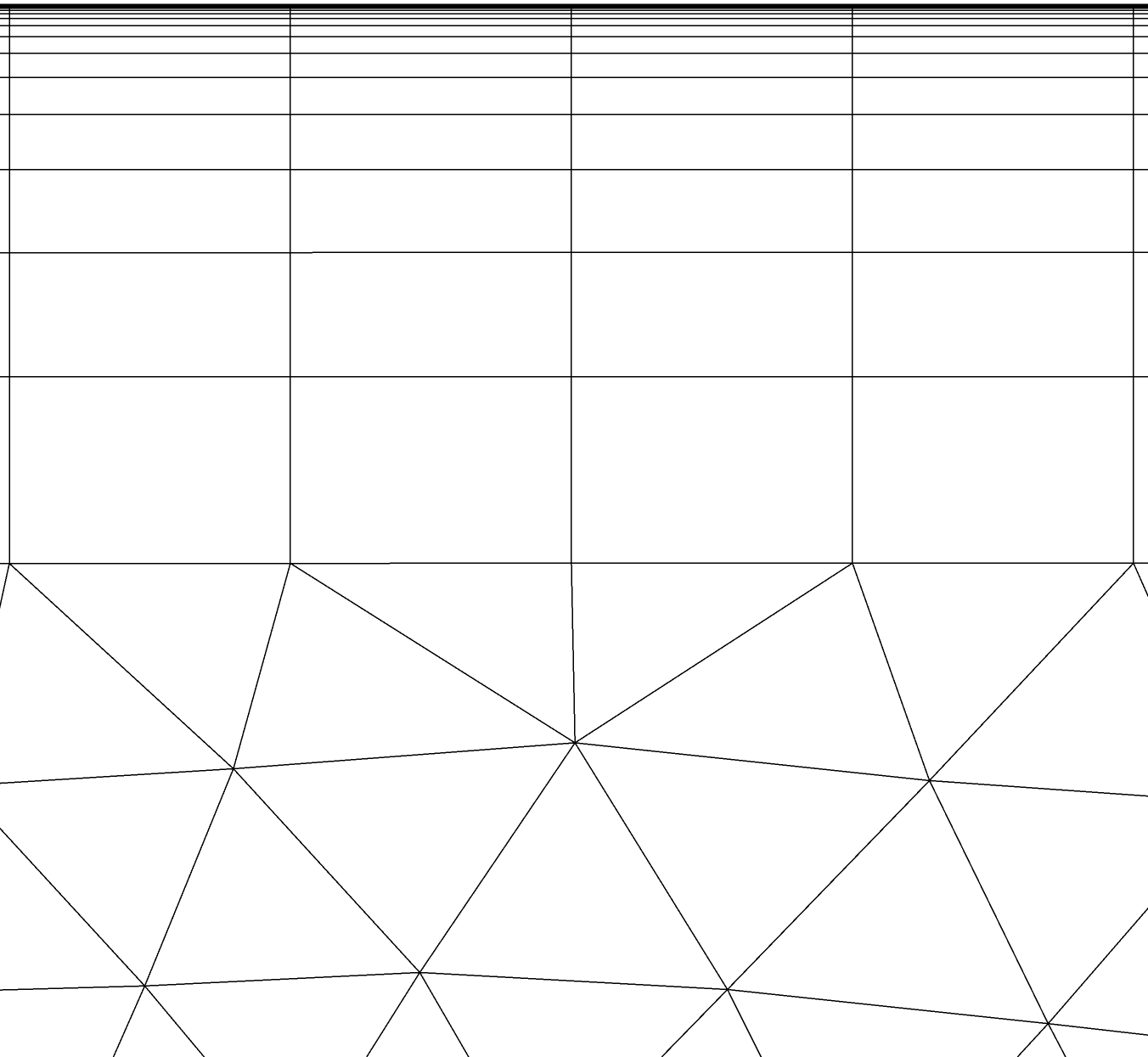}
\end{subfigure}
\quad
\begin{subfigure}[c]{0.618\textwidth}
\centering
\includegraphics*[width=\textwidth]{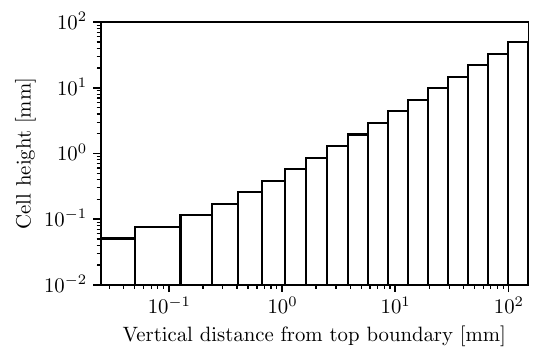}
\end{subfigure}
\caption[]{Top boundary mesh. (Left): Inflation layer with unstructured grid below. (Right): Nominal height of the cells in the inflation layer and the first tetrahedral cell. The first millimeter from the boundary is divided into 6~cell layers of increasing size.}
\label{fig:meshinflationlayer}
\end{figure}

\paragraph{Geometry and mesh}
The TPC geometry deployed for the heat transfer rate determination is further simplified to reduce the computational complexity. For this the reflector panels are removed and the region around the wire grid frame is cut, i.e.~the fluid volume is abstracted to a right orthogonal prism, with all boundary surfaces made of PMMA. This choice will impact the shape of the convection cell because the reflector panels are a barrier for flow and heat transfer. However, the objective of this section is to obtain a robust upper limit on the heat transfer across the liquid-gas interface. The heat dissipation over the TPC walls towards the veto volume increases if the reflector panels are removed. Thus, we expect to generally overestimate the liquid-gas interface heat transfer rate. The impact of removing the wire grid frame on the heat transfer is estimated below. We use a mostly unstructured tetrahedral fluid mesh without defeaturing with a minimum (maximum) side length of $\SI{9e-3}{mm}$ ($\SI{100}{mm}$) and a growth rate of $1.2$ between cells (typical cell side length: $\SI{50}{mm}$). To represent the liquid-gas boundary layer with appropriate granularity, an inflation layer with a structured grid consisting of 17 layers of prisms with a first-layer thickness of $\SI{5.07e-2}{mm}$ and a growth rate of $1.5$ is used (see figure~\ref{fig:meshinflationlayer}). The geometry is represented by $\num{1.6e6}$ fluid cells. This results in an orthogonality of $0.74\pm 0.13\,\mathrm{(std.)}$ (min.~$0.09$) and a skewness of $0.25\pm 0.13\,\mathrm{(std.)}$ (max.~$0.90$). To limit the total element count, the aspect ratios of the thin cells in the inflation layer are high with a mean of $\num{\sim200}$. However, the vertical top boundary heat transfer and the vertical flow simulation benefit from a fine fragmentation in the vertical direction. As we see below, the stability of the energy solution is not effected by this choice, cf.~figure~\ref{fig:heattransferconvergenceboundarylayertemperaturecurve}~(left).

\paragraph{Physics and solver setup}
While maintaining the same settings as in the previous section, the thermal boundary conditions here necessitate the use of the transient solver for good convergence. The solver adapts typical time steps in the range of $\SIrange{3}{4}{s}$, resulting in good numerical behaviour given the typical flow speeds and cell dimensions on the order of $\SI{1}{mm/s}$ and $\SI{10}{mm}$, respectively.

\paragraph{Operating and boundary conditions}

\begin{table}
\caption[Thermal Boundary Conditions]{Thermal boundary conditions for the liquid-phase TPC simulation with bottom and top heat.}
\centering
\begin{tabularx}{\textwidth}{>{\setlength{\baselineskip}{0.85\baselineskip}}Xc >{\setlength{\baselineskip}{0.85\baselineskip}}X >{\setlength{\baselineskip}{0.85\baselineskip}}X}
\toprule
\textbf{Condition}   & \textbf{Value}  & \textbf{Setup}   & \textbf{Comment}\\
\midrule
\textbf{Boundary}&&&\\
Nominal cathode window underside temperature & $\SI{89}{K}$ & Wall with fixed temperature and thickness &\\
&&&\\[-10pt]
Liquid-gas interface temperature & $\SI{88.60}{K}$ & Wall with fixed temperature & Saturation temperature at gas pocket pressure\\
&&&\\[-10pt]
Temperature of other TPC outer surfaces & $\SI{87.87}{K}$ & Walls with fixed temperature and thicknesses & Saturation temperature at ullage pressure\\
\bottomrule
\end{tabularx}
\label{tab:conditions3}
\end{table}

We repeat the liquid-phase simulation of the preceding section but change the temperature boundary condition of the top surface (liquid-gas interface). This boundary is a zero thickness wall at a saturation temperature of $\SI{88.60}{K}$ corresponding to the gas pocket pressure of $\SI{1.160}{bara}$. All other boundary conditions remain unchanged, see table~\ref{tab:conditions3}. The thermal boundary condition at the top surface allows us to model the liquid-phase flow of a dual-phase TPC using a single-phase simulation. It therefore accounts only for flows driven by heat transfer across this surface, while neglecting those resulting from mass transfer due to condensation. However, the condensation rate we determine below implies a downward flow with a velocity on the order of $\SI{e-8}{m/s}$, which is negligible compared to the convective flows present in the detector.

\subsection{Heat transfer rate across the liquid-gas interface}
\label{subsec:heat_transfer_lquid-gas_interface}

As noted above, the liquid bulk of the TPC is generally sub-cooled relative to the saturation temperature corresponding to the gas pocket pressure, as it is governed by the lower ullage pressure in the inner veto volume. However, the liquid-gas interface is necessarily at saturation temperature of the gas pocket as required for the phase change. This vertical temperature gradient induces a downward heat flux. The liquid removes the latent heat from the gas through convection and heat conduction, enabling condensation. This heat transfer limits the possible condensation rate. Figure~\ref{fig:Temperature_spatial_distribution_with_heat}~(top left) shows that the liquid temperature right above the cathode at nominal bottom photoelectronics heat is $\SI{88.17}{K}$. This is lower than the top temperature boundary of the liquid-gas interface considered here. Thus, the temperature (mass density) increases (decreases) in the upward direction. In the absence of other boundary conditions, this temperature inversion (compared to the case with bottom heat only) would lead to a quiescent thermally stratified flow. Without convective currents, the sole heat transfer process is heat conduction. The Nusselt number, which describes the ratio of the total heat transfer (heat transfer coefficient $h$) and the conductive heat transfer (thermal conductivity $k$), is then unity~\cite{Bergman} (section~9.8.1):
\begin{equation}
    \mathrm{Nu}_L\coloneq\frac{h}{k/L}=1 \quad .
    \label{eq:NusseltStratifiedFlow}
\end{equation}
Here, $L$ denotes the characteristic length of the geometry, in this case the distance of the warm top and the cold bottom boundaries of the horizontal enclosure. This equation is valid for all Rayleigh numbers $Ra_{L}$, which characterise the fluid's flow regime, in laminar and turbulent, for buoyancy-driven flow. The Rayleigh number is the product of two dimensionless numbers: the Grashof number $Gr_{L}$ and the Prandtl number $Pr$. The former is the ratio of buoyancy and viscous forces, the latter is the ratio of momentum and thermal diffusivity.

The characteristic length of the problem reduces to the thickness of a boundary layer if the steady condition is disturbed by forced currents from the recirculation flow or convective currents due to other temperature boundaries. For this reason, it is essential to derive the boundary layer thickness from simulation to evaluate equation~\eqref{eq:NusseltStratifiedFlow}.
\begin{figure}[t]
\centering
\begin{subfigure}[b]{0.48\textwidth}
\centering
\includegraphics*[width=\textwidth]{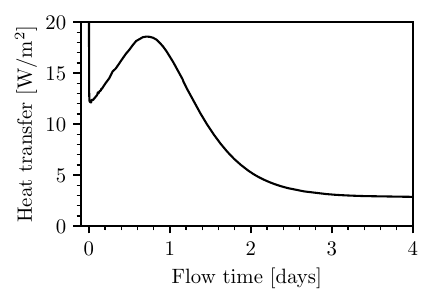}
\end{subfigure}
\quad
\begin{subfigure}[b]{0.48\textwidth}
\centering
\includegraphics*[width=\textwidth]{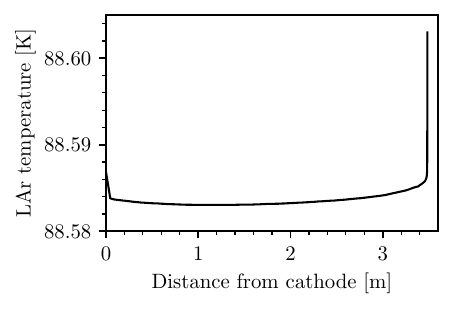}
\end{subfigure}
\caption[]{(Left): Heat flux from the top boundary to the liquid as a function of flow time after the start of the solving process. (Right): Vertical temperature profile close to the centre of the TPC.}
\label{fig:heattransferconvergenceboundarylayertemperaturecurve}
\end{figure}
Equilibrium conditions are only established slowly when initialising the problem with a uniform temperature of $\SI{87.87}{K}$ but the transient solver shows good convergence. Monitoring the convergence, we observe the formation of a thin boundary layer. The overall temperature increase of the bulk quickly reduces the heat flux from the top boundary, starting from an initial value of $\SI{187.4}{W/m^2}$ (see figure~\ref{fig:heattransferconvergenceboundarylayertemperaturecurve}~(left)). We obtain a well-converged mean heat flux of $\SI{2.9}{W/m^2}$ after $\SI{4}{d}$ of simulated time, which is approximately one tenth of the time required for a full turnover. This value corresponds to a total heat load of $\SI{30.6}{W}$ when integrated over the liquid surface or to the latent heat transfer rate required to liquefy a $\SI{1.8}{L/min}$ flow of cold, saturated gas. The vertical temperature profile in the TPC centre is shown in figure~\ref{fig:heattransferconvergenceboundarylayertemperaturecurve}~(right). The boundary layer at the top and the higher temperature in proximity of the cathode are clearly visible. In the region of the highest temperature gradient, we estimate an approximate typical thickness of the top boundary layer of $L=\SI{1}{mm}$\footnote{The mesh of the inflation layer is sufficiently refined to accurately represent this thickness (cf.~figure~\ref{fig:meshinflationlayer}).}. At this depth below the top boundary we find an average temperature difference of $\Delta T= \SI{14}{mK}$ from the top boundary condition. Using Newton's law of cooling with the interface area $A_{S}$, 
\begin{equation}
    \frac{\dot Q}{A_{S}} = h \cdot \Delta T \quad ,
    \label{eq:NewtonLawCooling}
\end{equation}
and equation~\eqref{eq:NusseltStratifiedFlow}, we obtain a conductive heat flux of $\dot Q/A_{S}=\SI{1.7}{W/m^2}$, which is approximately $\SI{60}{\%}$ of the simulated value. Thus, conductive heat transfer is indeed a significant contributor to the total heat transfer. However, the flow is evidently not entirely quiescent at the top boundary. 

\begin{figure}[t]
\centering
\begin{subfigure}[b]{0.48\textwidth}
\centering
\includegraphics*[width=\textwidth]{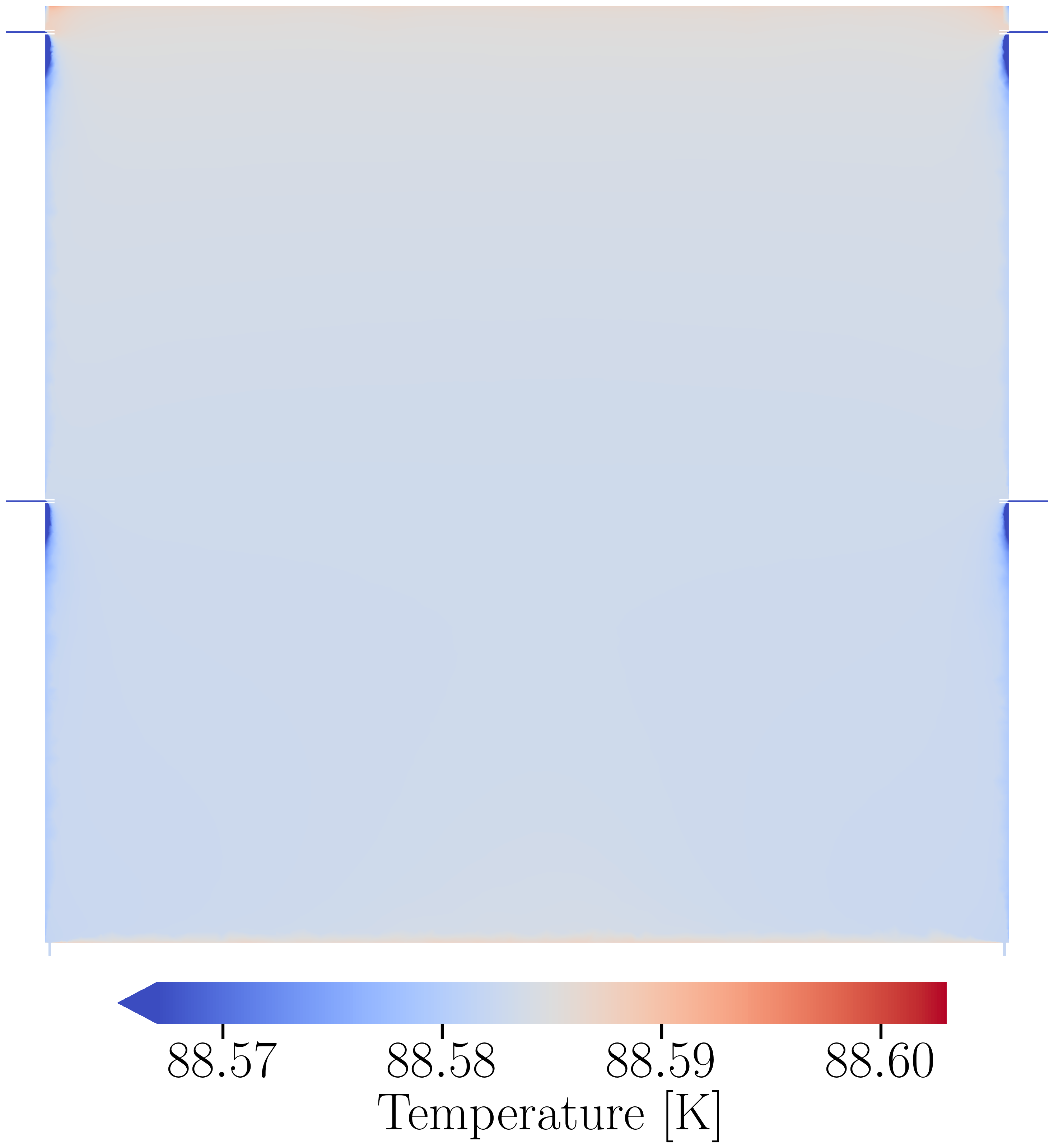}
\end{subfigure}
\quad
\begin{subfigure}[b]{0.48\textwidth}
\centering
\includegraphics*[width=\textwidth]{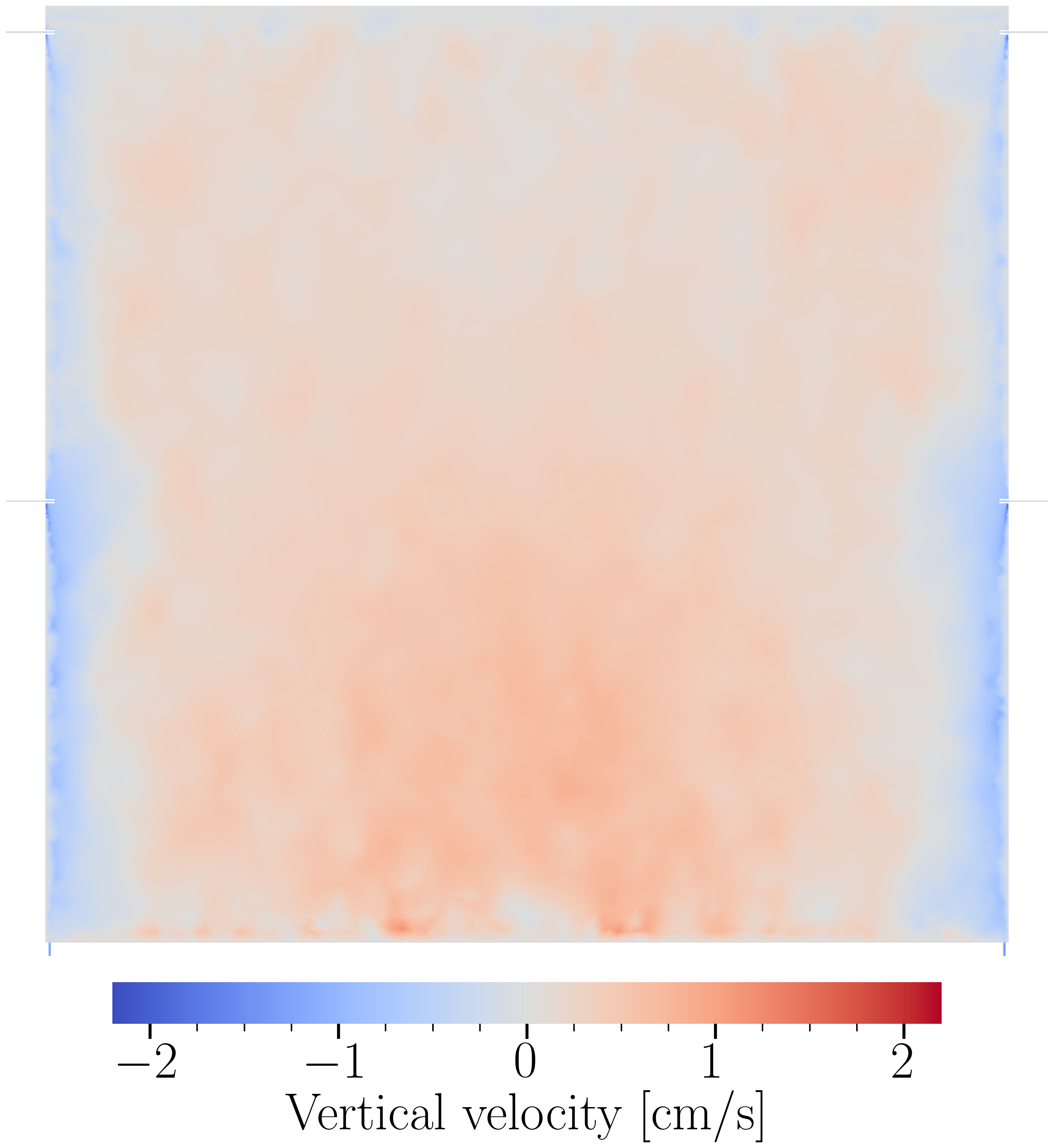}
\end{subfigure}
\caption[]{Simulation results with the top boundary interface at the gas pocket saturation temperature and with nominal bottom photoelectronics heat. (Left): Temperature map. (Right): Vertical velocity.}
\label{fig:LiquidInterface}
\end{figure}

The temperature map and the altered flow pattern are shown in figure~\ref{fig:LiquidInterface}. While the vertical flow speed is reduced towards the top, the flow pattern still features the convection cell induced by the bottom photoelectronics in the lower half of the TPC. This strong convective current in the bulk indicates good mixing and enduring calibration prospects with internal sources. The temperature in the liquid bulk is widely uniform and only $\SI{\sim 20}{mK}$ below the top boundary temperature, leading to the relatively small heat transfer from the top boundary. This demonstrates the strong thermal insulation provided by the thick PMMA walls, as heat dissipation from the TPC to the veto volume is limited by conduction through the TPC walls. Based on the boundary surfaces and the wall thicknesses, we can calculate the heat transfer via conduction from the warm TPC volume to the cold veto at $\SI{87.87}{K}$. For the moment, we assume a TPC temperature of $\SI{88.60}{K}$, equal to the temperature of the top boundary. We obtain the values shown in table~\ref{tab:heatbudget} and find good agreement with the simulated values. In particular, we calculate a heat transfer rate of $\SI{32.0}{W}$ from the top interface assuming that the system is in equilibrium and that total energy is conserved. This value is only $\SI{4}{\%}$ higher than the simulated result. Using a more accurate temperature of $\SI{88.58}{K}$ as simulated in the TPC centre, this value becomes $\SI{30.3}{W}$. In summary, the heat transfer from the top interface, i.e.~the maximum possible heat load from the gas pocket onto the liquid bulk is limited by the heat dissipation to the veto volume. Note that in this model we have removed the interface to the wire grid frame, which would increase the heat transfer rate from the top interface by $\SI{30.9}{W}$, providing a best-estimate value of $\SI{62}{W}$ for the heat transfer.

\begin{table}
\caption[]{Heat budget of the TPC model with bottom photoelectronics and gas pocket heat.
The values in parentheses are based on energy conservation and the assumption that the model is in thermal equilibrium, in particular that the outlet flow is at the bulk temperature.}
\centering
\begin{tabularx}{0.618\textwidth}{>{\setlength{\baselineskip}{0.85\baselineskip}}Xcc}
\toprule
\textbf{Boundary}   & \textbf{Calculation [W]}  & \textbf{Simulation [W]}\\
\midrule
Walls & $-\num{30.9}$ & $-\num{29.9}$  \\
Cathode & $+\num{11.1}$ & $+\num{11.6}$  \\
Inlet & $-\num{12.2}$ & $(-\num{12.3})$  \\
Top interface & $(+\num{32.0})$ & $+\num{30.6}$  \\
\bottomrule
\end{tabularx}
\label{tab:heatbudget}
\end{table} 

This result unmasks a weakness of this model: the heat load from the gas pocket onto the bulk could be significantly higher in case other heat transfer mechanisms between the TPC and the veto volume exist. For instance, it is expected that gaps of the order of $\SI{1}{mm}$ between the TPC components will emerge during cool-down due to thermal contraction. For this reason, we will assess the maximum possible convective heat transfer from the liquid-gas interface to the bulk. To this end, we study an isolated isothermal downward-facing horizontal plate, with dimensions of the liquid-gas interface, inside a LAr bath which is at infinite distance quiescent and at a constant lower temperature than the plate. The average Nusselt number of the resulting natural convection process in this model can be described by~\cite{Jaffer:2023}:

\begin{equation}
\label{eq:Nusselt_horizontal_plate_below}
    \overline{Nu}_{L}= 0.341 + 0.550 \cdot \left[ \frac{Ra_{L}}{\Xi(Pr)} \right]^{1/5} \quad , \quad \Xi(Pr) \coloneq \left( 1+\left( 0.5/Pr \right)^{\sqrt{1/3}} \right)^{\sqrt{3}} \quad .
\end{equation}
This correlation is experimentally tested up to high Rayleigh numbers of $Ra_{L} \sim \num{e12}$~\cite{FUJII1972755}. The characteristic length of this model is the surface area of the liquid-gas interface divided by its wetted perimeter, i.e.~one quarter of the hydraulic diameter~\cite{Bergman} (section~9.6.2). The fluid density below the plate decreases in the vertical direction and the effect of the buoyancy force is virtually blocked by the plate itself. This self-obstruction leads to a laminar flow around the plate towards its edges up to very high Rayleigh numbers~\cite{Rohsenow} (chapter~4, page~4.17f.). For a plate temperature of $\SI{88.60}{K}$ and a bath temperature of $\SI{87.87}{K}$, we find a Prandtl number of $Pr=2.23$ and a Rayleigh number of $Ra_{L}=\num{1.29e12}$. We use the fluid properties at the film temperature, i.e.~the mean temperature of the plate and the bath. Utilising equation~\eqref{eq:Nusselt_horizontal_plate_below} this yields an average Nusselt number of $\overline{Nu}_{L}=129$ and thus, with the definition in equation~\eqref{eq:NusseltStratifiedFlow} and equation~\eqref{eq:NewtonLawCooling}, a total heat transfer rate of $\SI{144}{W}$. This heat transfer rate from the liquid-gas interface to the liquid bulk represents an upper limit, allowing for unobstructed convection, i.e.~assuming no walls between the inner veto and the TPC volume. It corresponds to the latent heat of a flow of $\SI{8.3}{L/min}$ of cold, saturated gas.

\subsection{Dual-phase mass transfer rate}
\label{subsec:dual-phase_mass_transfer}  

\paragraph{Geometry and mesh}

\begin{figure}[t]
\centering
\includegraphics[width=1\textwidth]{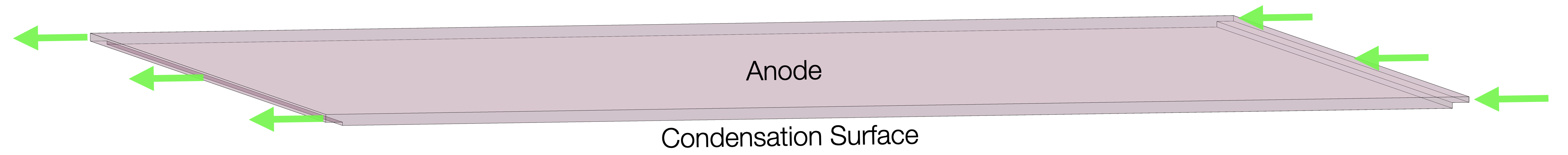}
\caption{Mock-up fluid geometry used to simulate the condensation rate in a dual-phase model. The arrows indicate the flow direction. Figure~\ref{fig:dual-phase} presents a cross section of the outlet for improved visibility.}
\label{fig:mock-up_geometry}
\end{figure}

To simulate the condensation rate in a dual-phase model, we use a mock-up geometry, consisting of a flat $\SI{1000}{mm} \times \SI{1000}{mm} \times \SI{10}{mm}$ volume, with a mass-flow gas inlet covering the top half of the volume's height in cross section, and a pressure outlet covering the top $\SI{7}{mm}$ of the height, thus allowing for a gas pocket of $\SI{7}{mm}$ thickness (see figure~\ref{fig:mock-up_geometry}). The mock-up model consists of $\SI{1.0e6}{}$ hexahedral fluid cells with a structured grid of good quality (orthogonality: $1$; skewness: $0$; aspect ratio: $10$).

\paragraph{Physics and solver setup}
The dual-phase simulation is performed with the pseudo-transient steady pressure-based solver, which enables the use of multiphase models, along with the coupled pressure-velocity scheme and second-order spatial discretisation~\cite{UserGuide,TheoryGuide}. We use the PRESTO! pressure and the modified HRIC volume fraction scheme to obtain sharp interface solutions. The Eulerian multiphase model is used with the multi-fluid VOF (volume of fluid) model option enabled for sharp liquid-gas interfaces, along with the implicit volume fraction formulation. Interfaces between phases are modelled with the type sharp and interfacial anti-diffusion. We use a symmetric drag coefficient because the two phases occupy separate regions in the model. Surface tension is described with the continuum surface stress model using a coefficient of $\SI{0.0122}{N/m}$~\cite{Stansfield_1958}. We use a two-resistance heat transfer coefficient based on the Ranz-Marshall model for both phases. Further phase interactions include a mass transfer mechanism via evaporation-condensation based on the thermal phase change model with the saturation temperature as input. This method, unlike models that are based on the Hertz-Knudsen-Schrage equation, enables mass transfer modelling without mass accommodation coefficients and is exclusive to the Eulerian multiphase model~\cite{TheoryGuide}. The interfacial area is estimated with the symmetric option, which considers the volume fractions of both phases. The fluid properties, including enthalpy, are implemented based on isobaric but temperature-dependent data from the NIST Standard Reference Database~\cite{NIST}.

\paragraph{Operating and boundary conditions}

\begin{table}
\caption[Operating and Boundary Conditions]{Operating and boundary conditions for the mock-up dual-phase simulation.}
\centering
\begin{tabularx}{\textwidth}{>{\setlength{\baselineskip}{0.85\baselineskip}}Xc >{\setlength{\baselineskip}{0.85\baselineskip}}X >{\setlength{\baselineskip}{0.85\baselineskip}}X}
\toprule
\textbf{Condition}   & \textbf{Value}  & \textbf{Setup}   & \textbf{Comment}\\
\midrule
\textbf{Operating}&&&\\
Gravity     &   $\SI{-9.81}{m/s^2}$ & Downward &\\
&&&\\[-10pt]
Static pressure & $\SI{1.160}{bara}$ & & Gas pocket pressure \\
\textbf{Boundary}&&&\\
Mass flow rate & $\SI{0.1}{g/s}$ & Mass flow inlet, $\SI{5}{\%}$ turbulent intensity at $\SI{1}{mm}$ length scale & $\hat{=} \, \SI{3.4}{slpm}$ \\
&&&\\[-10pt]
Outlet pressure & $\SI{0}{barg} $ & Pressure outlet &  \\
&&&\\[-10pt]
Inlet temperature & $\SI{88.60}{K}$ & Inlet with fixed temperature & Saturation temperature at gas pocket pressure\\
&&&\\[-10pt]
Condensation surface heat flux & $\SI{-2.9}{W/m^2}$ & Wall with heat flux & From liquid-phase TPC simulation\\
&&&\\[-10pt]
Heat flux of other walls & $\SI{0}{W/m^2}$ & Adiabatic walls & \\
\bottomrule
\end{tabularx}
\label{tab:conditions4}
\end{table}

The operating and boundary conditions of the mock-up simulation are found in table~\ref{tab:conditions4}. The cooling power (heat flux: $\SI{-2.9}{W/m^2}$) of the liquid obtained from the TPC simulation in the previous subsection is applied to the bottom surface (cf.~figure~\ref{fig:mock-up_geometry}) while all other walls are adiabatic. Gas is injected at a mass flow rate of $\SI{0.1}{g/s}$, corresponding to $\SI{3.4}{slpm}$, at saturation temperature, meaning that gas cooling is not considered here.

\paragraph{Result}

\begin{figure}[t]
\centering
\includegraphics[width=\textwidth]{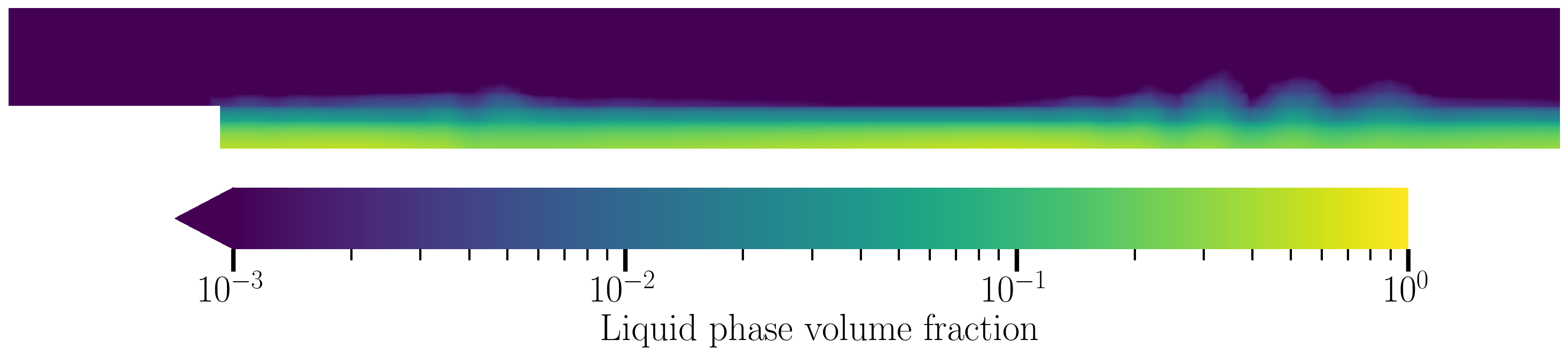}
\caption{Liquid phase volume fraction from the dual-phase simulation with the mock-up model. Shown is a vertical cross section through the centre of the model with the outlet on the leftmost edge. The liquid pools in the pocket above the condensation surface, leaving a gas phase of $\SI{7}{mm}$ on top. This thickness is defined by the outlet height as excess liquid overflows the step on the left and escapes through the outlet.}
\label{fig:dual-phase}
\end{figure}

The dual-phase result is presented in figure~\ref{fig:dual-phase}. We find a simulated mass transfer rate from the gas to the liquid phase of $\SI{-1.797e-2}{g/s}$, which is $\SI{18}{\%}$ of the inlet flow, via a volumetric integration in the model domain. This result matches the expectation based on the latent heat within $\SI{0.1}{\%}$ and remains constant over a wide range of inlet flows, up to at least $\SI{1}{g/s}$.

\subsection{Condensation on the anode}
\label{subsec:condensation_anode}

Condensation of GAr in the gas pocket cannot only happen on the liquid surface but also on the detector solids. In particular, the gas dissipates heat via conduction through the anode window to the liquid in the veto volume on top. If the resulting surface temperature $T_{S}$ of the anode is equal or lower than the saturation temperature, a liquid layer could form on the anode, and liquid droplets could eventually drip down onto the liquid surface. This might have an impact on the uniformity and resolution of the S2 signal. Thus, it is desirable to inject gas at a slightly higher temperature into the gas pocket to keep $T_{S}$ above saturation. We outline the modelling of this case in the following.

In the absence of phase transition, no external gas flow is required. In this hypothetical case, we can consider a stable gas pocket and study the convection inside the enclosure between two horizontal plates. If the anode is warmer than the liquid surface, the gas is thermally stratified and heat transfer from the top to the bottom is solely realised via heat conduction (see equation~\eqref{eq:NusseltStratifiedFlow}). If instead the liquid surface is warmer than the anode, we can distinguish two regimes: (1) up to a critical Rayleigh number of $\num{1708}$, heat conduction dominates as the primary heat transfer mechanism since viscous forces prevail over buoyancy forces, impeding convection~\cite{Bergman} (section~9.8.1); and (2) above this critical number, advection through Rayleigh-B\'{e}nard cells occurs. These phenomena can be collectively described through~\cite{HOLLANDS1975879}:

\begin{equation}
Nu_{L} = 1 + 1.44 \cdot \left[ 1 -\frac{1708}{Ra_{L}} \right]^{+} + \left[ \left( \frac{Ra_{L}}{5830} \right)^{1/3} -1 \right]^{+} \quad ,
\label{eq:NusseltConvectionTopPlateCold}
\end{equation}
where $[\, \cdot \,]^{+} \coloneq \mathrm{max}(\, \cdot \,,0)$ denotes the positive part of the expression in brackets. The characteristic length $L$ here is the gas pocket thickness of $\SI{7}{mm}$.

In addition to the interaction with the liquid surface, forced convection from an applied gas pocket flow can be considered. Assuming a sufficiently thin boundary layer at the solid-gas interface, i.e.~considering low fluid speeds, we can approximate the anode as isolated flat plate in an external parallel flow and use the average Nusselt number~\cite{Bergman} (section~7.2):

\begin{equation}
\overline{Nu}_{L} = 0.664 \cdot Re_{L}^{1/2} \cdot Pr^{1/3} \quad .
\label{eq:nusselt_forced_convection}
\end{equation}
Here, $Re_{L}$ denotes the Reynolds number, which characterises the fluid's flow regime for pressure-driven flows based on the ratio of the inertial and viscous forces. Equation~\eqref{eq:nusselt_forced_convection} is applicable for $Pr \geq 0.6$, a condition that is fulfilled in the considered case.

\begin{figure}[t]
\centering
\includegraphics[width=1\textwidth]{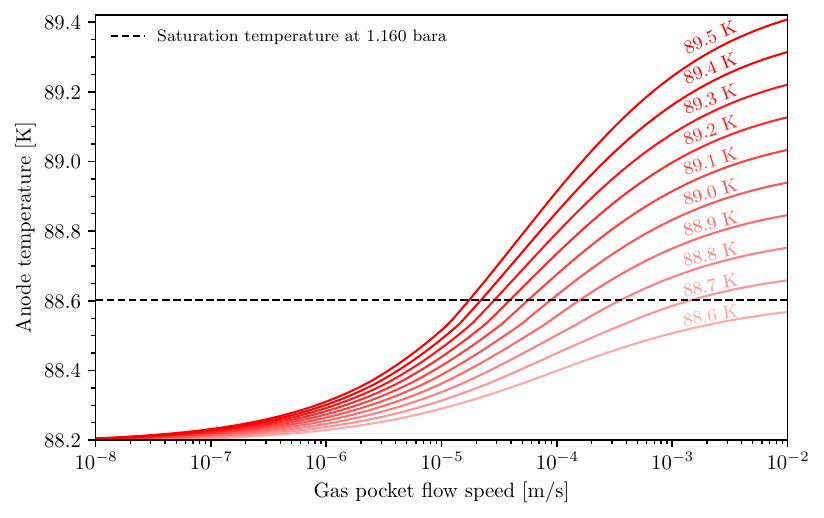}
\caption{Temperature of the anode versus gas pocket flow speed at gas inlet temperatures in the range of $\SIrange{88.6}{89.5}{K}$. Flow speeds of in the range of $\SIrange{e-4}{e-2}{m/s}$ correspond to flow rates of $\SIrange{1}{100}{slpm}$.}
\label{fig:anode_condensation}
\end{figure}

We can now combine the discussed processes. As an approximation, the heat transfers from both free and natural convection are treated as superimposed. In fact, depending on the relative size of buoyancy and inertial forces, we are in the regime of mainly forced convection ($Gr_{L}/Re_{L}^{2} \ll 1$), free convection ($Gr_{L}/Re_{L}^{2} \gg 1$) or mixed convection ($Gr_{L}/Re_{L}^{2} \approx 1$)~\cite{Bergman} (section~9.4). Since heat is dissipated via conduction through the anode window, the temperature of its top surface is required. For simplicity, we assume that it is fixed to $\SI{87.87}{K}$, i.e.~we do not model the PMMA-to-LAr heat transfer in the inner veto volume and assume that the liquid temperature is not elevated by the top photoelectronics\footnote{This is a conservative approach, yielding the maximum heat transfer rate through the anode window. The highest possible temperature of the anode window's top surface is the local saturation temperature of $\SI{88.5}{K}$.}. In equilibrium, energy conservation dictates that the combined positive heat transferred to the anode is equal to the heat conducted through the anode window. Solving the resulting equation for the anode temperature yields figure~\ref{fig:anode_condensation}. For this we use the argon fluid properties at the film temperatures of the processes. For free (forced) convection, the film temperature is the mean of the temperatures of the anode and the liquid-gas interface (the anode and the gas inlet). For flow rates of $\SIrange{1}{100}{slpm}$, we find that a minimum inlet temperature of $\SI{89}{K}$ is required to keep the anode above saturation temperature. If the anode temperature is below, the gas-anode heat transfer is determined by equations~\eqref{eq:NusseltConvectionTopPlateCold} and~\eqref{eq:nusselt_forced_convection}, and solely governed by free convection at very low flow speeds. If above, the gas-anode heat transfer is determined by equation~\eqref{eq:nusselt_forced_convection}, while the relative contribution of the gas heat conduction (equation~\eqref{eq:NusseltStratifiedFlow}) becomes less significant towards high flow speeds. At a gas inlet temperature of $\SI{89}{K}$ and flow rates of $\SIrange{1}{100}{slpm}$, the upward heat conduction rate through the anode window is $\SIrange{23}{33}{W}$.

Note that the simplified model presented here assumes an isothermal anode as well as a linear uniform flow. It is expected that a temperature gradient over the surface is developed. This would lead to an increased conductive heat transfer rate through the anode window in proximity of the warm gas inlets. This process depends on the specific geometry of the gas pocket inlet and outlet but generally increases the required flow at a given inlet temperature to maintain the minimum anode temperature above saturation everywhere. We further note that heat transfers calculated with the correlation in equation~\eqref{eq:nusselt_forced_convection} may deviate up to $\SI{25}{\%}$ from experimental results~\cite{Bergman} (section~7.2.6).
\section{Conclusions}
\label{Sec:Conclusions}

We have studied the hydrodynamic and thermal properties of the argon volume in the DS"~20k TPC, a next-generation WIMP detector under construction at LNGS, using a simplified yet realistic geometry. We have characterised the liquid mixing behaviour for various candidate inlet/outlet configurations based on their outlet RTD properties and selected a preferred geometry with stable performance over the relevant range of flow rates, measured by means of its RTD metrics. It has been found that the TPC has a good mixing-behaviour on timescales of the turnover time of $\SI{\sim 40}{\day}$. No regions with mean residence times higher than the turnover time have been found. The absence of such dead zones implies that the purified argon reaches everywhere in the TPC on timescales of the turnover time. The liquid flow pattern has been studied with and without convection induced by the heat from the bottom photoelectronics. We have further evaluated the reach of the internally distributed calibration source \ce{^{83\text{m}}Kr} by means of the RTD on a cylinder surface in the centre of the TPC. It has been demonstrated that a fraction of $\SI{93}{\%}$ of the injected source atoms are expected to reach the horizontal centre within $2\cdot T_{1/2}$. While the first source atoms are expected to arrive after $\SI{1.5}{min}$, they have the highest probability to arrive $0.12 \cdot T_{1/2}$ ($\SI{13}{min}$) after injection. This confirms also a good mixing behaviour on timescales of the \ce{^{83\text{m}}Kr} half-life and excellent prospects for the deployment of \ce{^{83\text{m}}Kr} in DS"~20k with homogeneously distributed events. The paper has further assessed the condensation rate of gas in the gas pocket on the liquid surface. We have estimated the heat transfer rate over the entire liquid-gas interface with a best-estimate value of $\SI{62}{W}$ and an upper limit of $\SI{144}{W}$. Finally, the case of a potential condensation of gas on the anode has been considered in a simple model. For typical gas flow rates of $\SIrange{1}{100}{slpm}$, a minimum gas inlet temperature of $\SI{89}{K}$ is necessary to avoid condensation. This leads to a heat dissipation through the anode window of $\SIrange{23}{33}{W}$ from the gas pocket to the veto volume for the specified flow regime. Recent updates of the TPC design and their impact on the results are discussed in appendix~\ref{app:designchanges}.

The presented modelling techniques are applicable to other liquid noble gas detectors to assist the design process. This becomes increasingly important as detectors reach target masses of tens of tonnes and beyond. In the absence of buoyancy-driven convective currents, effective mixing in geometries with relatively low surface-to-volume ratios, such as the DS"~20k TPC, can be achieved on timescales comparable to the turnover time, provided the recirculation flow rate is small relative to the fluid volume and the inlets and outlets are positioned at appropriate distances. However, in such cases, the flow speeds are insufficient for adequate homogeneous mixing on timescales much smaller than the turnover time. Hence, free convection is necessary for the efficient homogeneous distribution of short-lived calibration sources, such as \ce{^{83\text{m}}Kr}, within a DS"~20k-sized volume.

\acknowledgments{We acknowledge support by CERN which provided the computational infrastructure required for this work, such as the ANSYS\textsuperscript{\textregistered}~Fluent licence and access to big memory nodes on the cluster. This paper is based upon work supported by the U.~S.~National Science Foundation~(NSF) (grants No.~PHY-0919363, No.~PHY-1004054, No.~PHY-1004072, No.~PHY-1242585, No.~PHY-1314483, No.~PHY-1314507, No.~PHY-1622337, No.~PHY-1812482, No.~PHY-1812547, No.~PHY-2310046, No.~PHY-2310091, associated collaborative grants, No.~PHY-1211308, No.~PHY-1314501, No.~PHY-1455351 and No.~PHY-1606912, as well as Major Research Instrumentation grant No.~MRI-1429544), the Italian Istituto Nazionale di Fisica Nucleare~(INFN) (grants from Italian Ministero dell’Università e della Ricerca Progetto Premiale 2013 and Commissione Scientifica Nazionale~II), the Natural Sciences and Engineering Research Council of Canada, SNOLAB, and the Arthur B.~McDonald Canadian Astroparticle Physics Research Institute.
We also acknowledge the financial support by LabEx UnivEarthS (ANR-10-LABX-0023 and ANR18-IDEX-0001), Chinese Academy of Sciences (113111KYSB20210030) and National Natural Science Foundation of China (12020101004).
This work has also been supported by the S\~{a}o Paulo Research Foundation~(FAPESP) grant No.~2021/11489-7 and by the National Council for Scientific and Technological Development~(CNPq).
Support is acknowledged by the Deutsche Forschungsgemeinschaft (DFG, German Research Foundation) under Germany's Excellence Strategy -- EXC 2121: Quantum Universe -- 390833306.
Some authors were also supported by the Spanish Ministry of Science and Innovation~(MICINN) through grant No.~PID2022-138357NB-C22, the ``Atraccion de Talento'' grant No.~2018-T2/TIC-10494, the Polish NCN grant No.~UMO-2023/51/B/ST2/02099 and No.~UMO-2022/47/B/ST2/02015, the Polish Ministry of Science and Higher Education~(MNiSW) grant No.~6811/IA/SP/2018, the International Research Agenda Programme AstroCeNT grant No.~MAB/2018/7 funded by the Foundation for Polish Science from the European Regional Development Fund, the European Union’s Horizon~2020 research and innovation program under grant agreement No.~952480 (DarkWave), the Science and Technology Facilities Council, part of the United Kingdom Research and Innovation, and The Royal Society (United Kingdom), and IN2P3-COPIN consortium grant No.~20-152. We also wish to acknowledge the support from Pacific Northwest National Laboratory, which is operated by Battelle for the U.~S.~Department of Energy under Contract No.~DE-AC05-76RL01830.
This research was supported by the Fermi National Accelerator Laboratory~(Fermilab), a U.~S.~Department of Energy, Office of Science, HEP User Facility. Fermilab is managed by Fermi Research Alliance, LLC (FRA), acting under contract No.~DE-AC02-07CH11359.}

\appendix
\section{RTD-based optimisation of the inlet/outlet design}
\label{app:configurations}

\begin{figure}[ht!]
\centering
\includegraphics*[width=1\textwidth]{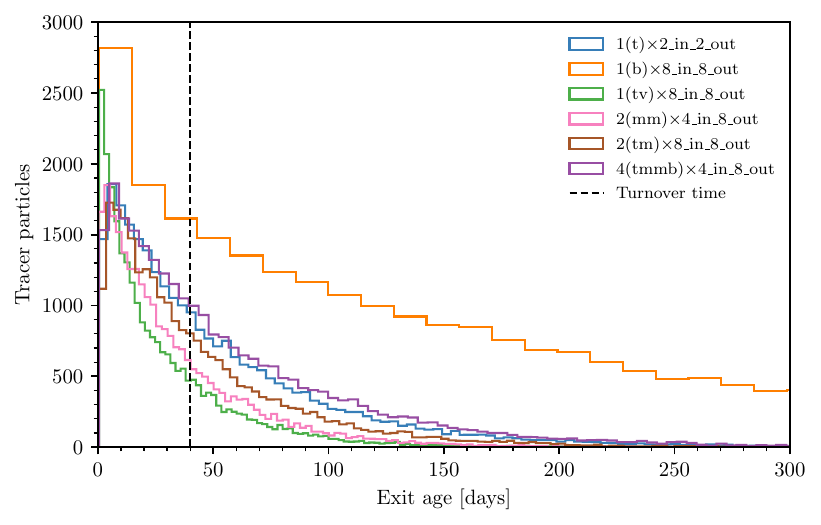}
\caption[]{RTDs of selected inlet/outlet configurations.}
\label{fig:RTDs_no_heat}
\end{figure}

\begin{figure}[ht]
\centering
\includegraphics[width=0.618\textwidth]{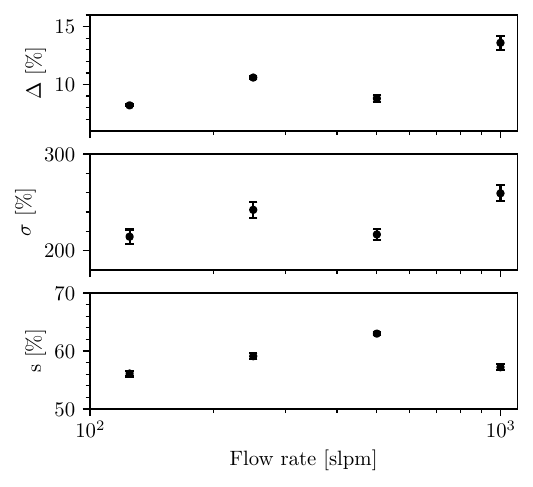}
\caption[]{RTD properties of the configuration \textit{2(tm)$\times$8\textunderscore in\textunderscore 8\textunderscore out\textunderscore double} as a function of recirculation flow rate. Uncertainties reflect the histogram binning and Poisson fluctuations.}
\label{fig:RTD_properties_2x8_top_double}
\end{figure}

The inlet/outlet configurations presented in table~\ref{tab:configurations} are studied in the isothermal flow-only case. Inlets are placed in the middle of the TPC walls. Top~(t) inlet rings are located $\SI{100}{mm}$ below the liquid surface. Middle~(m) inlet rings are located $\SI{100}{mm}$ below the horizontal centre plane of the liquid volume. Bottom~(b) inlet rings are located $\SI{100}{mm}$ above the cathode. Mid-top and Mid-bottom~(mm) inlet rings are located at $\SI{75}{\%}$ and $\SI{25}{\%}$ of the height of the liquid volume, respectively. The inlet flow direction is normal to the TPC walls, except in one configuration that is meant to induce a vortex~(v) flow in the TPC. Outlets are situated on the cathode window at a radius of $\SI{1773.5}{mm}$, i.e.~in between the reflector panels and the TPC walls. The inlet/outlet configurations are compared based on the mixing they induce in the TPC. As noted in section~\ref{Sec:Motivation}, mixing can be quantified by means of the RTD at the outlet. RTDs of a representative subset of studied configurations for the nominal TPC recirculation flow rate of $\SI{500}{slpm}$ can be found in figure~\ref{fig:RTDs_no_heat} (figure~\ref{fig:RTD_2x8_double_no_heat_MeanRT_spatial_distribution}~(left) shows the RTD of the configuration \textit{2(tm)$\times$8\textunderscore in\textunderscore 8\textunderscore out\textunderscore double}). Some RTDs feature long tails (e.g.~\textit{1(b)$\times$8\textunderscore in\textunderscore 8\textunderscore out}), indicating recirculation regions which trap particles, or fast arrival of a large fraction of particles, indicating channelling. In configuration \textit{1(tv)$\times$8\textunderscore in\textunderscore 8\textunderscore out} a shell of rotating fluid passes quickly from the inlets to the outlets, while the bulk centre of the TPC remains almost at rest. However, we find that the RTDs generally resemble the expectation of systems with good mixing behaviour and a delayed arrival of tracers with the most probable residence time. For this reason and since the timescales of the RTD at the outlet allow for a characterisation of the purification prospects of the system, we optimise the flow pattern for purification. To this end, we compare the inlet/outlet configurations based on the following metrics:
\begin{itemize}
    \item{Most probable residence time: the peak of the distribution should be as close as possible to the turnover time. This accounts for the fact that channelling, i.e.~fast arrival of low-age tracers at the outlet, will have a negative impact on the purification. We denote the most probable residence time divided by the turnover time by $\Delta$.} 
    \item{Width of the RTD: to reduce particle trapping, configurations with narrow RTDs are preferred. This quantity is related to the second moment of the RTD function. We denote the full width at tenth maximum~(FWTM) divided by the turnover time by $\sigma$.}
    \item{Symmetry/skewness of the RTD: an overly uneven distribution of tracers around the mean can indicate particle channelling or trapping. The ratio of particles with residence times lower than the turnover time should be close to $\SI{50}{\%}$. Note that it is $1-1/e=\SI{63.2}{\%}$ for a mixed flow reactor. This quantity is related to the third moment of the RTD function. We denote the fraction of entries below the turnover time by $s$.}
\end{itemize}
For a quantitative assessment we define a discriminator: $D \coloneq \frac{\Delta}{\sigma}\left( 1-2\lvert s-0.5 \rvert \right)$. The properties of the RTDs of all configurations and corresponding discriminator values are listed in table~\ref{tab:RTDs_no_heat}. With the exception of a few distinctive examples, the differences among most of the configurations are small and likely further diluted when free convection is added. However, the configuration \textit{2(tm)$\times$8\textunderscore in\textunderscore 8\textunderscore out\textunderscore double} shows the best performance. This configuration serves as the specimen for this paper. In figure~\ref{fig:RTD_properties_2x8_top_double}, we show the insensitivity of the RTD properties of this preferred configuration for flow rate changes as no systematic trend is observed in the three metrics.

\begin{table}[ht!]
\caption[Inlet/outlet configurations]{Inlet/outlet configurations. The total cross section of the inlet/outlets with nominal diameter equals the cross section of a circle with $\SI{1}{inch}$ diameter, approximately matching the cross section of the $\SI{1}{inch}$ outer diameter tubing used in the UAr cryogenics system~\cite{DarkSide-20k:UArCryo}. The inlet/outlet configurations with altered diameter are scaled to \textit{quarter/half/double} the nominal total cross section. The configuration marked by an asterisk is selected as specimen for this paper.}
\centering
\begin{tabularx}{\textwidth}{lcc >{\setlength{\baselineskip}{0.85\baselineskip}}Xc}
\toprule
\textbf{Configuration}& \textbf{Type} & \textbf{Qty.} & \textbf{Position} & $\boldsymbol{\varnothing}$ \textbf{[mm]} \\
\midrule
\textbf{Nominal Diameter}&&&&\\
\textit{1(t)$\times$2\textunderscore in\textunderscore 2\textunderscore out} & in & 2 & t~ring, at $\SI{180}{\degree}$ offset  & $\SI{18.0}{}$ \\
& out & 2 & at $\SI{180}{\degree}$ offset, $\SI{90}{\degree}$ rotated to inlets & $\SI{18.0}{}$ \\
\midrule
\textit{1(t)$\times$4\textunderscore in\textunderscore 4\textunderscore out} & in & 4 & t~ring, at $\SI{90}{\degree}$ offset  & $\SI{12.7}{}$ \\
& out & 4 & at $\SI{90}{\degree}$ offset, $\SI{45}{\degree}$ rotated to inlets & $\SI{12.7}{}$ \\
\midrule
\textit{1(t)$\times$8\textunderscore in\textunderscore 8\textunderscore out} & in & 8 & t~ring, at $\SI{45}{\degree}$ offset  & $\SI{9.0}{}$ \\
& out & 8 & at $\SI{45}{\degree}$ offset & $\SI{9.0}{}$ \\
\midrule
\textit{1(tv)$\times$8\textunderscore in\textunderscore 8\textunderscore out} & in & 8 & t~ring, at $\SI{45}{\degree}$ offset, $\SI{45}{\degree}$ clockwise angled & $\SI{9.0}{}$ \\
& out & 8 & at $\SI{45}{\degree}$ offset & $\SI{9.0}{}$ \\
\midrule
\textit{1(b)$\times$8\textunderscore in\textunderscore 8\textunderscore out} & in & 8 & b~ring, at $\SI{45}{\degree}$ offset  & $\SI{9.0}{}$ \\
& out & 8 & at $\SI{45}{\degree}$ offset & $\SI{9.0}{}$ \\
\midrule
\textit{2(tm)$\times$4\textunderscore in\textunderscore 8\textunderscore out} & in & 8 & tm~rings with 4 inlets each at $\SI{90}{\degree}$ offset, rings $\SI{45}{\degree}$ rotated to each other  & $\SI{9.0}{}$ \\
& out & 8 & at $\SI{45}{\degree}$ offset & $\SI{9.0}{}$ \\
\midrule
\textit{2(mm)$\times$4\textunderscore in\textunderscore 8\textunderscore out} & in & 8 & mm~rings with 4 inlets each at $\SI{90}{\degree}$ offset, rings $\SI{45}{\degree}$ rotated to each other  & $\SI{9.0}{}$ \\
& out & 8 & at $\SI{45}{\degree}$ offset & $\SI{9.0}{}$ \\
\midrule
\textit{2(tm)$\times$8\textunderscore in\textunderscore 8\textunderscore out} & in & 16 & tm~rings with 8 inlets each at $\SI{45}{\degree}$ offset  & $\SI{6.35}{}$ \\
& out & 8 & at $\SI{45}{\degree}$ offset & $\SI{9.0}{}$ \\
\midrule
\textit{2(mm)$\times$8\textunderscore in\textunderscore 8\textunderscore out} & in & 16 & mm~rings with 8 inlets each at $\SI{45}{\degree}$ offset  & $\SI{6.35}{}$ \\
& out & 8 & at $\SI{45}{\degree}$ offset & $\SI{9.0}{}$ \\
\midrule
\textit{4(tmmb)$\times$2\textunderscore in\textunderscore 8\textunderscore out} & in & 8 & tmmb~rings with 2 inlets each on opposite sides, rings $\SI{90}{\degree}$ rotated to each other  & $\SI{9.0}{}$ \\
& out & 8 & at $\SI{45}{\degree}$ offset & $\SI{9.0}{}$ \\
\midrule
\textit{4(tmmb)$\times$4\textunderscore in\textunderscore 8\textunderscore out} & in & 16 & tmmb~rings with 4 inlets each at $\SI{90}{\degree}$ offset, rings $\SI{45}{\degree}$ rotated to each other  & $\SI{6.35}{}$ \\
& out & 8 & at $\SI{45}{\degree}$ offset & $\SI{9.0}{}$ \\
\midrule
\textbf{Altered Diameter}&&&&\\
\textit{1(t)$\times$2\textunderscore in\textunderscore 2\textunderscore out\textunderscore half} & in & 2 & t~ring, at $\SI{180}{\degree}$ offset  & $\SI{12.7}{}$ \\
& out & 2 & at $\SI{180}{\degree}$ offset, $\SI{90}{\degree}$ rotated to inlets & $\SI{12.7}{}$ \\
\midrule
\textit{1(t)$\times$2\textunderscore in\textunderscore 2\textunderscore out\textunderscore quarter} & in & 2 & t~ring, at $\SI{180}{\degree}$ offset  & $\SI{9.0}{}$\\
& out & 2 & at $\SI{180}{\degree}$ offset, $\SI{90}{\degree}$ rotated to inlets & $\SI{9.0}{}$ \\
\midrule
\textit{2(tm)$\times$8\textunderscore in\textunderscore 8\textunderscore out\textunderscore double\textsuperscript{*}} & in & 16 & tm~rings with 8 inlets each at $\SI{45}{\degree}$ offset & $\SI{9.0}{}$ \\
& out & 8 & at $\SI{45}{\degree}$ offset & $\SI{12.7}{}$ \\
\midrule
\textit{2(tm)$\times$8\textunderscore in\textunderscore 8\textunderscore out\textunderscore half} & in & 16 & tm~rings with 8 inlets each at $\SI{45}{\degree}$ offset & $\SI{4.5}{}$ \\
& out & 8 & at $\SI{45}{\degree}$ offset & $\SI{6.35}{}$ \\
\bottomrule
\end{tabularx}
\label{tab:configurations}
\end{table}

\clearpage

\begin{table}[ht!]
\caption[Properties of RTDs]{Properties of the RTDs. Uncertainties reflect the histogram binning and Poisson fluctuations. The configuration marked by an asterisk is selected as specimen for this paper.}
\centering
\begin{tabularx}{\textwidth}{lXXXr}
\toprule
\textbf{Configuration}   & $\boldsymbol{\Delta}$  \textbf{[\%]}  & $\boldsymbol{\sigma}$ \textbf{[\%]}   & $\mathbf{s}$ \textbf{[\%]} &   $\mathbf{D \cdot 1000}$\\
\midrule
\textbf{Nominal Diameter}&&&&\\
\textit{1(t)$\times$2\textunderscore in\textunderscore 2\textunderscore out} & $7.7 \pm 0.3$ &  $307 \pm 10$ & $52.3 \pm 0.4$ & 24 \\
\textit{1(t)$\times$4\textunderscore in\textunderscore 4\textunderscore out} & $5.2 \pm 0.3$ & $268 \pm 8$  & $60.3 \pm 0.5$  & 16\\
\textit{1(t)$\times$8\textunderscore in\textunderscore 8\textunderscore out} & $4.8 \pm 0.3$ & $226 \pm 9$  & $59.6 \pm 0.5$  & 17\\
\textit{1(tv)$\times$8\textunderscore in\textunderscore 8\textunderscore out} & $4.0 \pm 0.3$ & $133 \pm 6$  & $73.5 \pm 0.5$  & 16\\
\textit{1(b)$\times$8\textunderscore in\textunderscore 8\textunderscore out} & $4.1 \pm 0.3$ & $890 \pm 36$  & $21.9 \pm 0.3$  & 2\\
\textit{2(tm)$\times$4\textunderscore in\textunderscore 8\textunderscore out} & $6.7 \pm 0.3$ & $256 \pm 8$  & $59.9 \pm 0.5$  & 21\\
\textit{2(mm)$\times$4\textunderscore in\textunderscore 8\textunderscore out} & $4.4 \pm 0.3$ & $207 \pm 6$  & $68.5 \pm 0.5$  & 13\\
\textit{2(tm)$\times$8\textunderscore in\textunderscore 8\textunderscore out} & $6.7 \pm 0.3$ & $262 \pm 8$  & $58.7 \pm 0.5$  & 21\\
\textit{2(mm)$\times$8\textunderscore in\textunderscore 8\textunderscore out} & $5.1 \pm 0.3$ & $248 \pm 8$  & $59.6 \pm 0.5$  & 17\\
\textit{4(tmmb)$\times$2\textunderscore in\textunderscore 8\textunderscore out} & $4.7 \pm 0.3$ & $201 \pm 6$  & $67.1 \pm 0.5$  & 15\\
\textit{4(tmmb)$\times$4\textunderscore in\textunderscore 8\textunderscore out} & $6.0 \pm 0.3$ & $347 \pm 11$  &  $48.1 \pm 0.4$ & 15\\
\midrule
\textbf{Altered Diameter}&&&&\\
\textit{1(t)$\times$2\textunderscore in\textunderscore 2\textunderscore out\textunderscore half}& $7.4 \pm 0.3$ &  $276 \pm 11$ & $57.3 \pm 0.7$ & 23 \\
\textit{1(t)$\times$2\textunderscore in\textunderscore 2\textunderscore out\textunderscore quarter}& $6.2 \pm 0.3$ &  $194 \pm 9$ & $70.6 \pm 1.0$ & 19 \\
\textit{2(tm)$\times$8\textunderscore in\textunderscore 8\textunderscore out\textunderscore double\textsuperscript{*}}& $8.8 \pm 0.3$ &  $217 \pm 6$ & $63.0 \pm 0.3$ & 30 \\
\textit{2(tm)$\times$8\textunderscore in\textunderscore 8\textunderscore out\textunderscore half}& $5.4 \pm 0.3$ &  $222 \pm 9$ & $63.9 \pm 0.7$ & 18 \\
\bottomrule
\end{tabularx}
\label{tab:RTDs_no_heat}
\end{table}

\section{Flow pattern in the isothermal case}
\label{app:flowpattern1}

\begin{figure}[t]
\centering
\begin{subfigure}[b]{0.48\textwidth}
\centering
\includegraphics*[width=\textwidth]{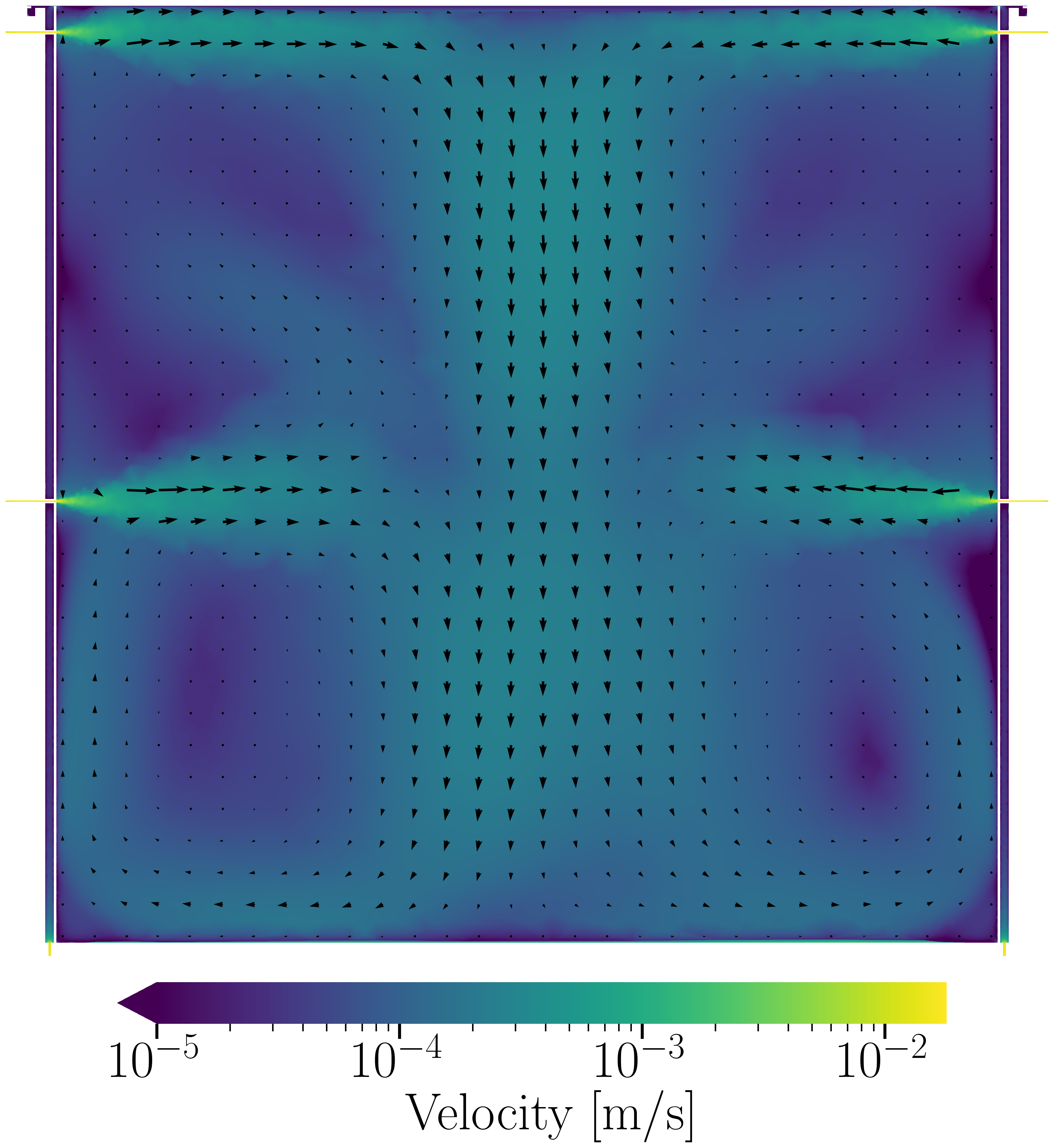}
\end{subfigure}
\quad
\begin{subfigure}[b]{0.48\textwidth}
\centering
\includegraphics*[width=\textwidth]{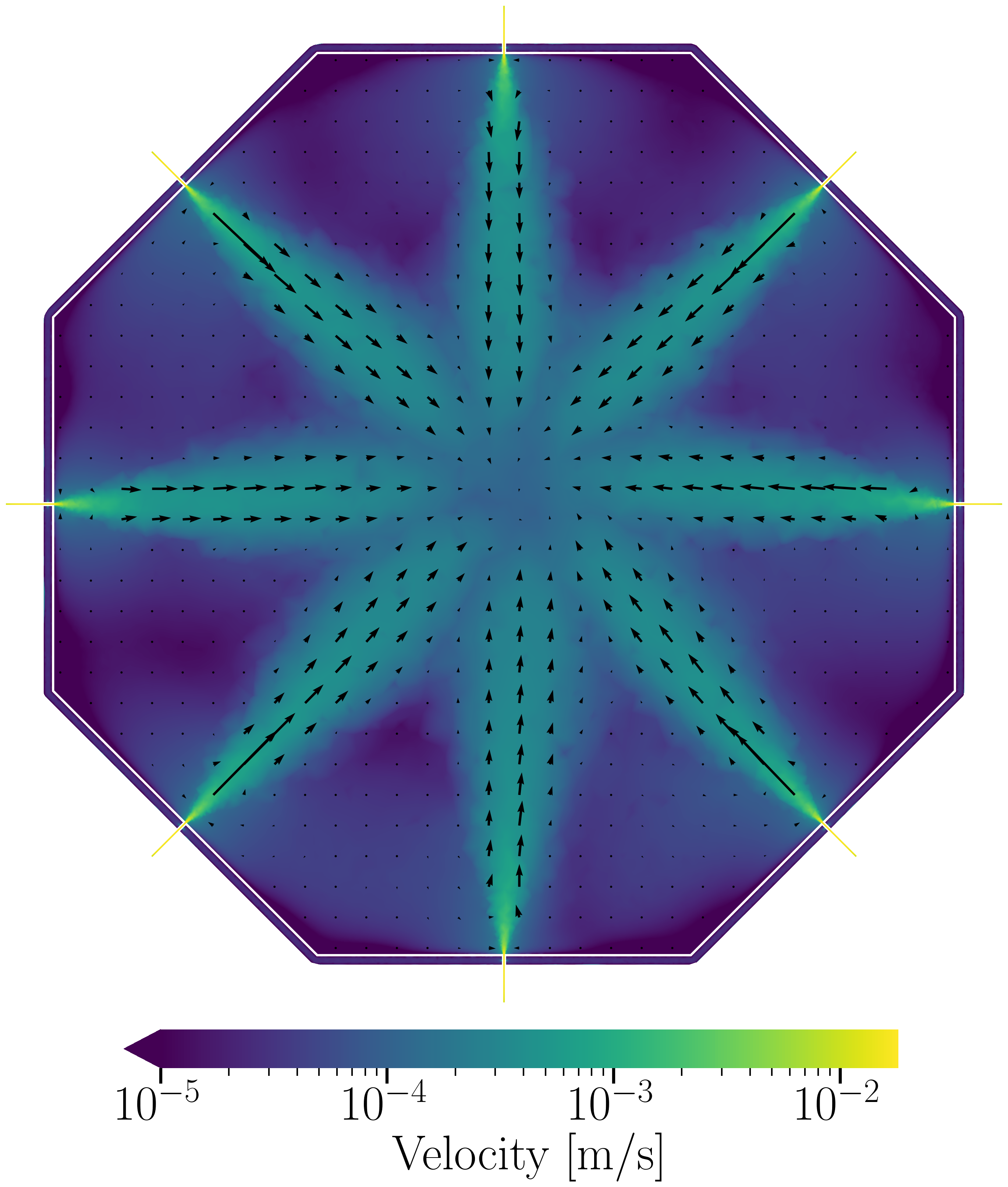}
\end{subfigure}
\quad
\begin{subfigure}[b]{0.48\textwidth}
\vspace{0.3cm}
\centering
\includegraphics*[width=\textwidth]{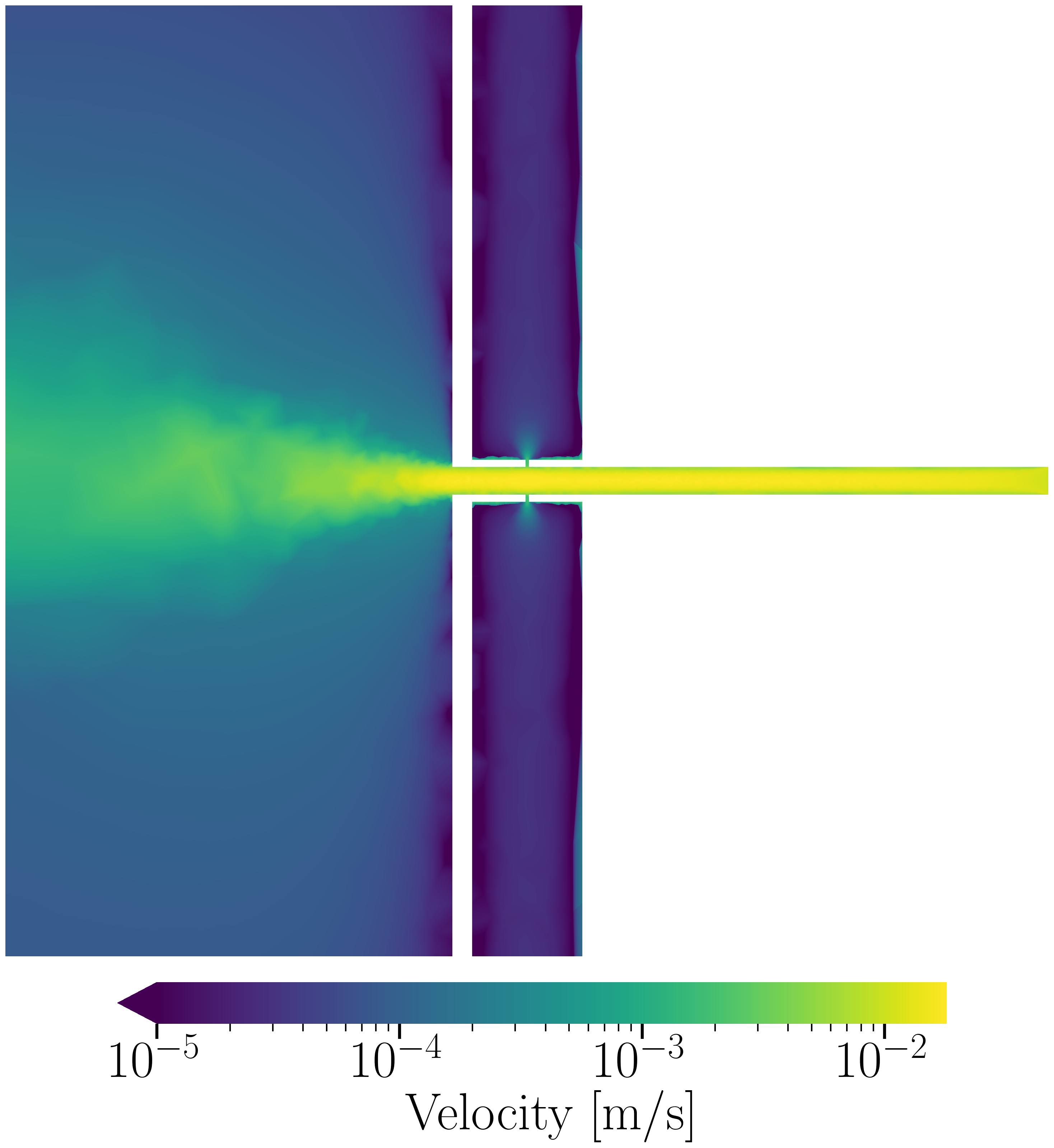}
\end{subfigure}
\quad
\begin{subfigure}[b]{0.48\textwidth}
\centering
\includegraphics*[width=\textwidth]{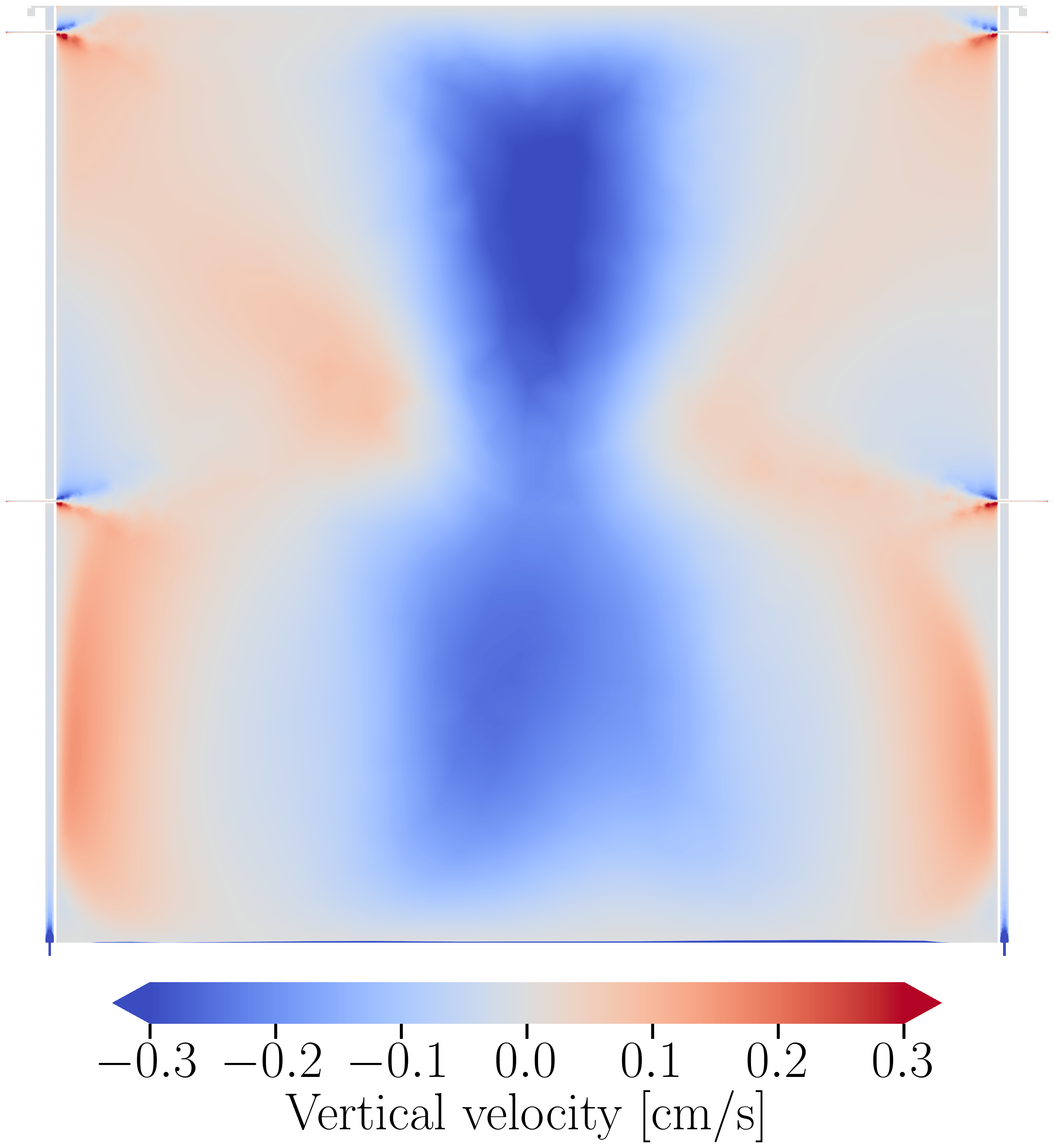}
\end{subfigure}
\caption[]{Flow velocity maps of the isothermal flow-only case. (Top left): Vertical cut view through the inlets and outlets. (Top right): Horizontal cut view through the lower row of inlets. (Bottom left): Detail of an inlet in a horizontal cut view (cf.~figure~\ref{fig:CFD_geometry}~(detail~2)). Note the small flow fraction that enters the intermediate space between the reflector panels and the TPC walls. (Bottom right): Vertical cut view through the inlets and outlets. Shown is the velocity in vertical direction with red indicating upward and blue indicating downward flow.}
\label{fig:Velocity_spatial_distribution_no_heat}
\end{figure}

Figure~\ref{fig:Velocity_spatial_distribution_no_heat} presents the flow velocity maps of the CFD solution of the isothermal flow-only case. The developed flow pattern is distinct and clearly dominated by the inlet streams since no other driving force such as convection is present. Upon meeting in the centre, the inlet streams of the top row sink down since the volume is bounded at the top. This induces a downward fluid movement that is adapted also by the inlet streams of the middle row. Upon arriving at the cathode, the fluid is deflected to larger radii and flows back up in proximity of the reflector panels, inducing a global circulation cell. The flow behind the reflector panels is moving strictly downward with a narrow velocity distribution around $\SI{3e-5}{m/s}$ at a distance from the outlets. This region is fed by the small side-inlets shown in figure~\ref{fig:CFD_geometry}~(detail 2). Due to Bernoulli's principle the pressure in the inlet tube is reduced compared to static fluid, which could potentially lead to backflow into the tube through the side holes. However, the fluid velocity close to the inlet tube wall is lower than in the centre. It has been verified with dedicated simulations that the pressure decrease does not overcome the internal overpressure of the inlets at flows smaller than the fivefold nominal flow. Thus, the supply of the region behind the reflector panels through the inlet side holes is assured for the operating conditions of interest. The regions with the highest speeds are the inlets and outlets with $\SI{\sim 1}{cm/s}$, while the slowest moving regions are found in proximity of the wire grid frame and the TPC surfaces. However, the volume fraction moving slower than $\SI{1}{\micro m/s}$ is $< 2\times 10^{-3}$, corresponding to a volume of $\SI{62}{L}$. The majority of the fluid volume ($\SI{95}{\%}$) moves at speeds in the range $\SIrange{0.01}{0.3}{mm/s}$. The volume-averaged RMS speed is $\SI{0.11}{mm/s}$. The standard error of this weighted RMS is $\SI{4e-4}{mm/s}$.

\section{TPC design update}
\label{app:designchanges}

In the final stages of this project, the collaboration has updated the TPC design. While the overall conclusions remain unchanged, these updates are expected to slightly alter the quantitative results. In the latest design, the wire grid frame no longer extends into the veto volume, leaving the TPC exterior surfaces composed solely of PMMA. This modification reduces the heat transfer rate from the TPC to the veto volume and decreases the cooling of the adjacent liquid (cf.~figure~\ref{fig:Temperature_spatial_distribution_with_heat}~(top right)). In subsection~\ref{subsec:heat_transfer_lquid-gas_interface} we calculated this value to be $\SI{30.9}{W}$. A second update involves the gap between the cathode and the reflector panels, which is now estimated to be larger, up to $\SI{5}{mm}$. This region previously exhibited the highest fluid speeds (cf.~figure~\ref{fig:SpeedDistribution_with_heat}), which are expected to decrease by approximately a factor of three. In subsection~\ref{subsec:heat_transfer_lquid-gas_interface} we also discussed vertical gaps forming between the TPC walls due to thermal contraction. For mechanical reasons, these gaps are initially set at $\SI{5}{mm}$ at room temperature. However, additional PMMA corner brackets are attached externally, effectively sealing the gaps along the vertical dimension. This modification significantly limits the open cross-sectional area between the TPC and the veto volume. Consequently, the calculated upper limit for the heat transfer rate over the liquid-gas interface can be considered conservative.


\bibliographystyle{jhep}
\bibliography{main}

\providecommand{\href}[2]{#2}\begingroup\raggedright\begin{thebibliography}{10}

\bibitem{Majumdar:2021llu}
K.~Majumdar and K.~Mavrokoridis, \emph{{Review of Liquid Argon Detector
  Technologies in the Neutrino Sector}},
  \href{http://dx.doi.org/10.3390/app11062455}{\emph{Appl. Sciences} {\bfseries
  11} (2021) 2455}, [\href{https://arxiv.org/abs/2103.06395}{{\ttfamily
  2103.06395}}].

\bibitem{EXO-200:2019rkq}
{\scshape EXO-200} collaboration, G.~Anton et~al., \emph{{Search for
  Neutrinoless Double-$\beta$ Decay with the Complete EXO-200 Dataset}},
  \href{http://dx.doi.org/10.1103/PhysRevLett.123.161802}{\emph{Phys. Rev.
  Lett.} {\bfseries 123} (2019) 161802},
  [\href{https://arxiv.org/abs/1906.02723}{{\ttfamily 1906.02723}}].

\bibitem{McDonald:2024osu}
A.~B. McDonald, \emph{{Dark matter detection with liquid argon}},
  \href{http://dx.doi.org/10.1016/j.nuclphysb.2024.116436}{\emph{Nucl. Phys. B}
  {\bfseries 1003} (2024) 116436}.

\bibitem{Baudis:2023pzu}
L.~Baudis, \emph{{Dual-phase xenon time projection chambers for~rare-event
  searches}}, \href{http://dx.doi.org/10.1098/rsta.2023.0083}{\emph{Phil.
  Trans. Roy. Soc. Lond. A} {\bfseries 382} (2023) 20230083},
  [\href{https://arxiv.org/abs/2311.05320}{{\ttfamily 2311.05320}}].

\bibitem{Cennini:1993abz}
P.~Cennini et~al., \emph{{Argon purification in the liquid phase}},
  \href{http://dx.doi.org/10.1016/0168-9002(93)91209-6}{\emph{Nucl. Instrum.
  Meth. A} {\bfseries 333} (1993) 567--570}.

\bibitem{Plante:2022khm}
G.~Plante, E.~Aprile, J.~Howlett and Y.~Zhang, \emph{{Liquid-phase purification
  for multi-tonne xenon detectors}},
  \href{http://dx.doi.org/10.1140/epjc/s10052-022-10832-w}{\emph{Eur. Phys. J.
  C} {\bfseries 82} (2022) 860},
  [\href{https://arxiv.org/abs/2205.07336}{{\ttfamily 2205.07336}}].

\bibitem{Vogl:2023hbg}
C.~Vogl, M.~Schwarz, P.~Krause, G.~Zuzel and S.~Sch\"onert, \emph{{A
  liquid-phase loop-mode argon purification system}},
  \href{http://dx.doi.org/10.1088/1748-0221/19/03/C03030}{\emph{JINST}
  {\bfseries 19} (2024) C03030},
  [\href{https://arxiv.org/abs/2311.18697}{{\ttfamily 2311.18697}}].

\bibitem{XENON100:2017gsw}
{\scshape XENON100} collaboration, E.~Aprile et~al., \emph{{Online$^{222}$ Rn
  removal by cryogenic distillation in the XENON100 experiment}},
  \href{http://dx.doi.org/10.1140/epjc/s10052-017-4902-x}{\emph{Eur. Phys. J.
  C} {\bfseries 77} (2017) 358},
  [\href{https://arxiv.org/abs/1702.06942}{{\ttfamily 1702.06942}}].

\bibitem{XENON:2021fkt}
{\scshape XENON} collaboration, E.~Aprile et~al., \emph{{Application and
  modeling of an online distillation method to reduce krypton and argon in
  XENON1T}}, \href{http://dx.doi.org/10.1093/ptep/ptac074}{\emph{PTEP}
  {\bfseries 2022} (2022) 053H01},
  [\href{https://arxiv.org/abs/2112.12231}{{\ttfamily 2112.12231}}].

\bibitem{Murra:2022mlr}
M.~Murra, D.~Schulte, C.~Huhmann and C.~Weinheimer, \emph{{Design, construction
  and commissioning of a high-flow radon removal system for XENONnT}},
  \href{http://dx.doi.org/10.1140/epjc/s10052-022-11001-9}{\emph{Eur. Phys. J.
  C} {\bfseries 82} (2022) 1104},
  [\href{https://arxiv.org/abs/2205.11492}{{\ttfamily 2205.11492}}].

\bibitem{ABE201250}
{\scshape XMASS} collaboration, K.~Abe et~al., \emph{Radon removal from gaseous
  xenon with activated charcoal},
  \href{http://dx.doi.org/https://doi.org/10.1016/j.nima.2011.09.051}{\emph{Nucl.
  Instrum. Methods Phys. Res.} {\bfseries 661} (2012) 50--57}.

\bibitem{Pushkin:2018wdl}
K.~Pushkin et~al., \emph{{Study of radon reduction in gases for rare event
  search experiments}},
  \href{http://dx.doi.org/10.1016/j.nima.2018.06.076}{\emph{Nucl. Instrum.
  Meth. A} {\bfseries 903} (2018) 267--276},
  [\href{https://arxiv.org/abs/1805.11306}{{\ttfamily 1805.11306}}].

\bibitem{XENON:2024lbh}
{\scshape XENON} collaboration, E.~Aprile et~al., \emph{{Offline tagging of
  radon-induced backgrounds in XENON1T and applicability to other liquid xenon
  time projection chambers}},
  \href{http://dx.doi.org/10.1103/PhysRevD.110.012011}{\emph{Phys. Rev. D}
  {\bfseries 110} (2024) 012011},
  [\href{https://arxiv.org/abs/2403.14878}{{\ttfamily 2403.14878}}].

\bibitem{LZCollaboration:2024lux}
{\scshape LZ} collaboration, J.~Aalbers et~al., \emph{{Dark Matter Search
  Results from 4.2 Tonne-Years of Exposure of the LUX-ZEPLIN (LZ) Experiment}},
   \href{https://arxiv.org/abs/2410.17036}{{\ttfamily 2410.17036}}.

\bibitem{DarkSide-20k:UArCryo}
{\scshape DarkSide-20k} collaboration, F.~Acerbi et~al., \emph{{Benchmarking
  the design of the cryogenics system for the underground argon in
  DarkSide-20k}},
  \href{http://dx.doi.org/10.1088/1748-0221/20/02/P02016}{\emph{JINST}
  {\bfseries 20} (2025) P02016},
  [\href{https://arxiv.org/abs/2408.14071}{{\ttfamily 2408.14071}}].

\bibitem{DUNE:2020txw}
{\scshape DUNE} collaboration, B.~Abi et~al., \emph{{Deep Underground Neutrino
  Experiment (DUNE), Far Detector Technical Design Report, Volume IV: Far
  Detector Single-phase Technology}},
  \href{http://dx.doi.org/10.1088/1748-0221/15/08/T08010}{\emph{JINST}
  {\bfseries 15} (2020) T08010},
  [\href{https://arxiv.org/abs/2002.03010}{{\ttfamily 2002.03010}}].

\bibitem{Tu:2023nyz}
S.~Z. Tu, C.~Jiang, T.~R. Junk and T.~Yang, \emph{{A numerical solver for
  investigating the space charge effect on the electric field in liquid argon
  time projection chambers}},
  \href{http://dx.doi.org/10.1088/1748-0221/18/06/P06022}{\emph{JINST}
  {\bfseries 18} (2023) P06022}.

\bibitem{Aalbers:2022dzr}
J.~Aalbers et~al., \emph{{A next-generation liquid xenon observatory for dark
  matter and neutrino physics}},
  \href{http://dx.doi.org/10.1088/1361-6471/ac841a}{\emph{J. Phys. G}
  {\bfseries 50} (2023) 013001},
  [\href{https://arxiv.org/abs/2203.02309}{{\ttfamily 2203.02309}}].

\bibitem{XLZD:2024nsu}
{\scshape XLZD} collaboration, J.~Aalbers et~al., \emph{{The XLZD Design Book:
  Towards the Next-Generation Liquid Xenon Observatory for Dark Matter and
  Neutrino Physics}},  \href{https://arxiv.org/abs/2410.17137}{{\ttfamily
  2410.17137}}.

\bibitem{DarkSide-20k:2017zyg}
{\scshape DarkSide-20k} collaboration, C.~E. Aalseth et~al.,
  \emph{{DarkSide-20k: A 20 tonne two-phase LAr TPC for direct dark matter
  detection at LNGS}},
  \href{http://dx.doi.org/10.1140/epjp/i2018-11973-4}{\emph{Eur. Phys. J. Plus}
  {\bfseries 133} (2018) 131},
  [\href{https://arxiv.org/abs/1707.08145}{{\ttfamily 1707.08145}}].

\bibitem{Zani:2024ybb}
{\scshape DarkSide-20k} collaboration, A.~Zani, \emph{{The DarkSide-20k
  experiment}},
  \href{http://dx.doi.org/10.1088/1748-0221/19/03/C03058}{\emph{JINST}
  {\bfseries 19} (2024) C03058},
  [\href{https://arxiv.org/abs/2402.07566}{{\ttfamily 2402.07566}}].

\bibitem{DarkSide:2015}
{\scshape DarkSide} collaboration, P.~Agnes et~al., \emph{{Results From the
  First Use of Low Radioactivity Argon in a Dark Matter Search}},
  \href{http://dx.doi.org/10.1103/PhysRevD.93.081101}{\emph{Phys. Rev. D}
  {\bfseries 93} (2016) 081101},
  [\href{https://arxiv.org/abs/1510.00702}{{\ttfamily 1510.00702}}].

\bibitem{DUNE:2021hwx}
{\scshape DUNE} collaboration, A.~A. Abud et~al., \emph{{Design, construction
  and operation of the ProtoDUNE-SP Liquid Argon TPC}},
  \href{http://dx.doi.org/10.1088/1748-0221/17/01/P01005}{\emph{JINST}
  {\bfseries 17} (2022) P01005},
  [\href{https://arxiv.org/abs/2108.01902}{{\ttfamily 2108.01902}}].

\bibitem{ANSYS}
\mbox{ANSYS, Inc.}, ``{ANSYS Fluent 2020 R2,
  \href{https://www.ansys.com/products/fluids/ansys-fluent}{https://www.ansys.com/products/fluids/ansys-fluent},
  accessed 2023, September 4}.''.

\bibitem{Levenspiel1999}
O.~Levenspiel, \emph{Chemical Reaction Engineering}.
\newblock John Wiley \& Sons, New York, 3rd~ed., 1999.

\bibitem{vanUffelen:2023tkl}
{\scshape DarkSide-20k} collaboration, M.~van Uffelen, \emph{{The simulation of
  the DarkSide-20k calibration}},  in \emph{{57th Rencontres de Moriond on
  Electroweak Interactions and Unified Theories}}, 2023.
\newblock \href{https://arxiv.org/abs/2311.07437}{{\ttfamily 2311.07437}}.

\bibitem{Lippincott:2009ea}
W.~H. Lippincott, S.~B. Cahn, D.~Gastler, L.~W. Kastens, E.~Kearns, D.~N.
  McKinsey et~al., \emph{{Calibration of liquid argon and neon detectors with
  $^{83}Kr^m$}},
  \href{http://dx.doi.org/10.1103/PhysRevC.81.045803}{\emph{Phys. Rev. C}
  {\bfseries 81} (2010) 045803},
  [\href{https://arxiv.org/abs/0911.5453}{{\ttfamily 0911.5453}}].

\bibitem{McCutchan:2015vcl}
E.~A. McCutchan, \emph{{Nuclear Data Sheets for A = 83}},
  \href{http://dx.doi.org/10.1016/j.nds.2015.02.002}{\emph{Nucl. Data Sheets}
  {\bfseries 125} (2015) 201--394}.

\bibitem{DANCKWERTS:1953}
P.~V. Danckwerts, \emph{Continuous flow systems: Distribution of residence
  times},
  \href{http://dx.doi.org/https://doi.org/10.1016/0009-2509(53)80001-1}{\emph{Chem.
  Eng. Sci.} {\bfseries 2} (1953) 1--13}.

\bibitem{Fogler}
H.~S. Fogler, \emph{Elements of chemical reaction engineering}.
\newblock Prentice-Hall international series in the physical and chemical
  engineering sciences. Prentice Hall PTR, Upper Saddle River, N.J., 3rd~ed.,
  1999.

\bibitem{UserGuide}
\mbox{ANSYS, Inc.}, \emph{{ANSYS Fluent User's Guide}}, 15.0~ed., 2013.

\bibitem{TheoryGuide}
\mbox{ANSYS, Inc.}, \emph{{ANSYS Fluent Theory Guide}}, 15.0~ed., 2013.

\bibitem{Li:2010}
G.~Li, A.~Mukhopadhyay, C.-Y. Cheng and Y.~Dai, \emph{{Various Approaches to
  Compute Fluid Residence Time in Mixing Systems}},  vol.~ASME 2010 3rd Joint
  US-European Fluids Engineering Summer Meeting: Volume 1, Symposia – Parts
  A, B, and C of \emph{Fluids Engineering Division Summer Meeting},
  pp.~295--304, 2010.
\newblock \href{http://dx.doi.org/10.1115/FEDSM-ICNMM2010-30771}{DOI}.

\bibitem{UDFManual}
\mbox{ANSYS, Inc.}, \emph{{ANSYS Fluent UDF Manual}}, 14.0~ed., 2011.

\bibitem{Cini-Castagnoli:1960}
G.~Cini‐Castagnoli and F.~P. Ricci, \emph{{Self‐Diffusion in Liquid
  Argon}}, \href{http://dx.doi.org/10.1063/1.1700899}{\emph{J. Chem. Phys.}
  {\bfseries 32} (1960) 19--20}.

\bibitem{Fisher:1972}
R.~A. {Fisher} and R.~O. {Watts}, \emph{{Calculated self-diffusion coefficients
  for liquid argon}}, \href{http://dx.doi.org/10.1071/PH720529}{\emph{Aust. J.
  Phys.} {\bfseries 25} (1972) 529}.

\bibitem{Baleo:2000}
J.-N. Baléo and P.~Le~Cloirec, \emph{Validating a prediction method of mean
  residence time spatial distributions},
  \href{http://dx.doi.org/https://doi.org/10.1002/aic.690460403}{\emph{AIChE
  Journal} {\bfseries 46} (2000) 675--683}.

\bibitem{Liu:2010}
M.~Liu and J.~N. Tilton, \emph{Spatial distributions of mean age and higher
  moments in steady continuous flows},
  \href{http://dx.doi.org/https://doi.org/10.1002/aic.12151}{\emph{AIChE
  Journal} {\bfseries 56} (2010) 2561--2572}.

\bibitem{NIST}
P.~J. Linstrom and W.~G. Mallard, eds., \emph{{NIST Chemistry WebBook, NIST
  Standard Reference Database Number 69}}, vol.~20899.
\newblock {National Institute of Standards and Technology, Gaithersburg MD,
  20899}, 2023,
  \href{http://dx.doi.org/https://doi.org/10.18434/T4D303}{https://doi.org/10.18434/T4D303}.

\bibitem{Gaur:1982}
U.~Gaur, S.~Lau, B.~B. Wunderlich and B.~Wunderlich, \emph{{Heat Capacity and
  Other Thermodynamic Properties of Linear Macromolecules VI. Acrylic
  Polymers}}, \href{http://dx.doi.org/10.1063/1.555671}{\emph{J. Phys. Chem.
  Ref. Data} {\bfseries 11} (1982) 1065--1089}.

\bibitem{Pyda2014:sm_ptd_athas_0010}
``{Poly(methyl methacrylate) (PMMA) Heat Capacity, Enthalpy, Entropy, Gibbs
  Energy: Datasheet from ``The Advanced THermal Analysis System (ATHAS)
  Databank -- Polymer Thermodynamics'' Release 2014 in SpringerMaterials
  (https://materials.springer.com/polymerthermodynamics/docs/athas{\_}0010)}.''

\bibitem{STEPHENS:1972}
R.~Stephens, G.~Cieloszyk and G.~Salinger, \emph{Thermal conductivity and
  specific heat of non-crystalline solids: Polystyrene and polymethyl
  methacrylate},
  \href{http://dx.doi.org/https://doi.org/10.1016/0375-9601(72)90483-5}{\emph{Phys.
  Lett. A} {\bfseries 38} (1972) 215--217}.

\bibitem{hartwig:1982}
G.~Hartwig and D.~Evans, \emph{Nonmetallic Materials and Composites at Low
  Temperature}.
\newblock Cryogenic Materials Series. Springer New York, 1986,
  \href{http://dx.doi.org/https://doi.org/10.1007/978-1-4899-2010-2}{https://doi.org/10.1007/978-1-4899-2010-2}.

\bibitem{Bradley:2013}
P.~Bradley and R.~Radebaugh, \emph{Properties of Selected Materials at
  Cryogenic Temperatures}.
\newblock CRC Press, Boca Raton, FL, 2013.

\bibitem{DUNE-doc-17481}
G.~J. Michna, S.~P. Gent, D.~Pederson and C.~Streff, \emph{{CFD analysis of the
  fluid, heat, and impurity flows in ProtoDUNE single phase detector}},
  {DUNE-doc-17481-v1 (2019)}.

\bibitem{PhysRevFluids.6.090502}
O.~Shishkina, \emph{{Rayleigh-B\'enard convection: The container shape
  matters}},
  \href{http://dx.doi.org/10.1103/PhysRevFluids.6.090502}{\emph{Phys. Rev.
  Fluids} {\bfseries 6} (Sep, 2021) 090502}.

\bibitem{XENON:2016rze}
{\scshape XENON} collaboration, E.~Aprile et~al., \emph{{Results from a
  Calibration of XENON100 Using a Source of Dissolved Radon-220}},
  \href{http://dx.doi.org/10.1103/PhysRevD.95.072008}{\emph{Phys. Rev. D}
  {\bfseries 95} (2017) 072008},
  [\href{https://arxiv.org/abs/1611.03585}{{\ttfamily 1611.03585}}].

\bibitem{Malling:2014oxk}
D.~C. Malling, \emph{{Measurement and Analysis of WIMP Detection Backgrounds,
  and Characterization and Performance of the Large Underground Xenon Dark
  Matter Search Experiment}}.
\newblock PhD thesis, Brown U., 2014.
\newblock 10.7301/Z0057D9F.

\bibitem{Vaartstra:2022}
G.~Vaartstra, Z.~Lu, J.~H. Lienhard and E.~N. Wang, \emph{{Revisiting the
  Schrage Equation for Kinetically Limited Evaporation and Condensation}},
  \href{http://dx.doi.org/10.1115/1.4054382}{\emph{J. Heat Transfer} {\bfseries
  144} (2022) 080802}.

\bibitem{Zhang_2017}
L.~Zhang, Y.~R. Li, L.~Q. Zhou and C.~M. Wu, \emph{Comparison study on the
  calculation formula of evaporation mass flux through the plane vapour-liquid
  interface}, \href{http://dx.doi.org/10.1088/1742-6596/925/1/012019}{\emph{J.
  Phys. Conf. Ser.} {\bfseries 925} (2017) 012019}.

\bibitem{Yasuoka:1994}
K.~Yasuoka, M.~Matsumoto and Y.~Kataoka, \emph{{Evaporation and condensation at
  a liquid surface. I. Argon}},
  \href{http://dx.doi.org/10.1063/1.468216}{\emph{J. Chem. Phys.} {\bfseries
  101} (1994) 7904--7911}.

\bibitem{Bergman}
T.~L. Bergman and A.~S. Lavine, \emph{Fundamentals of Heat and Mass Transfer}.
\newblock John Wiley \& Sons, Inc., 8th~ed., 2017.

\bibitem{Jaffer:2023}
A.~Jaffer, \emph{Natural convection heat transfer from an isothermal plate},
  \href{http://dx.doi.org/10.3390/thermo3010010}{\emph{Thermo} {\bfseries 3}
  (2023) 148--175}, [\href{https://arxiv.org/abs/2201.02612}{{\ttfamily
  2201.02612}}].

\bibitem{FUJII1972755}
T.~Fujii and H.~Imura, \emph{Natural-convection heat transfer from a plate with
  arbitrary inclination},
  \href{http://dx.doi.org/https://doi.org/10.1016/0017-9310(72)90118-4}{\emph{Int.
  J. Heat Mass Transf.} {\bfseries 15} (1972) 755--767}.

\bibitem{Rohsenow}
W.~M. Rohsenow, J.~P. Hartnett and Y.~I. Cho, eds., \emph{Handbook of Heat
  Transfer}.
\newblock McGraw-Hill Campanies, Inc., 3rd~ed., 1998.

\bibitem{Stansfield_1958}
D.~Stansfield, \emph{{The Surface Tensions of Liquid Argon and Nitrogen}},
  \href{http://dx.doi.org/10.1088/0370-1328/72/5/321}{\emph{Proc. Phys. Soc.}
  {\bfseries 72} (1958) 854}.

\bibitem{HOLLANDS1975879}
K.~Hollands, G.~Raithby and L.~Konicek, \emph{Correlation equations for free
  convection heat transfer in horizontal layers of air and water},
  \href{http://dx.doi.org/https://doi.org/10.1016/0017-9310(75)90179-9}{\emph{Int.
  J. Heat Mass Transf.} {\bfseries 18} (1975) 879--884}.

\end{thebibliography}\endgroup

\end{document}